\documentclass[pra,superscriptaddress,preprint,showpacs,nofootinbibfloatfix,amsmath,amsfonts,amssymb]{revtex4-2}
\usepackage{amsmath,amsfonts,amssymb,color}
\usepackage{amsthm}
\usepackage{leftidx}
\usepackage{graphicx}
\usepackage{xcolor}
\usepackage{dcolumn}
\usepackage{bm}
\usepackage{epstopdf}
\usepackage{epsfig}
\usepackage{environ}
\usepackage{pdfcomment}

\usepackage{float}
\usepackage[T1]{fontenc}
\usepackage[latin9]{inputenc}
\usepackage{setspace}
\usepackage{esint}
\usepackage{ulem}
\usepackage{physics}

\begin{document}
	

\title{Non-Hermitian Floquet Topological Matter -- A Review}

\author{Longwen Zhou}
\email{zhoulw13@u.nus.edu}
\affiliation{%
	College of Physics and Optoelectronic Engineering, Ocean University of China, Qingdao 266100, China
}
\affiliation{%
	Key Laboratory of Optics and Optoelectronics, Qingdao 266100, China
}
\affiliation{%
	Engineering Research Center of Advanced Marine Physical Instruments and Equipment, Ministry of Education, Qingdao 266100, China
}

\author{Da-Jian Zhang}
\email{
	zdj@sdu.edu.cn}
\affiliation{%
	Department of Physics, Shandong University, Jinan 250100, China
}

\date{\today}

\begin{abstract}
The past few years have witnessed a surge of interest in non-Hermitian Floquet topological matters due to their exotic properties resulting from the interplay between driving fields and non-Hermiticity. The present review sums up our studies on non-Hermitian Floquet topological matters in one and two spatial dimensions. We first give a bird's-eye view of the literature for clarifying the physical significance of non-Hermitian Floquet systems. We then introduce, in a pedagogical manner, a number of useful tools tailored for the study of non-Hermitian
Floquet systems and their topological properties. With the aid of these tools, we present typical examples
of non-Hermitian Floquet topological insulators, superconductors, and
quasicrystals, with a focus on their topological invariants, bulk-edge
correspondences, non-Hermitian skin effects, dynamical properties,
and localization transitions. We conclude this review by summarizing
our main findings and presenting our vision of future directions.
\end{abstract}

\maketitle

\newpage{}

\tableofcontents{}

\newpage{}

\section{Introduction\label{sec:Int}}

When a physical system is driven periodically in time, its properties
could be drastically modified, leading to new phases and phenomena
beyond the static limit \cite{FloBok01,FloBok02,FloBok03,FloBok04,FloBok05,FloBok06}.
One such example, which can be traced back to the early history of
human civilization, is the Archimedes screw pump. Under periodic operations,
the Archimedes screw could transfer water from low-lying rivers into
high-lying irrigation ditches, rather than letting the water to follow
its natural flowing tendency. Another example is the tide caused by
the combined effects of the gravitational forces exerted by the Moon
and its periodic orbiting around the Earth. In the quantum domain,
rich nonequilibrium features have been identified in periodically
driven systems over the past decades, such as the Rabi oscillation
\cite{Rabi1937,RabiBok1,RabiBok2}, stimulated Raman adiabatic passage
\cite{STIRAP1,STIRAP2,STIRAP3}, dynamical localization \cite{DL1,DL2,DL3},
Thouless pump \cite{Thouless1983,Thouless1984,ThoulessRev}, time
crystal \cite{DTC1,DTC2,DTC3}, Floquet topological phase \cite{FTP1,FTP2,FTP3}
and integer quantum Hall effect from chaos \cite{ChaosIQHE1,ChaosIQHE2,ChaosIQHE3}
(for more information, see the reviews \cite{FloRev01,FloRev02,FloRev03,FloRev04,FloRev05,FloRev06,FloRev07,FloRev08,FloRev09,FloRev10,FloRev11,FloRev12,FloRev13,FloRev14,FloRev15,FloRev16,FloRev17,FloRev18,FloRev19,FloRev20,FloRev21,FloRev22,FloRev23,FloRev24,FloRev25,FloRev26,FloRev27,FloRev28,FloRev29,DTCRev1,DTCRev2,DTCRev3,DTCRev4}
and the references therein). Among all these discoveries, the Floquet
topological matter stands out as one pivotal impetus in the study
of periodically driven quantum systems. Over the past twenty years,
it has attracted great attention in the context of quantum dynamics
\cite{FloRev06,FloRev09,FloRev10}, quantum simulation \cite{FloRev07,FloRev08,FloRev16},
condensed matter physics \cite{FloRev11,FloRev12,FloRev15}, and so
on.

The topological properties of a periodically driven system are mainly
coded in its Floquet states, which are eigenstates of the evolution
operator over a driving period. This one-period propagator is called
the Floquet operator. For a system described by the Hamiltonian $\hat{H}(t)=\hat{H}(t+T)$,
the Floquet operator can be expressed as $\hat{U}=\hat{\mathsf{T}}e^{-i\int_{t_{0}}^{t_{0}+T}\hat{H}(t)dt}$.
Here $t$ denotes time, $T$ is the driving period, $t_{0}$ denotes
the initial time of the period, $\hat{\mathsf{T}}$ performs the time
ordering, and we have set the Planck constant $\hbar=1$. Solving
the eigenvalue equation $\hat{U}|\psi\rangle=e^{-iE}|\psi\rangle$
gives us the Floquet eigenstate $|\psi\rangle$ and the quasienergy
$E$. The latter is a phase factor and defined modulus $2\pi$. Its
range $E\in[-\pi,\pi)$ is usually referred to as the Floquet quasienergy
Brillouin zone. Moreover, the quasienergy is stroboscopically conserved
and plays the role of ``energy'' in driven systems with only discrete-time
translational symmetries. If the system also obeys the spatial translational
symmetry, the quasienergy dispersion $E({\bf k})$ with respect to
the conserved quasimomentum ${\bf k}$ could also group into bands,
which are thus called Floquet bands. Under appropriate conditions,
these quasienergy bands could show nontrivial topological properties,
which can be captured by topological invariants of the corresponding
Floquet-Bloch states or the Floquet operator itself. Periodically
driven quantum systems could thus support a new class of nonequilibrium
phases, which is nowadays known as the Floquet topological matter.

Similar to static topological phases, Floquet topological phases could
also possess symmetry-protected edge states under open boundary conditions
and exhibit quantized dynamical or transport signals. However, three
key features distinguish Floquet topological phases from their static
counterparts. First, suitable driving fields could break the symmetry
dynamically and open gaps around the touching points of static energy
bands, yielding band inversions and topologically nontrivial bandstructures
\cite{FTP1}. This aspect of driving can usually be taken into account
by the Floquet effective Hamiltonians obtained from high-frequency
expansions of the driving potential \cite{FloHeff1,FloHeff2,FloHeff3,FloHeff4}.
Second, as the quasienergy $E$ is bounded from above ($E=\pi$) and
below ($E=-\pi$), Floquet bands could meet with each other at $E=\pm\pi$
and develop nontrivial windings around the whole quasienergy Brillouin
zone $E\in[-\pi,\pi)$. These spectral windings result in unique states
of matter in periodically driven systems, such as Floquet semimetals
with Floquet band holonomy \cite{FSM1,FSM2,FSM3}, degenerated edge
modes at $E=\pm\pi$ (anomalous Floquet $\pi$ modes) \cite{FPiMod1,FPiMod2,FPiMod3,FPiMod4}
and anomalous chiral edge states in Floquet topological insulators
\cite{AFTI1,AFTI2,AFTI3}, which have no counterparts in static systems.
Third, the driving field could assist in the formation of long-ranged,
and even spatially non-decaying coupling among different degrees of
freedom in a lattice \cite{FTP3}, leading to Floquet phases with
large topological invariants, rich topological transitions, and a substantial number of
topological edge states \cite{FTPLg01,FTPLg02,FTPLg03,FTPLg04,FTPLg05,FTPLg07,FTPLg08}.
These phases and boundary states go beyond the description of any
static tight-binding models in typical situations (i.e., with finite
ranged or spatially decaying hopping amplitudes). The investigation
of the above-mentioned features have not only led to new classification
schemes of Floquet matter in theories \cite{FTPCls1,FTPCls2,FTPCls3,FTPCls4}
but also promoted experimental realizations of numerous Floquet topological
phases in solid-state and quantum simulator setups \cite{FTPExp01,FTPExp02,FTPExp03,FTPExp04,FTPExp05,FTPExp06,FTPExp07,FTPExp08,FTPExp09,FTPExp10,FTPExp11,FTPExp12,FTPExp13,FTPExp14,FTPExp15,FTPExp16},
bringing new hope for applications in ultrafast electronics \cite{FloRev11}
and topological quantum computing \cite{FloQC1,FloQC2,FloQC3}.

Non-Hermitian physics deals with classical and open quantum systems
subject to measurements, dissipation, gain and loss or nonreciprocal
effects \cite{NHBok1,NHBok2,NHBok3,NHBok4}. A non-Hermitian system
can usually be modeled by a Hamiltonian $\hat{H}$ that is not self-adjoint,
i.e., $\hat{H}\neq\hat{H}^{\dagger}$. The spectrum of such a non-Hermitian
Hamiltonian is complex in general and the resulting dynamics is non-unitary.
In recent years, the study of topological phases in non-Hermitian
systems have attracted much attention from both theoretical and experimental
sides (see  Refs.~\cite{NHTPRev01,NHTPRev02,NHTPRev03,NHTPRev04,NHTPRev05,NHTPRev06,NHTPRev07,NHTPRev08,NHTPRev09,NHTPRev10,NHTPRev11,NHTPRev12,NHTPRev13,NHTPRev14,NHTPRev15,NHTPRev16,NHTPRev17,NHTPRev18,NHTPRev19,NHTPRev20,NHTPRev21,NHTPRev22,NHTPRev23,NHTPRev24,NHTPRev25}
for reviews). The investigation of these nonequilibrium phases may
help us deepening our understanding of topological matter in open
systems and optimize the design of noise-resilient quantum devices.

\begin{figure}
	\begin{centering}
		\includegraphics[scale=0.70]{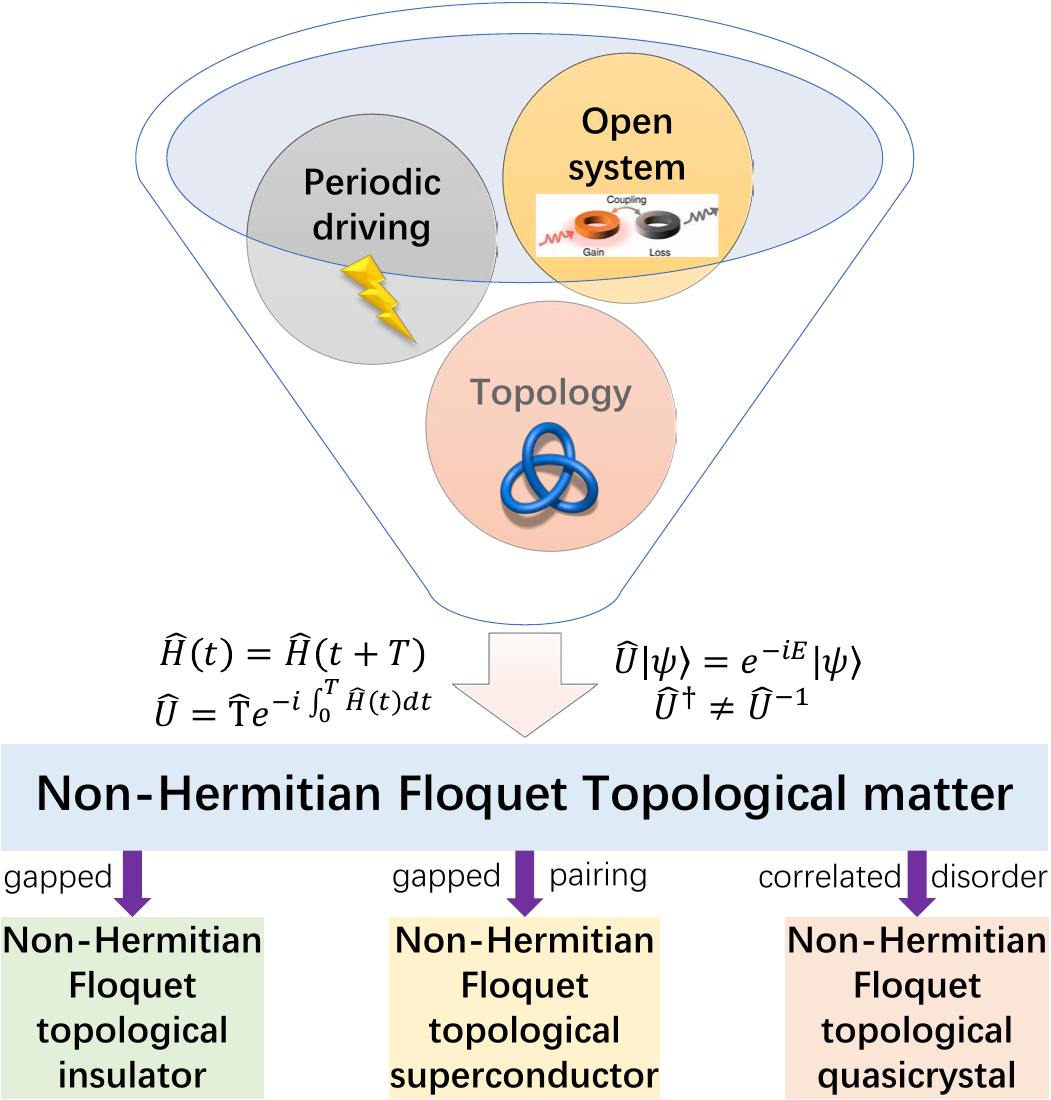}
		\par\end{centering}
	\caption{\textbf{Non-Hermitian Floquet topological matter}: a schematic diagram
		to illustrate the concepts and definitions of several typical phases. \label{fig:NHFTP}}
\end{figure}

The recent booming of non-Hermitian topological matter is driven by
a couple of key concepts, including the parity and time-reversal (PT) symmetry, the exceptional
point (EP), and the non-Hermitian skin effect (NHSE). The PT symmetry can appear in open systems with
balanced gain and loss. A PT-symmetric, non-Hermitian Hamiltonian
has a real spectrum in the PT unbroken regime \cite{Bender2002,Zhang2019}. The PT-symmetry
was thus viewed as a possible way to lift the restriction of Hermiticity
for Hamiltonians in its early studies \cite{NHBok4}.
Recently, it was found that the topological and transport properties
of a system may also change when it undergoes a PT-symmetry-breaking
transition, yielding phases unique to non-Hermitian Hamiltonians \cite{PTExp1,PTExp2,PTExp3}.
Models with the PT symmetry thus become focuses in the study of non-Hermitian
topological matter \cite{NHTPRev03,NHTPRev04,NHTPRev05,NHTPRev07,NHTPRev08,NHTPRev11,NHTPRev17,NHTPRev25}.
The combination of PT and other symmetries further results in various
classification schemes for non-Hermitian topological phases that go
beyond their Hermitian counterparts \cite{NHCls01,NHCls02,NHCls03,NHCls04,NHCls05,NHCls06,NHCls07,NHCls08,NHCls09}.
The EP is a class of level degeneracy point unique to non-Hermitian
operators. At an EP, the geometric and algebraic multiplicities of
a non-Hermitian matrix are different, leading to the breakdown of its
diagonalizability and the coalescence of its eigenvectors \cite{NHBok1,NHTPRev01,NHTPRev02}.
In recent years, various gapless band structures (nodal points, lines,
loops, surfaces, knots etc.) formed by EPs were discovered, giving rise to rich non-Hermitian topological (semi-)metallic phases with intriguing
transport properties \cite{NHTPRev09,NHTPRev11,NHTPRev15,NHTPRev16,NHTPRev18,XuPRL2017}.
Besides, EPs were also found to play key roles in the topological
energy transfer \cite{NHTET1,NHTET2,NHTET3}, high-precision sensing
\cite{EPSense1,EPSense2,EPSense3,EPSense4}, topological lasers \cite{EPLas1,EPLas2,EPLas3}
and strongly correlated phases \cite{NHTPRev15,NHTPRev16}. The NHSE
refers to the accumulation of bulk states around the edges of
an open-boundary non-Hermitian lattice. It highlights the extreme
sensitivity of non-Hermitian physics to the boundary condition of
a system \cite{NHTPRev13,NHTPRev19,NHTPRev24}. This phenomenon not
only blurs the distinction between bulk and edge states in non-Hermitian
models but also breaks the well-established bulk-boundary correspondence
in Hermitian topological matter \cite{NHSE01,NHSE02,NHSE03,NHSE04,NHSE05}.
Over the past few years, a couple of theoretical frameworks have been
introduced to characterize the NHSE and its related topological phenomena
\cite{NHSE06,NHSE07,NHSE08,NHSE09,NHSE10,NHSE11,NHSE12,NHSE13,NHSE14,NHSE15,NHSE16},
which are accompanied by the experimental observations of NHSE in
AMO systems, electrical circuits and metamaterials \cite{SkinExp01,SkinExp02,SkinExp03,SkinExp04,SkinExp05,SkinExp06,SkinExp07}.
Entanglement transitions associated with the NHSE was also identified
in a recent study \cite{SkinESEE}. Further progresses have
been made in the study of non-Hermitian topological matter in randomly
disordered \cite{NHDsod01,NHDsod02,NHDsod03,NHDsod04,NHDsod05}, quasiperiodic
\cite{NHQC01,NHQC02,NHQC03,NHQC04,NHQC05} and many-body systems \cite{NHMBP01,NHMBP02,NHMBP03,NHMBP04,NHMBP05,NHMBP06,NHMBP07,NHMBP08,NHMBP09,NHMBP10,NHMBP11,NHMBP12,NHMBP13,NHMBP14,NHMBP15}.

With all these developments, a natural follow-up is to consider the
system in a more general situation, in which it is subject to both
time-periodic drivings and non-Hermitian effects. Such non-Hermitian
Floquet systems may possess exotic dynamical phenomena and topological
phases with no static or Hermitian analogies. On the theoretical side,
the investigation of driven non-Hermitian systems may lead to the
discovery of new topological states and bring about the improvement
of classification schemes for nonequilibrium phases of matter in general.
On the practical side, the exploration of non-Hermitian Floquet matter
is helpful to the design of new approaches for preparing or stabilizing
topologically nontrivial states and controlling material properties.
It also stimulates new ideas for realizing quantum devices and quantum
computing protocols that are robust to perturbations caused by the
environment. Though still at the early stage, many progresses have
been made in the realization and characterization of non-Hermitian
Floquet phases \cite{NHFTP01,NHFTP02,NHFTP03,NHFTP04,NHFTP05,NHFTP06,NHFTP07,NHFTP08,NHFTP09,NHFTP10,NHFTP11,NHFTP12,NHFTP13,NHFTP14,NHFTP15,NHFTP16,NHFTP17,NHFTP18,NHFTP19,NHFTP20,NHFTP21,NHFTP22,NHFTP23,NHFTP24,NHFTP25,NHFTP26,NHFTP27,NHFTP28,NHFTP29,NHFTP30,NHFTP31,NHFTP32,NHFTP33,NHFTP34,NHFTP35,NHFTP352,NHFTP36,NHFTP37,NHFTP38,NHFTP39,NHFTP40,NHFTP41,NHFTP42,NHFTP43,NHFTP44,NHFTP45}.
In this review, we limit our scope to the discussion of a number of typical
topological phases we discovered in non-Hermitian Floquet systems
\cite{ZhouNHFTP01,ZhouNHFTP02,ZhouNHFTP03,ZhouNHFTP04,ZhouNHFTP05,ZhouNHFTP06,ZhouNHFTP07,ZhouNHFTP08,ZhouNHFTP09,ZhouNHFTP10,ZhouNHFTP11}.
In Sec.~\ref{sec:Bac}, we give a pedagogical introduction to some
key aspects of Floquet systems, including their dynamical and topological
characterizations. In Sec.~\ref{sec:Res}, we present typical examples
of non-Hermitian Floquet topological insulators, superconductors,
quasicrystals and summarize their main physical properties, with a
focus on the features that are unique to driven non-Hermitian systems.
In Sec.~\ref{sec:Sum}, we conclude this review, briefly mention some
relevant studies and discuss potential future work.
\section{Backgrounds\label{sec:Bac}}

We start with a recap of the basic description of a non-Hermitian
Floquet system. The Hamiltonian of such a system satisfies $\hat{H}(t)=\hat{H}(t+T)$,
and there exists $t\in[0,T]$ such that $\hat{H}(t)\neq[\hat{H}(t)]^{\dagger}$.
Here $t$ denotes time and $T$ is the driving period. The state of the system evolves according to the Schr\"odinger
equation
\begin{equation}
	i\frac{\partial}{\partial t}|\Psi(t)\rangle=\hat{H}(t)|\Psi(t)\rangle,\label{eq:Seq}
\end{equation}
where we have set $\hbar=1$. We first show that this equation can
be solved by Floquet states even though $\hat{H}(t)$ is non-Hermitian.
This is followed by different ways of obtaining the Floquet states
in general situations, in the high-frequency regime, and in the adiabatic
regime. We next discuss the symmetry, topological invariants, and dynamical
characterizations of non-Hermitian Floquet states, with a focus on
the types of physical systems explored in our previous studies. We conclude
this section by presenting some tools for characterizing the spectrum properties
and localization transitions in non-Hermitian Floquet disordered systems.

\subsection{Floquet theorem\label{subsec:FloThem}}

We sketch a proof of the Floquet theorem in this subsection \cite{ZhouThesis2015}.
It follows the proof of the Bloch theorem for waves in one-dimensional periodic lattices \cite{GrossoBook}.
Applying the Fourier expansion to our time-periodic Hamiltonian $\hat{H}(t)$
in the time-frequency domain, we find
\begin{equation}
	\hat{H}(t)=\sum_{n}\hat{H}(\omega_{n})e^{i\omega_{n}t},\qquad\omega_{n}=n\omega=n\frac{2\pi}{T},\qquad n\in\mathbb{Z}.
\end{equation}
Here $\omega$ is the driving frequency. For an infinite system, the
set of plane waves $\{|\psi_{\varepsilon}(t)\rangle=|\varphi({\bf r})\rangle e^{-i\varepsilon t}\}$
can be chosen as a suitable basis. We can write down a matrix expression
for $\hat{H}(t)$ in the orthonormal and complete basis $\{|\psi_{\varepsilon}(t)\rangle\}$.
Acting $\hat{H}(t)$ on $|\psi_{\varepsilon}(t)\rangle$, we obtain
\begin{equation}
	\hat{H}(t)|\psi_{\varepsilon}(t)\rangle=\sum_{n}\hat{H}(\omega_{n})|\varphi({\bf r})\rangle e^{-i(\varepsilon-\omega_{n})t}=\sum_{n}\hat{H}(\omega_{n})|\psi_{\varepsilon-\omega_{n}}(t)\rangle.
\end{equation}
For any given $n\in\mathbb{Z}$, the state $|\psi_{\varepsilon-\omega_{n}}(t)\rangle$
belongs to the subspace
\begin{equation}
	S_{\varepsilon}=\{|\psi_{\varepsilon}(t)\rangle,|\psi_{\varepsilon\pm\omega}(t)\rangle,|\psi_{\varepsilon\pm2\omega}(t)\rangle,...,|\psi_{\varepsilon\pm n\omega}(t)\rangle,...\}.
\end{equation}
It is clear that any two subspaces $S_{\varepsilon}$ and $S_{\varepsilon'}$
$(\varepsilon'\neq\varepsilon)$ are decoupled under the action of
$\hat{H}(t)$ if Re$(\varepsilon-\varepsilon')\in(-\omega,\omega)$. Moreover,
$S_{\varepsilon}$ and $S_{\varepsilon'}$ are equivalent if there
exists an $n\in\mathbb{Z}$ such that $\varepsilon-\varepsilon'=n\omega$.
Therefore, we can study the dynamics of the system separately in each
subspace $S_{\varepsilon}$ for ${\rm Re}(\varepsilon)\in[-\pi/T,\pi/T)$,
which is usually called the first quasienergy Brillouin zone (BZ).
The quasienergy $\varepsilon$ is thus a conserved quantity due to
the discrete-time translational symmetry of $\hat{H}(t)$, similar
to the conserved quasimomentum due to the discrete-space translational
symmetry of a static Hamiltonian. In the subspace $S_{\varepsilon}$,
a solution of the Schr\"odinger equation can be written as
\begin{equation}
	|\Psi_{\varepsilon}(t)\rangle=\sum_{n}c(\varepsilon+\omega_{n})|\psi_{\varepsilon+\omega_{n}}(t)\rangle=e^{-i\varepsilon t}\sum_{n}c(\varepsilon+\omega_{n})|\varphi({\bf r})\rangle e^{-i\omega_{n}t},\label{eq:PsiEt}
\end{equation}
where $\{c(\varepsilon+\omega_{n})\}$ are complex coefficients. It
is clear that Eq.~(\ref{eq:PsiEt}) possesses a time-periodic
component
\begin{equation}
	|u_{\varepsilon}(t)\rangle\equiv\sum_{n}c(\varepsilon+\omega_{n})|\varphi({\bf r})\rangle e^{-i\omega_{n}t}=|u_{\varepsilon}(t+T)\rangle.
\end{equation}
Therefore, we can express the general solution $|\Psi_{\varepsilon}(t)\rangle$
as the product of an oscillating phase factor $e^{-i\varepsilon t}$
and a time-periodic Floquet mode $|u_{\varepsilon}(t)\rangle$, i.e.,
\begin{equation}
	|\Psi_{\varepsilon}(t)\rangle=e^{-i\varepsilon t}|u_{\varepsilon}(t)\rangle.\label{eq:FloStat1}
\end{equation}
It also implies that 
\begin{equation}
	|\Psi_{\varepsilon}(t+T)\rangle=e^{-i\varepsilon T}|\Psi_{\varepsilon}(t)\rangle.\label{eq:FloStat2}
\end{equation}
The latter equation indicates that the only change of the state $|\Psi_{\varepsilon}(t)\rangle$
after undergoing a one-period evolution is to pick up an exponential
factor $e^{-i\varepsilon T}$. We refer to the set $\{|\Psi_{\varepsilon}(t)\rangle|{\rm Re}(\varepsilon)\in[-\pi/T,\pi/T)\}$
as Floquet eigenstates of the system. They form an orthonormal and complete basis at
each instant of time $t$.

Since the evolution from $|\Psi_{\varepsilon}(t)\rangle$ to $|\Psi_{\varepsilon}(t+T)\rangle$
is governed by the Schr\"odinger equation, we can also
express Eq.~(\ref{eq:FloStat2}) as 
\begin{equation}
	\hat{U}(t+T,t)|\Psi_{\varepsilon}(t)\rangle=e^{-i\varepsilon T}|\Psi_{\varepsilon}(t)\rangle,\label{eq:FE1}
\end{equation}
where $\hat{U}(t+T,t)=\hat{\mathsf{T}}e^{-i\int_{t}^{t+T}\hat{H}(t')dt'}$
is nothing but the Floquet operator (evolution operator over one driving
period) of the system. When we are only concerned with the stroboscopic
dynamics, the initial time dependence of Eq.~(\ref{eq:FE1}) is not
important. In this case, we can set $t=0$ in Eq.~(\ref{eq:FE1})
and express the Floquet eigenvalue equation as
\begin{equation}
	\hat{U}|\Psi_{E}\rangle=e^{-iE}|\Psi_{E}\rangle,\label{eq:FE2}
\end{equation}
where $\hat{U}=\hat{\mathsf{T}}e^{-i\int_{0}^{T}\hat{H}(t)dt}$ and
we have introduced $E=\varepsilon T$ as the dimensionless quasienergy,
whose first BZ is given by $[-\pi,\pi)$. To sum up, we find that
the solution of Eq.~(\ref{eq:Seq}) with a time-periodic $\hat{H}(t)$
has the form of Eq.~(\ref{eq:FloStat1}) or (\ref{eq:FloStat2}),
where $|\Psi_{\varepsilon}(t)\rangle$ is an eigenstate of the Floquet
operator of the system. For stroboscopic observations, all the Floquet
eigenstates can thus be obtained by solving the eigenvalue equation
(\ref{eq:FE2}) of Floquet operator $\hat{U}$. Due to the completeness
of Floquet eigenstate basis $\{|\Psi_{E}\rangle\}$, we can expand
an arbitrary initial state of the system as $|\Psi(0)\rangle=\sum_{E}c_{E}|\Psi_{E}\rangle.$
The resulting state after an evolution over $n$ driving periods is
then given by
\begin{equation}
	|\Psi(t)\rangle=\sum_{E}c_{E}e^{-inE}|\Psi_{E}\rangle.\label{eq:SeqSol}
\end{equation}
Note that for a non-Hermitian $\hat{H}(t)$, $\hat{U}$ is generally
non-unitary and the quasienergy $E$ may have a nonvanishing imaginary
part. In this case, the real part of $E$ still belongs to the range
of $[-\pi,\pi)$ and our arguments leading to the general solution
(\ref{eq:SeqSol}) can be satisfied.

\subsection{Floquet eigenvalue equation\label{subsec:Feq}}
In the most general situations, we can solve Eq.~(\ref{eq:FE2}) numerically
by the split-operator method \cite{FloBok03}. Dividing the evolution
periodic $T$ into $N$ segments with a large enough $N$,
we can express the Floquet operator $\hat{U}$ approximately as
\begin{equation}
	\hat{U}\simeq e^{-i\hat{H}((N-1)\Delta t)\Delta t}e^{-i\hat{H}((N-2)\Delta t)\Delta t}\cdots e^{-i\hat{H}(\Delta t)\Delta t}e^{-i\hat{H}(0)\Delta t},\label{eq:Uapp}
\end{equation}
where $\Delta t=T/N$. Over each small time interval $\Delta t$,
$\hat{H}(t)$ is approximately time-independent and we can diagonalize
it numerically at $t=\ell\Delta t$ as
\begin{equation}
	\hat{H}(\ell\Delta t)=V_{\ell}D_{\ell}V_{\ell}^{-1}.
\end{equation}
Each column of $V_{\ell}$ represents a right eigenvector of $\hat{H}(\ell\Delta t)$.
The evolution operator over the one-time interval $\Delta t$ then takes
the form
\begin{equation}
	e^{-i\hat{H}(\ell\Delta t)\Delta t}=V_{\ell}e^{-iD_{\ell}\Delta t}V_{\ell}^{-1},\qquad\ell=0,1,...,N-1.
\end{equation}
The multiplication of all the $V_{\ell}e^{-iD_{\ell}\Delta t}V_{\ell}^{-1}$
from right to left for $\ell=0$ to $N-1$ yields Eq.~(\ref{eq:Uapp}),
which further converges to the exact Floquet operator in the limit
$N\rightarrow\infty$. We can thus numerically solve the Floquet eigenvalue
equation by diagonalizing the approximated $\hat{U}$ in Eq.~(\ref{eq:Uapp}).
This approach works in principle for systems with any individual or multiple driving frequencies.
But it may become time-consuming in practice for certain continuously or slowly driven systems.

When the driving field takes the form of periodic kicking or quenching,
the series in Eq.~(\ref{eq:Uapp}) can be greatly simplified and even
obtained exactly. Here we give several examples. Consider a time-periodic
Hamiltonian of the form
\begin{equation}
	\hat{H}(t)=\hat{H}_{0}+\sum_{\ell\in\mathbb{Z}}\delta(t/T-\ell)\hat{H}_{1},\label{eq:KickH}
\end{equation}
where $\delta(t/T-\ell)$ is the delta function peaked at $t=\ell T$,
i.e., each integer multiple of the driving period. The dynamics over
each driving period thus constitutes a free evolution part controlled
by $\hat{H}_{0}$ and a delta kicking force controlled by $\hat{H}_{1}$.
The quantum kicked rotor is one representative example of such a system
\cite{FloRev01}. Focusing on the one-period evolution from $t=\ell T-0^{+}$
to $t=(\ell+1)T-0^{+}$, we find the Floquet operator of $\hat{H}(t)$
to be
\begin{alignat}{1}
	\hat{U}= & e^{-i\int_{\ell T+0^{+}}^{(\ell+1)T-0^{+}}\hat{H}(t)dt}e^{-i\int_{\ell T-0^{+}}^{\ell T+0^{+}}\hat{H}(t)dt}\nonumber \\
	= & e^{-i\int_{\ell T+0^{+}}^{(\ell+1)T-0^{+}}\hat{H}_{0}dt}e^{-i\int_{\ell T-0^{+}}^{\ell T+0^{+}}\delta(t/T-\ell)\hat{H}_{1}dt}\nonumber \\
	= & e^{-i\hat{H}_{0}T}e^{-i\hat{H}_{1}T}.\label{eq:KickU}
\end{alignat}
Similarly, if there are two kicks separated by a time interval $\tau$
within each driving period, the Hamiltonian could take the form of
\begin{equation}
	\hat{H}(t)=\hat{H}_{0}+\sum_{\ell\in\mathbb{Z}}\delta(t/T-\tau/T-\ell)\hat{H}_{2}+\sum_{\ell\in\mathbb{Z}}\delta(t/T-\ell)\hat{H}_{1}.
\end{equation}
For the evolution from time $t=\ell T-0^{+}$ to $t=(\ell+1)T-0^{+}$,
the Floquet operator now takes the form of
\begin{equation}
	\hat{U}=e^{-i\hat{H}_{0}(T-\tau)}e^{-i\hat{H}_{2}T}e^{-i\hat{H}_{0}\tau}e^{-i\hat{H}_{1}T}.
\end{equation}
One typical example of such a system is the double-kicked quantum
rotor \cite{FTP3}. For a periodically quenched Hamiltonian in the
form of
\begin{equation}
	\hat{H}(t)=\begin{cases}
		\hat{H}_{1} & t\in[\ell T,\ell T+T_{1})\\
		\hat{H}_{2} & t\in[\ell T+T_{1},\ell T+T_{1}+T_{2})
	\end{cases},\label{eq:QuenchH}
\end{equation}
where $T=T_{1}+T_{2}$, we can also directly obtain the corresponding
Floquet operator from $t=\ell T-0^{+}$ to $t=(\ell+1)T-0^{+}$ as
\begin{alignat}{1}
	\hat{U}= & e^{-i\int_{\ell T+T_{1}}^{\ell T+T-0^{+}}\hat{H}(t)dt}e^{-i\int_{\ell T-0^{+}}^{\ell T+T_{1}-0^{+}}\hat{H}(t)dt}\nonumber \\
	= & e^{-i\int_{\ell T+T_{1}}^{\ell T+T-0^{+}}\hat{H}_{2}dt}e^{-i\int_{\ell T-0^{+}}^{\ell T+T_{1}-0^{+}}\hat{H}_{1}dt}\nonumber \\
	= & e^{-i\hat{H}_{2}T_{2}}e^{-i\hat{H}_{1}T_{1}}.\label{eq:QuenchU}
\end{alignat}
The discrete-time quantum walk can be viewed as one example of such
a periodically quenched system \cite{FloRev04}. Time-periodic quenches
are also frequently implemented in the study of discrete time crystals
\cite{DTC1,DTC2,DTC3}. When $[{\hat H}_1,{\hat H}_2]\neq0$, the quenches (or kicks)
may effectively generate long-range coupling in the system according to the
Baker-Campbell-Hausdorff formula, leading to Floquet phases with large topological
invariants and many boundary states. This point will be explicitly demonstrated by
the examples discussed in Sec.~\ref{sec:Res}.

For a continuously driven system, the solution of the Floquet eigenvalue
equation may also be obtained approximately in terms of the frequency
(Sambe) space formalism \cite{SambeSpace1,SambeSpace2,SambeSpace3}.
Inserting the Floquet state in Eq.~(\ref{eq:FloStat1}) into the Schr\"odinger
equation (\ref{eq:Seq}) and reorganizing the terms, we find
\begin{equation}
	\left[\hat{H}(t)-i\frac{\partial}{\partial t}\right]|u_{\varepsilon}(t)\rangle=\varepsilon|u_{\varepsilon}(t)\rangle.
\end{equation}
Using the Fourier expansion of $\hat{H}(t)$ and $|u_{\varepsilon}(t)\rangle=\sum_{n}e^{-i\omega_{n}t}|u_{\varepsilon}(\omega_{n})\rangle$,
we further obtain
\begin{equation}\label{eq:22}
	\sum_{m,n}\hat{H}(\omega_{m})e^{i(\omega_{m}-\omega_{n})t}|u_{\varepsilon}(\omega_{n})\rangle-\sum_{n}\omega_{n}e^{-i\omega_{n}t}|u_{\varepsilon}(\omega_{n})\rangle=\varepsilon\sum_{n}e^{-i\omega_{n}t}|u_{\varepsilon}(\omega_{n})\rangle.
\end{equation}
Multiplying $\frac{1}{T}e^{i\omega_{\ell}t}$ from the left on both
sides of Eq.~(\ref{eq:22}) and performing the integral over a driving period $T$, we
arrive at
\begin{equation}
	\sum_{m}(\hat{H}_{m-n}-\omega_{m}\delta_{m,n})|u_{\varepsilon}(\omega_{m})\rangle=\varepsilon|u_{\varepsilon}(\omega_{n})\rangle,\label{eq:FE3}
\end{equation}
where
\begin{equation}
	\hat{H}_{m-n}\equiv\hat{H}(\omega_{m-n})=\frac{1}{T}\int_{0}^{T}e^{-i\omega_{m-n}t}\hat{H}(t)dt,
\end{equation}
and $\omega_{m-n}=(m-n)\omega=(m-n)\frac{2\pi}{T}$. Equation (\ref{eq:FE3})
is an infinite-dimensional matrix equation of the form
\begin{equation}
	\begin{bmatrix}\ddots & \ddots & \ddots & \ddots & \ddots & \ddots & \ddots\\
		\ddots & \hat{H}_{0}+2\omega & \hat{H}_{1} & \hat{H}_{2} & \hat{H}_{3} & \hat{H}_{4} & \ddots\\
		\ddots & \hat{H}_{-1} & \hat{H}_{0}+\omega & \hat{H}_{1} & \hat{H}_{2} & \hat{H}_{3} & \ddots\\
		\ddots & \hat{H}_{-2} & \hat{H}_{-1} & \hat{H}_{0} & \hat{H}_{1} & \hat{H}_{2} & \ddots\\
		\ddots & \hat{H}_{-3} & \hat{H}_{-2} & \hat{H}_{-1} & \hat{H}_{0}-\omega & \hat{H}_{1} & \ddots\\
		\ddots & \hat{H}_{-4} & \hat{H}_{-3} & \hat{H}_{-2} & \hat{H}_{-1} & \hat{H}_{0}-2\omega & \ddots\\
		\ddots & \ddots & \ddots & \ddots & \ddots & \ddots & \ddots
	\end{bmatrix}\begin{bmatrix}\vdots\\
		|u_{\varepsilon}(\omega_{-2})\rangle\\
		|u_{\varepsilon}(\omega_{-1})\rangle\\
		|u_{\varepsilon}(\omega_{0})\rangle\\
		|u_{\varepsilon}(\omega_{1})\rangle\\
		|u_{\varepsilon}(\omega_{2})\rangle\\
		\vdots
	\end{bmatrix}=\varepsilon\begin{bmatrix}\vdots\\
		|u_{\varepsilon}(\omega_{-2})\rangle\\
		|u_{\varepsilon}(\omega_{-1})\rangle\\
		|u_{\varepsilon}(\omega_{0})\rangle\\
		|u_{\varepsilon}(\omega_{1})\rangle\\
		|u_{\varepsilon}(\omega_{2})\rangle\\
		\vdots
	\end{bmatrix}.\label{eq:FE4}
\end{equation}
Note that each $\hat{H}_{m-n}$ has the same Hilbert space dimension
$d$ as the original Hamiltonian $\hat{H}(t)$. As an example, for a harmonically
driven system described by the Hamiltonian
\begin{equation}
	\hat{H}(t)=\hat{H}_{0}+\hat{V}e^{i\omega t}+\hat{W}e^{-i\omega t},
\end{equation}
the above equation reduces to the following block tridiagonal form
\begin{equation}
	\begin{bmatrix}\ddots & \ddots & \ddots & \ddots & \ddots & \ddots & \ddots\\
		\ddots & \hat{H}_{0}+2\omega & \hat{V} & 0 & 0 & 0 & \ddots\\
		\ddots & \hat{W} & \hat{H}_{0}+\omega & \hat{V} & 0 & 0 & \ddots\\
		\ddots & 0 & \hat{W} & \hat{H}_{0} & \hat{V} & 0 & \ddots\\
		\ddots & 0 & 0 & \hat{W} & \hat{H}_{0}-\omega & \hat{V} & \ddots\\
		\ddots & 0 & 0 & 0 & \hat{W} & \hat{H}_{0}-2\omega & \ddots\\
		\ddots & \ddots & \ddots & \ddots & \ddots & \ddots & \ddots
	\end{bmatrix}\begin{bmatrix}\vdots\\
		|u_{\varepsilon}(\omega_{-2})\rangle\\
		|u_{\varepsilon}(\omega_{-1})\rangle\\
		|u_{\varepsilon}(\omega_{0})\rangle\\
		|u_{\varepsilon}(\omega_{1})\rangle\\
		|u_{\varepsilon}(\omega_{2})\rangle\\
		\vdots
	\end{bmatrix}=\varepsilon\begin{bmatrix}\vdots\\
		|u_{\varepsilon}(\omega_{-2})\rangle\\
		|u_{\varepsilon}(\omega_{-1})\rangle\\
		|u_{\varepsilon}(\omega_{0})\rangle\\
		|u_{\varepsilon}(\omega_{1})\rangle\\
		|u_{\varepsilon}(\omega_{2})\rangle\\
		\vdots
	\end{bmatrix}.
\end{equation}
In practical calculations, one should truncate the infinite-dimensional
matrix in Eq.~(\ref{eq:FE4}) at a sufficiently high harmonics $N\omega$,
leading to an $Nd\times Nd$ dimensional Floquet effective Hamiltonian,
whose eigenvalue problem can be numerically solved. Assuming the characteristic
energy scale of $\hat{H}_{m-n}$ for all $m-n$ to be $\Omega$, we
can take a smaller $N$ to do the truncation for a larger ratio of
$\omega/\Omega$, i.e., for a high-frequency driving field. 
Instead, for a resonantly or slowly driven system, more harmonics should be kept during the truncation.
All the discussions presented in this subsection hold for both Hermitian
and non-Hermitian Hamiltonians $\hat{H}(t)$.

\subsection{Floquet effective Hamiltonian and high-frequency expansion\label{subsec:HFA}}
From the Floquet operator $\hat{U}$ of a periodically driven system,
one can formally define its Floquet effective Hamiltonian as
\begin{equation}
	\hat{H}_{{\rm eff}}=\frac{i}{T}\ln\hat{U}\Leftrightarrow\hat{U}=e^{-i\hat{H}_{{\rm eff}}T}.\label{eq:Heff}
\end{equation}
Taking into account the fact that the quasienergies are defined modulus
$2\pi/T$, the $\hat{H}_{{\rm eff}}$ contains the same physical information
as $\hat{U}$. Yet, it provides us with more room to treat the properties
of Floquet systems in analogy with static Hamiltonian models. For
a continuously driven system, the explicit form of $\hat{H}_{{\rm eff}}$
is usually involved. This can be inspected from Eq.~(\ref{eq:Uapp}),
as the $\hat{H}(\ell\Delta t)$ and $\hat{H}(m\Delta t)$
do not commute for $\ell\neq m$ in general. When the frequency of the driving
field is high enough, an approximate series expression for $\hat{H}_{{\rm eff}}$
can be obtained via high-frequency expansion methods \cite{FloHeff1,FloHeff2,FloHeff3,FloHeff4}.
Here we recap one such method in its full generality, which is applicable
to both Hermitian and non-Hermitian systems.

We first assume that the time-periodic Hamiltonian $\hat{H}(t)$ can be decomposed into
a static part $\hat{H}_{0}$ and a periodically modulated part $\hat{V}(t)$, i.e.,
\begin{equation}
	\hat{H}(t)=\hat{H}_{0}+\hat{V}(t),\qquad\hat{V}(t)=\hat{V}(t+T).
\end{equation}
Here $T=2\pi/\omega$ is the driving period with $\omega$ denoting the driving
frequency. Next, we apply a similarity transformation to the evolved
state $|\Psi(t)\rangle$ in the Schr\"odinger equation
(\ref{eq:Seq}), yielding a rotated state
\begin{equation}
	|\Phi(t)\rangle=\hat{{\cal U}}(t)|\Psi(t)\rangle=e^{i\hat{K}(t)}|\Psi(t)\rangle.
\end{equation}
Here $\hat{K}(t)$ is sometimes called the kick operator. It encodes
the information regarding the micromotion dynamics of the system.
Plugging $|\Psi(t)\rangle=e^{-i\hat{K}(t)}|\Phi(t)\rangle$ into 
Eq.~(\ref{eq:Seq}) leads to the transformed Schr\"odinger
equation 
\begin{equation}
	i\frac{\partial}{\partial t}|\Phi(t)\rangle=\hat{H}_{{\rm eff}}|\Phi(t)\rangle,
\end{equation}
where
\begin{equation}
	\hat{H}_{{\rm eff}}=e^{i\hat{K}(t)}\hat{H}(t)e^{-i\hat{K}(t)}+i\frac{de^{i\hat{K}(t)}}{dt}e^{-i\hat{K}(t)}.\label{eq:Heff2}
\end{equation}
The aim of the high-frequency expansion method is to find a time-independent
$\hat{H}_{{\rm eff}}$ by transferring all time-dependent terms into
the kick operator $\hat{K}(t)$. When such a purpose is formally achieved,
we can express the Floquet evolution of a system as
\begin{equation}
	\hat{U}(t_{1},t_{0})|\Psi(t_{0})\rangle=e^{-i\hat{K}(t_{1})}e^{-i\hat{H}_{{\rm eff}}(t_{1}-t_{0})}e^{i\hat{K}(t_{0})}|\Psi(t_{0})\rangle.\label{eq:Ut1t0}
\end{equation}
That is, the system is subject to an initial kick associated with
the operator $\hat{K}(t_{0})$, then evolved under the time-independent
$\hat{H}_{{\rm eff}}$, and finally subject to a second kick carried
out by the operator $\hat{K}(t_{1})$. There is no need to perform any
time-ordered integral in the calculation of $\hat{U}(t_{1},t_{0})$.

Assuming the driving frequency $\omega$ to be large, we may expand
$\hat{H}_{{\rm eff}}$ and $\hat{K}(t)$ into power series of $1/\omega$
as
\begin{equation}
	\hat{H}_{{\rm eff}}=\sum_{n=0}^{\infty}\frac{1}{\omega^{n}}\hat{H}_{{\rm eff}}^{(n)},\qquad\hat{K}=\sum_{n=1}^{\infty}\frac{1}{\omega^{n}}\hat{K}^{(n)}.\label{eq:Heff3}
\end{equation}
In the meantime, we can apply Taylor expansions to the two terms in
Eq.~(\ref{eq:Heff2}) and express $\hat{H}_{{\rm eff}}$ as
\begin{alignat}{1}
	\hat{H}_{{\rm eff}}= & \sum_{n=0}^{\infty}\frac{i^{n}}{n!}[\stackrel{n\,{\rm of}\,\hat{K}}{\overbrace{\hat{K},[\hat{K},...,\hat{K},[\hat{K}},}\hat{H}]]]\nonumber \\
	+ & \sum_{n=1}^{\infty}i\frac{i^{n}}{n!}[\stackrel{(n-1)\,{\rm of}\,\hat{K}}{\overbrace{\hat{K},[\hat{K},...,\hat{K},[\hat{K}},}\partial_{t}\hat{K}]]].\label{eq:Heff4}
\end{alignat}
Note that we have concealed the explicit time-dependence of $\hat{K}(t)$
and $\hat{K}^{(n)}(t)$ in the above two equations for brevity. To
proceed, we impose two further requirements for $\hat{K}(t)$, 
\begin{equation}
	\hat{K}(t)=\hat{K}(t+T),\qquad\int_{0}^{T}\hat{K}(t)dt=0.
\end{equation}
Inserting Eq.~(\ref{eq:Heff3}) and the Fourier expansion
\begin{equation}
	\hat{H}(t)=\hat{H}_{0}+\sum_{m\neq0}\hat{V}_{m}e^{im\omega t}
\end{equation}
into Eq.~(\ref{eq:Heff4}), we arrive at
\begin{alignat}{1}
	& \sum_{n=0}^{\infty}\frac{1}{\omega^{n}}\hat{H}_{{\rm eff}}^{(n)}\nonumber \\
	= & \sum_{n=0}^{\infty}\frac{i^{n}}{n!}[\stackrel{n\,{\rm of}\,\hat{K}}{\overbrace{\sum_{p}\frac{1}{\omega^{p}}\hat{K}^{(p)},[\sum_{q}\frac{1}{\omega^{q}}\hat{K}^{(q)},...,\sum_{r}\frac{1}{\omega^{r}}\hat{K}^{(r)},[\sum_{s}\frac{1}{\omega^{s}}\hat{K}^{(s)}},}\left(\hat{H}_{0}+\sum_{m\neq0}\hat{V}_{m}e^{im\omega t}\right)]]]\nonumber \\
	+ & \sum_{n=1}^{\infty}\frac{i^{n+1}}{n!}[\stackrel{(n-1)\,{\rm of}\,\hat{K}}{\overbrace{\sum_{p}\frac{1}{\omega^{p}}\hat{K}^{(p)},[\sum_{q}\frac{1}{\omega^{q}}\hat{K}^{(q)},...,\sum_{r}\frac{1}{\omega^{r}}\hat{K}^{(r)},[\sum_{s}\frac{1}{\omega^{s}}\hat{K}^{(s)}},}\partial_{t}\sum_{m}\frac{1}{\omega^{m}}\hat{K}^{(m)}]]].\label{eq:Heff5}
\end{alignat}
The high-frequency expansions of $\hat{H}_{{\rm eff}}$ and $\hat{K}$
are obtained by equating both sides of Eq.~(\ref{eq:Heff5}) at each
order of $1/\omega$.

We can now carry out the calculations order-by-order to find the first few terms
in the series of $\hat{H}_{{\rm eff}}$ and $\hat{K}$.
At the zeroth order, we could obtain
\begin{equation}
	\hat{H}_{{\rm eff}}^{(0)}=\hat{H}_{0}+\hat{V}(t)-\frac{1}{\omega}\partial_{t}\hat{K}^{(1)}
\end{equation}
from Eq.~(\ref{eq:Heff5}). To ensure that $\hat{H}_{{\rm eff}}^{(0)}$
is time-independent, we must choose
\begin{equation}
	\frac{1}{\omega}\partial_{t}\hat{K}^{(1)}=\hat{V}(t)=\sum_{m\neq0}\hat{V}_{m}e^{im\omega t},
\end{equation}
such that 
\begin{equation}
	\hat{K}^{(1)}=\sum_{m\neq0}\frac{1}{im}\hat{V}_{m}e^{im\omega t}.\label{eq:K1t}
\end{equation}
Therefore, up to the zeroth order of $1/\omega$, we have the following
approximation for the effective Hamiltonian
\begin{equation}
	\hat{H}_{{\rm eff}}\approx\hat{H}_{{\rm eff}}^{(0)}=\hat{H}_{0}.
\end{equation}
Up to the first order of $1/\omega$, we have the following approximation
for the kick operator
\begin{equation}
	\hat{K}(t)\approx\frac{1}{\omega}\hat{K}^{(1)}=\frac{1}{\omega}\sum_{m\neq0}\frac{\hat{V}_{m}}{im}e^{im\omega t}.
\end{equation}
At the first order, we could find from Eq.~(\ref{eq:Heff5}) that
\begin{alignat}{1}
	\hat{H}_{{\rm eff}}^{(1)}& = i\left[\hat{K}^{(1)},\hat{H}_{0}+\hat{V}(t)\right]-\frac{1}{\omega}\partial_{t}\hat{K}^{(2)}-\frac{i}{2}\left[\hat{K}^{(1)},\frac{1}{\omega}\partial_{t}\hat{K}^{(1)}\right]\nonumber \\
	& = i\left[\hat{K}^{(1)},\hat{H}_{0}+\hat{V}(t)\right]-\frac{1}{\omega}\partial_{t}\hat{K}^{(2)}-\frac{i}{2}\left[\hat{K}^{(1)},\hat{V}(t)\right]\nonumber \\
	& = i\left[\hat{K}^{(1)},\hat{H}_{0}\right]+\frac{i}{2}\left[\hat{K}^{(1)},\hat{V}(t)\right]-\frac{1}{\omega}\partial_{t}\hat{K}^{(2)}.
\end{alignat}
By definition, any non-vanishing elements of $i\left[\hat{K}^{(1)},\hat{H}_{0}\right]$
and $-\frac{1}{\omega}\partial_{t}\hat{K}^{(2)}$ are time-dependent.
The only time-independent term that could contribute to $\hat{H}_{{\rm eff}}^{(1)}$
comes from the term $\frac{i}{2}\left[\hat{K}^{(1)},\hat{V}(t)\right]$,
which can be written explicitly as
\begin{alignat}{1}
	\frac{i}{2}\left[\hat{K}^{(1)},\hat{V}(t)\right] & = \frac{i}{2}\left[\sum_{m\neq0}\frac{1}{im}\hat{V}_{m}e^{im\omega t},\sum_{m'\neq0}\hat{V}_{m'}e^{im'\omega t}\right]\nonumber \\
	& = \sum_{m,m'\neq0}\frac{e^{i(m+m')\omega t}}{2m}\left[\hat{V}_{m},\hat{V}_{m'}\right].
\end{alignat}
The time-independent terms are those with $m'=-m$. Collecting these
terms together, we find
\begin{equation}
	\hat{H}_{{\rm eff}}^{(1)}=\sum_{m\neq0}\frac{\left[\hat{V}_{m},\hat{V}_{-m}\right]}{2m}=\sum_{m\neq0}\frac{\hat{V}_{m}\hat{V}_{-m}}{m}.
\end{equation}
Therefore, up to the first order of $1/\omega$, we have the following
approximation for the effective Hamiltonian
\begin{equation}
	\hat{H}_{{\rm eff}}\approx\hat{H}_{{\rm eff}}^{(0)}+\frac{1}{\omega}\hat{H}_{{\rm eff}}^{(1)}=\hat{H}_{0}+\frac{1}{\omega}\sum_{m\neq0}\frac{\hat{V}_{m}\hat{V}_{-m}}{m}.
\end{equation}
The form of $\hat{K}^{(2)}$ is determined by the remaining time-dependent
terms. Performing the integration over time directly, we obtain
\begin{alignat}{1}
	\hat{K}^{(2)}= & \sum_{m\neq0}\frac{1}{m}\left[\hat{V}_{m},\hat{H}_{0}\right]\int\omega e^{im\omega t}dt+\sum_{m,m'\neq0,-m}\frac{\left[\hat{V}_{m},\hat{V}_{m'}\right]}{2m}\int\omega e^{i(m+m')\omega t}dt\nonumber \\
	= & \sum_{m\neq0}\frac{1}{im^{2}}\left[\hat{V}_{m},\hat{H}_{0}\right]e^{im\omega t}+\sum_{m,m'\neq0,-m}\frac{\left[\hat{V}_{m},\hat{V}_{m'}\right]}{2im(m+m')}e^{i(m+m')\omega t}.\label{eq:K2t}
\end{alignat}
Up to the second order of $1/\omega$, we can thus approximate the
kick operator by
\begin{alignat}{1}
	& \hat{K}(t)\approx\frac{1}{\omega}\hat{K}^{(1)}+\frac{1}{\omega^{2}}\hat{K}^{(2)}\nonumber \\
	& = \frac{1}{\omega}\sum_{m\neq0}\frac{\hat{V}_{m}}{im}e^{im\omega t}+\frac{1}{\omega^{2}}\sum_{m\neq0}\frac{1}{im^{2}}\left[\hat{V}_{m},\hat{H}_{0}\right]e^{im\omega t}\nonumber \\
	& + \frac{1}{\omega^{2}}\sum_{m,m'\neq0,-m}\frac{\left[\hat{V}_{m},\hat{V}_{m'}\right]}{2im(m+m')}e^{i(m+m')\omega t}.
\end{alignat}

We could continue these self-consistent calculations to obtain 
higher-order terms in the high-frequency expansion of $\hat{H}_{{\rm eff}}$
and $\hat{K}(t)$. For example, for the second-order component $\hat{H}_{{\rm eff}}^{(2)}$,
we have 
\begin{alignat}{1}
	\hat{H}_{{\rm eff}}^{(2)} & =\frac{-1}{2}[\hat{K}^{(1)},[\hat{K}^{(1)},\hat{H}_{0}+\hat{V}(t)]]+i[\hat{K}^{(2)},\hat{H}_{0}+\hat{V}(t)]\nonumber \\
	+ & \frac{-1}{\omega}\partial_{t}\hat{K}^{(3)}+\frac{-i}{2\omega}[\hat{K}^{(1)},\partial_{t}\hat{K}^{(2)}]+\frac{-i}{2\omega}[\hat{K}^{(2)},\partial_{t}\hat{K}^{(1)}]\nonumber \\
	+ & \frac{1}{6\omega}[\hat{K}^{(1)},[\hat{K}^{(1)},\partial_{t}\hat{K}^{(1)}].
\end{alignat}
Dropping the time-dependent terms, we are left with
\begin{equation}
	\hat{H}_{{\rm eff}}^{(2)}=\frac{1}{2}\sum_{m\neq0}\frac{[[\hat{V}_{m},\hat{H}_{0}],\hat{V}_{-m}]}{m^{2}}+\frac{1}{3}\sum_{l,m\neq0}\frac{[\hat{V}_{l},[\hat{V}_{m},\hat{V}_{-l-m}]]}{lm}.
\end{equation}
Therefore, up to the second-order correction in $1/\omega$, the effective
Hamiltonian $\hat{H}_{{\rm eff}}$ turns out to be
\begin{alignat}{1}
	& \hat{H}_{{\rm eff}}\approx\hat{H}_{0}+\frac{1}{\omega}\sum_{m\neq0}\frac{\hat{V}_{m}\hat{V}_{-m}}{m}\nonumber \\
	+ & \frac{1}{2\omega^{2}}\sum_{m\neq0}\frac{[[\hat{V}_{m},\hat{H}_{0}],\hat{V}_{-m}]}{m^{2}}+\frac{1}{3\omega^{2}}\sum_{l,m\neq0}\frac{[\hat{V}_{l},[\hat{V}_{m},\hat{V}_{-l-m}]]}{lm}.
\end{alignat}
This approximation holds for both Hermitian and non-Hermitian Floquet
systems under high-frequency driving fields. Note that for stroboscopic
evolution, we have $t_{1}-t_{0}=NT$ with $N\in\mathbb{Z}$ and $\hat{K}(t_{1})=\hat{K}(t_{0})$ in Eq.~(\ref{eq:Ut1t0}).
In this case, the term $\hat{K}(t_{0})$ in Eq.~(\ref{eq:Ut1t0})
describes an initial phase, which can be set to zero as the expansion
of $\hat{K}(t)$ at every order of $1/\omega$ is only determined
up to a constant {[}see Eqs.~(\ref{eq:K1t}) and (\ref{eq:K2t}){]}.
Therefore, for stroboscopic dynamics, we can simply use the $\hat{H}_{{\rm eff}}$
to capture most of the essential physics. In Sec.~\ref{sec:Res}, we will showcase
the application of the high-frequency methods presented here to the Floquet engineering of non-Hermitian
quasicrystals.

To be concrete, we illustrate the usage of $\hat{H}_{{\rm eff}}$ here with a simple
example. Consider a harmonically driven two-level Hamiltonian of the form
\begin{equation}
	{\hat H}(t)=(h_{x}+i\gamma/2)[\cos(\omega t)\sigma_{x}+\sin(\omega t)\sigma_{y}]+h_{z}\sigma_{z},
\end{equation}
where $h_{x},h_{z},\gamma\in\mathbb{R}$, and $\sigma_\alpha$ for $\alpha=x,y,z$ are
Pauli matrices. It is not hard to verify that by choosing the kick
operator as
\begin{equation}
	e^{i{\hat K}(t)}=\begin{pmatrix}1 & 0\\
		0 & e^{-i\omega t}
	\end{pmatrix},
\end{equation}
the Floquet effective Hamiltonian in the rotating frame can be exactly
obtained from Eq.~(\ref{eq:Heff2}), i.e.,
\begin{equation}
	{\hat H}_{{\rm eff}}=\frac{\omega}{2}\sigma_{0}+(h_{x}+i\gamma/2)\sigma_{x}+(h_{z}-\omega/2)\sigma_{z},\label{eq:HeffEP}
\end{equation}
where $\sigma_0$ denotes the $2\times2$ identity matrix.
We see that the two levels of $H(t)$ have the instantaneous eigenenergies
\begin{equation}
	E_{\pm}=\pm\sqrt{h_{x}^{2}+h_{z}^{2}-\gamma^{2}/4+ih_{x}\gamma},\label{eq:EeffEP}
\end{equation}
which are time-independent and complex in general. They could meet
with each other at a second-order EP with $E=0$ when $h_{x}=0$ and
$h_{z}=\pm\gamma/2$. Meanwhile, ${\hat H}_{{\rm eff}}$ in Eq.~(\ref{eq:HeffEP}) has two quasienergy
levels defined by
\begin{equation}
	\varepsilon_{\pm}=\frac{\omega}{2}\pm\sqrt{h_{x}^{2}+\left(h_{z}-\frac{\omega}{2}\right)^{2}-\gamma^{2}/4+ih_{x}\gamma}.\label{eq:QuasiE}
\end{equation}
They could meet with each other at a second-order Floquet EP with
$\varepsilon=\omega/2$ when $h_{x}=0$ and $h_{z}=\omega/2\pm\gamma/2$.
It is clear that the EP is shifted by the driving in both its energy
and location in the parameter space. If we define $\varepsilon T$
as the dimensionless quasienergy $E$, the Floquet EP of our two-level system appears
at the quasienergy $E=\pi$. This could lead to Floquet exceptional
topology and anomalous $\pi$ edge modes, which are unique to driven
non-Hermitian systems. 
We will give an explicit example to demonstrate this point in Sec.~\ref{sec:Res}.

\subsection{Adiabatic perturbation theory\label{subsec:APT}}
We now consider a non-Hermitian system that is subject to slow-in-time
and cyclic modulations. It can be viewed as the opposite limit of
a high-frequency driven system. Introducing a re-scaled and dimensionless
time variable $s=t/T$, we can express the Schr\"odinger
equation (\ref{eq:Seq}) as
\begin{equation}
	i\frac{\partial}{\partial s}|\Psi(s)\rangle=T\hat{H}(s)|\Psi(s)\rangle.\label{eq:SeqAPT}
\end{equation}
Our purpose is to find an expansion for $|\Psi(s)\rangle$ in the
series of $1/T$. It constitutes an adiabatic perturbation theory (APT)
of the system \cite{APT01}, providing that different energy levels
of $\hat{H}(s)$ are gapped for all $s\in[0,1]$.

The Hamiltonian $\hat{H}(s)$ is periodic in the scaled time $s$
with $\hat{H}(s)=\hat{H}(s+1)$. We denote its instantaneous right
and left biorthonormal eigenvectors as $\{|n(s)\rangle\}$ and $\{|\widetilde{n}(s)\rangle\}$
\cite{BioQM01}, such that
\begin{equation}
	\hat{H}(s)|n(s)\rangle=E_{n}(s)|n(s)\rangle,\qquad\langle\widetilde{n}(s)|\hat{H}(s)=\langle\widetilde{n}(s)|E_{n}(s),\label{eq:BioE}
\end{equation}
\begin{equation}
	\langle\widetilde{m}(s)|n(s)\rangle=\delta_{mn},\qquad\sum_{n}|n(s)\rangle\langle\widetilde{n}(s)|=1.\label{eq:BioBasis}
\end{equation}
Here $E_{n}(s)$ is the instantaneous eigenenergy associated with
$|n(s)\rangle$, which could be complex if $\hat{H}(s)\neq\hat{H}^{\dagger}(s)$.
To proceed, we introduce an ansatz solution for the time-evolved state
$|\Psi(s)\rangle$ as
\begin{equation}
	|\Psi(s)\rangle=\sum_{n}e^{-i\Omega_{n}(s)}c_{n}(s)|n(s)\rangle,\label{eq:PsiAPT}
\end{equation}
where
\begin{equation}
	\Omega_{n}(s)\equiv T\int_{0}^{s}E_{n}(s')ds'
\end{equation}
describes a dynamical phase accumulated over the time interval $[0, s]$. At $s=0$, we have $\Omega_{n}(0)=0$ and 
\begin{equation}
	c_{n}(0)=\langle\widetilde{n}(0)|\Psi(0)\rangle
\end{equation}
for the initial state $|\Psi(0)\rangle$. Plugging Eq.~(\ref{eq:PsiAPT}) into 
Eq.~(\ref{eq:SeqAPT}), we obtain
\begin{equation}
	\sum_{n}e^{-i\Omega_{n}(s)}\dot{c}_{n}(s)|n(s)\rangle=-\sum_{n}e^{-i\Omega_{n}(s)}c_{n}(s)|\dot{n}(s)\rangle,
\end{equation}
where $\dot{c}_{n}(s)=dc_{n}/ds$ and $|\dot{n}(s)\rangle=d|n(s)\rangle/ds$.
Acting $\langle\widetilde{m}(s)|$ from the left on both sides of
the above equation and using Eq.~(\ref{eq:BioBasis}), we find
\begin{equation}
	\dot{c}_{m}(s)=-\sum_{n\neq m}e^{i\Omega_{mn}(s)}c_{n}(s)\langle\widetilde{m}(s)|\dot{n}(s)\rangle,
\end{equation}
where $\Omega_{mn}(s)\equiv\Omega_{m}(s)-\Omega_{n}(s)$. We have
made the parallel transport gauge choice so that $\langle\widetilde{n}(s)|\dot{n}(s)\rangle=0$.
Performing the integration over $s$ and keeping the terms on the
right-hand-side up to the first order of $1/T$ (i.e., up to the first
order non-adiabatic correction), we arrive at
\begin{equation}
	c_{m}(s)=c_{m}(0)+\frac{1}{T}\sum_{n\neq m}\left.\left[\frac{i\langle\widetilde{m}(s')|\dot{n}(s')\rangle}{\Delta_{mn}(s')}e^{i\Omega_{mn}(s')}\right]\right|_{s'=0}^{s'=s}c_{n}(0),\label{eq:cms}
\end{equation}
where $\Delta_{mn}(s)=E_{m}(s)-E_{n}(s)$. To reach Eq.~(\ref{eq:cms}),
we have assumed $\Omega_{mn}(s)$ to be real for all $m\neq n$ and
$s\in[0,1]$. This can be achieved if every instantaneous energy level
possesses the same imaginary part $i\gamma(s)$. On the other hand,
Eq.~(\ref{eq:cms}) is valid if the $\hat{H}(s)$ is PT-invariant
at every $s$, such that $E_{n}(s)\in\mathbb{R}$ for all $n$.
The approximation in Eq.~(\ref{eq:cms}) does not hold if $\Delta_{mn}(s)\notin\mathbb{R}$,
which would cause the exponential amplification or decay of the amplitude
$c_{n}(s)$. 
So the APT developed here is more restrictive in applications than its Hermitian counterparts.
Inserting Eq.~(\ref{eq:cms}) into Eq.~(\ref{eq:PsiAPT})
results in the solution to Eq.~(\ref{eq:SeqAPT}) up to first
order non-adiabatic corrections, i.e.,
\begin{equation}
	|\Psi(s)\rangle=\sum_{m}e^{-i\Omega_{m}(s)}\left\{ c_{m}(0)+\frac{1}{T}\sum_{n\neq m}\left.\left[\frac{i\langle\widetilde{m}(s')|\dot{n}(s')\rangle}{\Delta_{mn}(s')}e^{i\Omega_{mn}(s')}\right]\right|_{s'=0}^{s'=s}c_{n}(0)\right\} |m(s)\rangle.\label{eq:PsiAPT2}
\end{equation}
Since we have employed the formalism of biorthonormal eigenvectors,
we may consider another ansatz solution for the left eigenvector
\begin{equation}
	\langle\widetilde{\Psi}(s)|=\sum_{n}e^{i\Omega_{n}(s)}c_{n}^{*}(s)\langle\widetilde{n}(s)|,
\end{equation}
which obeys a conjugate Schr\"odinger equation
\begin{equation}
	-i\frac{\partial}{\partial s}\langle\widetilde{\Psi}(s)|=\langle\widetilde{\Psi}(s)|\hat{H}.\label{eq:SeqAPT2}
\end{equation}
Following the same reasoning in deriving Eq.~(\ref{eq:PsiAPT2}), we find,
up to the first order of $1/T$, that
\begin{equation}
	\langle\widetilde{\Psi}(s)|=\sum_{m}e^{+i\Omega_{m}(s)}\left\{ c_{m}^{*}(0)+\frac{1}{T}\sum_{n\neq m}\left[\frac{i\langle\dot{\widetilde{n}}(s')|m(s')\rangle}{\Delta_{nm}(s')}e^{i\Omega_{nm}(s')}\right]_{s'=0}^{s'=s}c_{n}^{*}(0)\right\} \langle\widetilde{m}(s)|.\label{eq:PsiAPT3}
\end{equation}
Assuming that initially only the state with $m=\ell$ is occupied
such that $c_{m}(0)=\delta_{m\ell}$, we can further simplify
Eqs.~(\ref{eq:PsiAPT2}) and (\ref{eq:PsiAPT3}) to
\begin{equation}
	|\Psi_{\ell}(s)\rangle=e^{-i\Omega_{\ell}(s)}\left[|\ell(s)\rangle+\frac{1}{T}\sum_{m\neq\ell}\frac{i\langle\widetilde{m}(s)|\dot{\ell}(s)\rangle}{\Delta_{m\ell}(s)}|m(s)\rangle\right]-\frac{1}{T}\sum_{m\neq\ell}e^{-i\Omega_{m}(s)}\frac{i\langle\widetilde{m}(0)|\dot{\ell}(0)\rangle}{\Delta_{m\ell}(0)}|m(s)\rangle,
\end{equation}
\begin{equation}
	\langle\widetilde{\Psi}_{\ell}(s)|=e^{+i\Omega_{\ell}(s)}\left[\langle\widetilde{\ell}(s)|+\frac{1}{T}\sum_{m\neq\ell}\frac{i\langle\dot{\widetilde{\ell}}(s)|m(s)\rangle}{\Delta_{\ell m}(s)}\langle\widetilde{m}(s)|\right]-\frac{1}{T}\sum_{m\neq\ell}e^{+i\Omega_{m}(s)}\frac{i\langle\dot{\widetilde{\ell}}(0)|m(0)\rangle}{\Delta_{\ell m}(0)}\langle\widetilde{m}(s)|.
\end{equation}
For any observable $\hat{O}$, up to the correction of order $1/T$,
we can now express its biorthogonal average over the states $\{|\Psi_{\ell}(s)\rangle,|\widetilde{\Psi}_{\ell}(s)\rangle\}$
as
\begin{alignat}{1}
	\langle\widetilde{\Psi}_{\ell}(s)|\hat{O}|\Psi_{\ell}(s)\rangle & =\langle\widetilde{\ell}(s)|\hat{O}|\ell(s)\rangle+\frac{1}{T}\sum_{m\neq\ell}\frac{i\langle\widetilde{m}(s)|\dot{\ell}(s)\rangle}{\Delta_{m\ell}(s)}\langle\widetilde{\ell}(s)|\hat{O}|m(s)\rangle\nonumber \\
	& +\frac{1}{T}\sum_{m\neq\ell}\frac{i\langle\dot{\widetilde{\ell}}(s)|m(s)\rangle}{\Delta_{\ell m}(s)}\langle\widetilde{m}(s)|\hat{O}|\ell(s)\rangle.\label{eq:Oave}
\end{alignat}
We notice that this average does not contain any time-oscillating phase factors.

We now illustrate this APT with an application in the study of dynamical topological phenomena.
Let us consider noninteracting particles in a one-dimensional (1D)
periodic lattice, whose onsite potential is also varied slowly and
periodically in time. This is the typical situation encountered in
the topological Thouless pump \cite{Thouless1983,Thouless1984,ThoulessRev}.
The group velocity of the particle can be expressed as $\hat{v}=\partial_{k}\hat{H}$,
where $k$ is the quasimomentum. At the initial time $s=0$, we assume
that the band $\ell$ is uniformly filled, and it is separated from
the other bands at all $k$ and $s$. The pumped number of particles
over one adiabatic cycle due to this initially filled band is then
given by
\begin{equation}
	N_{\ell}=\int_{-\pi}^{\pi}\frac{dk}{2\pi}\int_{0}^{1}ds\langle\widetilde{\Psi}_{\ell}(s)|\partial_{k}\hat{H}|\Psi_{\ell}(s)\rangle.\label{eq:TPump}
\end{equation}
With the aid of Eq.~(\ref{eq:BioE}), it is not hard to identify that
\begin{equation}
	\langle\widetilde{\ell}(s)|\partial_{k}\hat{H}|\ell(s)\rangle=\partial_{k}E_{\ell}(k,s),\label{eq:HFT1}
\end{equation}
\begin{equation}
	\frac{\langle\widetilde{\ell}(s)|\partial_{k}\hat{H}|m(s)\rangle}{\Delta_{m\ell}(s)}=\langle\widetilde{\ell}(s)|\partial_{k}|m(s)\rangle,\qquad m\neq\ell,\label{eq:HFT2}
\end{equation}
where $E_{\ell}(k,s)$ denotes the energy dispersion of the $\ell$th
adiabatic Bloch band. Plugging Eqs.~(\ref{eq:HFT1}), (\ref{eq:HFT2}),
and (\ref{eq:Oave}) into Eq.~(\ref{eq:TPump}), we obtain
\begin{alignat}{1}
	N_{\ell}= & \int_{-\pi}^{\pi}\frac{dk}{2\pi}\int_{0}^{1}ds\partial_{k}E_{\ell}(k,s)\\
	+ & \sum_{m\neq\ell}\int_{-\pi}^{\pi}\frac{dk}{2\pi i}\int_{0}^{T}dt\langle\partial_{k}\widetilde{\ell}(k,t)|m(k,t)\rangle\langle\widetilde{m}(k,t)|\partial_{t}\ell(k,t)\rangle\nonumber \\
	- & \sum_{m\neq\ell}\int_{-\pi}^{\pi}\frac{dk}{2\pi i}\int_{0}^{T}dt\langle\partial_{t}\widetilde{\ell}(k,t)|m(k,t)\rangle\langle\widetilde{m}(k,t)|\partial_{k}\ell(k,t)\rangle.\nonumber 
\end{alignat}
Noting that $E_{\ell}(k=-\pi,s)=E_{\ell}(k=\pi,s)$ and $\sum_{m}|m(k,t)\rangle\langle\widetilde{m}(k,t)|=1$,
we can simplify the above equation and arrive at the pumped number
of particles over an adiabatic cycle
\begin{equation}
	N_{\ell}=\int_{-\pi}^{\pi}\frac{dk}{2\pi i}\int_{0}^{T}dt\left[\langle\partial_{k}\widetilde{\ell}(k,t)|\partial_{t}\ell(k,t)\rangle-\langle\partial_{t}\widetilde{\ell}(k,t)|\partial_{k}\ell(k,t)\rangle\right].\label{eq:Ch}
\end{equation}
Equation (\ref{eq:Ch}) describes nothing but the Chern number of the adiabatic
Bloch band $\ell$, whose energy dispersion is defined on a two dimensional
(2D) torus $(k,t)\in[-\pi,\pi)\times[0,T)$. The conditions for
Eq.~(\ref{eq:Ch}) to hold are as follows. First, the Hamiltonian
of the system should be quasi-Hermitian with a real spectrum (e.g.,
PT-invariant) in our considered parameter regime. Second, the band
$E_{\ell}$ should be well-gapped from the other bands throughout
the 2D torus $(k,t)\in[-\pi,\pi)\times[0,T)$, and $\hbar/T$ should
be much smaller than $|\Delta_{\ell m}|$ for all $m\neq\ell$ in order to guarantee the adiabatic condition.
Third,
the evolution of left and right vectors of the system should follow
Eqs.~(\ref{eq:SeqAPT}) and (\ref{eq:SeqAPT2}). Note here that
the expression of Chern number is not sensitive to the
choice of biorthonormal basis. We will get the same $N_{\ell}$ after
changing $\widetilde{\ell}\rightarrow\ell$ or $\ell\rightarrow\widetilde{\ell}$
in Eq.~(\ref{eq:Ch}), as proved before for non-Hermitian Chern
bands \cite{NHCls01}.

\subsection{Symmetry and topological characterization\label{subsec:Sym}}
Over the past few years, rich symmetry classifications and topological invariants
have been identified for non-Hermitian topological matter \cite{NHCls01,NHCls02,NHCls03,NHCls04,NHCls05,NHCls06,NHCls07,NHCls08,NHCls09}.
In this subsection, we mainly recap two symmetries together with their
associated topological numbers. They are the most relevant ones for
the characterization of non-Hermitian Floquet topological phases reviewed
in this work.

We first discuss the PT-symmetry, which is associated with the operator
${\cal PT}$. Here ${\cal P}$ denotes the parity operator and ${\cal T}$ denotes the time-reversal operator.
When the Hamiltonian of a non-Hermitian system $\hat{H}$
respects the PT-symmetry, we have $[{\cal PT},\hat{H}]=0$.
In this case, the system could have a real spectrum in the PT unbroken
regime, where the eigenstates $|\psi\rangle$ and ${\cal PT}|\psi\rangle$
of $\hat{H}$ are coincident up to a global phase. To see this, let
us consider a non-degenerate eigenstate $|\psi\rangle$ of $\hat{H}$
that satisfies the eigenvalue equation
\begin{equation}
	\hat{H}|\psi\rangle=E|\psi\rangle.
\end{equation}
The PT-symmetry of $\hat{H}$ then implies that
\begin{equation}
	\hat{H}({\cal PT}|\psi\rangle)=E^{*}({\cal PT}|\psi\rangle).
\end{equation}
Therefore, ${\cal PT}|\psi\rangle$ is also an eigenstate of $\hat{H}$
with the energy $E^{*}$. If $|\psi\rangle$ and $\hat{H}$ share
the same PT-symmetry, $|\psi\rangle$ should be the common eigenstate
of $\hat{H}$ and ${\cal PT}$. ${\cal PT}|\psi\rangle$ can thus
only differ from $|\psi\rangle$ up to a global phase, which means
that $E=E^{*}\in\mathbb{R}$. However, with the change of system parameters
(e.g., the strengths of gain and loss), the PT-symmetry of $|\psi\rangle$
could be spontaneously broken and the spectrum of $\hat{H}$ could
switch from real to complex after undergoing a PT-symmetry breaking
transition. A topological invariant, defined as \cite{NHQC01}
\begin{equation}
	w=\int_{0}^{2\pi}\frac{d\theta}{2\pi i}\partial_{\theta}\ln\det[\hat{H}(\theta)-E_{0}],\label{eq:EWN}
\end{equation}
might be employed to characterize such a real-to-complex spectral
transition. Here $E_{0}$ is a base energy chosen appropriately on
the complex plane. The parametrized Hamiltonian $\hat{H}(\theta)=\hat{H}(\theta+2\pi)$,
where $\theta$ can be viewed as the quasimomentum along an artificial
dimension. We have also taken the periodic boundary condition (PBC)
for $\hat{H}$ before implementing its $\theta$-parametrization.
The $w$ in Eq.~(\ref{eq:EWN}) thus depicts a spectral winding number
with respect to the base energy $E_{0}$, i.e., it counts the number
of times that the spectrum of $\hat{H}(\theta)$ winds around $E_{0}$
on the complex plane when the synthetic quasimomentum $\theta$ is
varied over a cycle. When the spectrum of $\hat{H}(\theta)$ is real,
we must have $w=0$, as a spectral loop cannot be formed on the complex
plane in this case. When the spectrum of $\hat{H}(\theta)$ is complex,
$w$ may take an integer-quantized value if $\hat{H}(\theta)$ possesses
spectral loops around $E_{0}$. A suitably chosen $E_{0}$ could then
yield a nonzero $w$ when the first spectral loop appears on the complex-$E$
plane, thereby detecting the topological changes in the spectrum
of $\hat{H}$ across the PT-breaking transition. 

For a Floquet system,
if the time-periodic Hamiltonian $\hat{H}(t)$ possesses the PT-symmetry
at each instant $t$, the resulting Floquet operator also has the
PT-symmetry. In this case, we can define the spectral winding number
$w$ as in Eq.~(\ref{eq:EWN}) for the stroboscopic Floquet effective
Hamiltonian so as to capture the PT-breaking transition in the quasienergy
spectrum. We will provide explicit examples for this usage in
Subsec.~\ref{subsec:TQC}, where we consider PT transitions, localization
transitions, and topological transitions in non-Hermitian Floquet quasicrystals.

We next consider the chiral (or sublattice) symmetry, whose associated
operator will be denoted by ${\cal S}$. When the Hamiltonian $\hat{H}$
of a system respects the chiral symmetry ${\cal S}$, we have
${\cal S}\hat{H}{\cal S}=-\hat{H}$, where ${\cal S}$ is both Hermitian
and unitary \cite{Tenfold01}. The implication of this symmetry on
the spectrum of $\hat{H}$ is as follows. Suppose that $|\psi\rangle$
is an eigenstate of a chiral-symmetric $\hat{H}$ with the energy
$E$, i.e., $\hat{H}|\psi\rangle=E|\psi\rangle$. We have 
\begin{equation}
	\hat{H}({\cal S}|\psi\rangle)=-E({\cal S}|\psi\rangle).
\end{equation}
Therefore, ${\cal S}|\psi\rangle$ is also an eigenstate of $\hat{H}$
with the energy $-E$. The eigenstates of a chiral-symmetric $\hat{H}$
should then come in pairs of $\{|\psi\rangle,{\cal S}|\psi\rangle\}$
with the energies $\{E,-E\}$ that are symmetric with respect to $E=0$.
This further leads to a chiral-symmetry protected degeneracy for any
eigenstate with $E=0$. When the spectrum of $\hat{H}$ is gapped
at $E=0$, we can group its energy levels into two clusters with Re$E<0$
and Re$E>0$. If these two clusters meet with each other at $E=0$
and then separate with the change of certain system parameters,
the system may undergo a phase transition. In one dimension, the change
of band topology of the system before and after such a transition
could be characterized by a winding number $w_{0}$, defined
as \cite{Tenfold01,WN01,WN02}
\begin{equation}
	w_{0}=\int_{-\pi}^{\pi}\frac{dk}{4\pi}{\rm Tr}[{\cal S}{\cal Q}(k)i\partial_{k}{\cal Q}(k)].\label{eq:W0Stat}
\end{equation}
Here the PBC has been assumed and $k\in[-\pi,\pi)$ denotes the quasimomentum.
The sign-resolved projector ${\cal Q}(k)$ is obtained from the spectral
decomposition of $\hat{H}=\sum_{k}|k\rangle H(k)\langle k|$ 
with $H(k)=\sum_{n}E_{n}(k)|n(k)\rangle\langle\widetilde{n}(k)|$
at each $k$ by attributing
$+1$ ($-1$) to every energy band $n$ with ${\rm Re}E_{n}>0$ (${\rm Re}E_{n}<0$).
${\cal Q}(k)$ can thus be expressed as \cite{ZhouNHFTP08}
\begin{equation}
	{\cal Q}(k)=\sum_{n}{\rm sgn}\{{\rm Re}[E_{n}(k)]\}|n(k)\rangle\langle\widetilde{n}(k)|.
\end{equation}
Here $H(k)|n(k)\rangle=E_{n}(k)|n(k)\rangle$ and $\langle\widetilde{n}(k)|H(k)=\langle\widetilde{n}(k)|E_{n}(k)$.
The set $\{|n(k)\rangle,|\widetilde{n}(k)\rangle\}$ of basis satisfies
the biorthonormal relations in Eq.~(\ref{eq:BioBasis}). Under the
PBC, the winding number $w_{0}$ could characterize the bulk topological
properties of a 1D chiral-symmetric Hamiltonian $\hat{H}$, either
Hermitian or non-Hermitian. It could further distinguish between different
bulk topological insulating phases by showing a quantized jump at
the transition point, where the two band clusters of $\hat{H}$ meet
with each other at $E=0$. However, under the open boundary condition
(OBC), due to the possible existence of NHSE, the $w_{0}$ defined
in Eq.~(\ref{eq:W0Stat}) may not be able to correctly predict
the gap-closing points of the spectrum in the parameter space and
determine the number of degenerate edge modes at $E=0$ in different
parameter regions. This non-Hermiticity induced breakdown of bulk-edge
correspondence may be recovered by introducing a real-space counterpart
of $w_{0}$, defined as \cite{NHSE10}
\begin{equation}
	W_{0}=-\frac{1}{L_{{\rm B}}}{\rm Tr}_{{\rm B}}({\cal S}{\cal Q}[{\cal Q},\hat{N}]).\label{eq:W0StatOpen}
\end{equation}
Here ${\cal S}$ is the chiral symmetry operator of $\hat{H}$, $\hat{N}$
is the position operator in real space, and ${\cal Q}$ is the flat
band projector defined as
\begin{equation}
	{\cal Q}=\sum_{|\psi_{j}\rangle\in{\rm bulk}}{\rm sgn}[{\rm Re}(E_{j})]|\psi_{j}\rangle\langle\widetilde{\psi}_{j}|,
\end{equation}
where the summation is now taken over all the bulk eigenstates $\{|\psi_{j}\rangle\}$
of $\hat{H}$ under the OBC. The whole lattice of length $L=L_{{\rm B}}+2L_{{\rm E}}$
is decomposed into three segments, with a bulk region of length $L_{{\rm B}}$
in the middle and two edge regions of the same length $L_{{\rm E}}$
at the left and right boundaries of the open chain. The trace ${\rm Tr}_{{\rm B}}(\cdot)$
is only taken over the bulk region, which excludes all possible interruptions
caused by the NHSE in the edge regions. The resulting $W_{0}$ was
found to be able to faithfully capture the topological phase transitions
and bulk-edge correspondence in 1D, chiral symmetric non-Hermitian
systems even in the presence of NHSE \cite{NHSE10}. It was also suggested
to be equivalent to the topological winding number defined through
the generalized Brillouin zone of non-Hermitian systems. By definition,
the winding number $W_{0}$ in Eq.~(\ref{eq:W0StatOpen}) is also
robust to perturbations induced by symmetry-preserved disorders and
impurities, making it applicable to more general situations. In the
clean, Hermitian, and thermodynamic limit, the $W_{0}$ in Eq.~(\ref{eq:W0StatOpen})
can be further reduced to the $w_{0}$ in Eq.~(\ref{eq:W0Stat}) \cite{Tenfold01,WN01,WN02}.

For a non-Hermitian Floquet system, we can state its chiral symmetry
as follows \cite{ZhouNHFTP01}. From Eq.~(\ref{eq:Heff}), we
can express the Floquet operator as $\hat{U}=e^{-i\hat{H}_{{\rm eff}}}$,
where we have set the driving period $T=1$ for brevity. Viewing the
$\hat{H}_{{\rm eff}}$ as a static Hamiltonian, we say that it respects
the chiral symmetry if there exists a unitary and Hermitian operator ${\cal S}$
such that ${\cal S}\hat{H}_{{\rm eff}}{\cal S}=-\hat{H}_{{\rm eff}}$.
At the level of $\hat{U}$, the chiral symmetry then implies that
\begin{equation}
	{\cal S}\hat{U}{\cal S}=\hat{U}^{-1}.
\end{equation}
As in the case of static systems, the chiral symmetry of $\hat{U}$
has a direct implication for the symmetry of its spectrum. If $|\Psi_{E}\rangle$
is an eigenstate of $\hat{U}$ with the quasienergy $E$, i.e., $\hat{U}|\Psi_{E}\rangle=e^{-iE}|\Psi_{E}\rangle$,
we immediately have
\begin{equation}
	\hat{U}({\cal S}|\Psi_{E}\rangle)=e^{-i(-E)}({\cal S}|\Psi_{E}\rangle),
\end{equation}
which means that ${\cal S}|\Psi_{E}\rangle$ is also an eigenstate
of $\hat{U}$ with the quasienergy $-E$. The Floquet spectrum of
$\hat{U}$ is then symmetric with respect to both the quasienergies
$E=0$ and $E=\pi$. The latter is because $-E$ and $E$ are identified
at the quasienergy $\pi$. When the spectrum is gapped at $E=0$ and
$\pi$, the quasienergy levels of $\hat{U}$ could be grouped into
two clusters. One of them has the quasienergy ${\rm Re}E\in(-\pi,0)$
and the other one has ${\rm Re}E\in(0,\pi)$. They could meet with
each other at either the quasienergy zero or $\pi$, leading to two
possible phase transitions. This implies that a complete topological
characterization of a chiral symmetric Floquet system should require
at least two winding numbers, which is rather different from the case
of static systems where a single winding number is sufficient. To
identify these winding numbers, let us consider the example of a periodically
kicked 1D system, whose Hamiltonian and Floquet operator take the
forms of Eqs.~(\ref{eq:KickH}) and (\ref{eq:KickU}). We also
assume that the two parts of Hamiltonians $\hat{H}_{0}$ and $\hat{H}_{1}$
in Eq.~(\ref{eq:KickH}) have the same chiral symmetry ${\cal S}$.
At the level of $\hat{U}=e^{-i\hat{H}_{0}}e^{-i\hat{H}_{1}}$ (assuming
$T=1$), it is not straightforward to identify the form of a chiral
symmetry. However, we can apply similarity transformations to $\hat{U}$
and express it in two symmetric time frames \cite{FPiMod2} as
\begin{equation}
	\hat{U}_{1}=e^{-\frac{i}{2}\hat{H}_{1}}e^{-i\hat{H}_{0}}e^{-\frac{i}{2}\hat{H}_{1}},\label{eq:KickU1}
\end{equation}
\begin{equation}
	\hat{U}_{2}=e^{-\frac{i}{2}\hat{H}_{0}}e^{-i\hat{H}_{1}}e^{-\frac{i}{2}\hat{H}_{0}}.\label{eq:KickU2}
\end{equation}
It is then clear that ${\cal S}\hat{U}_{\alpha}{\cal S}=\hat{U}_{\alpha}^{-1}$
for $\alpha=1,2$, that is, the Floquet operators $\hat{U}_{1}$ and
$\hat{U}_{2}$ in the two symmetric time frames respect the same chiral
symmetry ${\cal S}$. We can thus introduce a winding number for each
of them under the PBC as \cite{ZhouNHFTP01}
\begin{equation}
	w_{\alpha}=\int_{-\pi}^{\pi}\frac{dk}{4\pi}{\rm Tr}[{\cal S}{\cal Q}_{\alpha}(k)i\partial_{k}{\cal Q}_{\alpha}(k)],\label{eq:w12}
\end{equation}
where $k\in[-\pi,\pi)$, and
\begin{equation}
	{\cal Q}_{\alpha}(k)=\sum_{n}{\rm sgn}\{{\rm Re}[E_{n}(k)]\}|n_{\alpha}(k)\rangle\langle\widetilde{n}_{\alpha}(k)|.\label{eq:Qa}
\end{equation}
The $E_{n}(k)$ in Eq.~(\ref{eq:Qa}) now denotes the quasienergy of the Floquet eigenstate
$|n_{\alpha}(k)\rangle$ of $\hat{U}_{\alpha}$ at the quasimomentum
$k$ under the PBC. Using the $w_{1}$ and $w_{2}$, we can construct
another pair of winding numbers $w_{0}$ and $w_{\pi}$,
given by \cite{ZhouNHFTP01}
\begin{equation}
	w_{0}=\frac{w_{1}+w_{2}}{2},\qquad w_{\pi}=\frac{w_{1}-w_{2}}{2}.\label{eq:w0p}
\end{equation}

In Subsecs.~\ref{subsec:TI} and \ref{subsec:TSC}, we will demonstrate
with explicit examples that the $w_{0}$ ($w_{\pi}$) could correctly
capture the bulk topological transitions of non-Hermitian Floquet bands
through the gap closing/reopening at the quasienergy $E=0$ ($E=\pi$)
in various chiral symmetric, non-Hermitian Floquet insulating and
superconducting models. Furthermore, in the absence of NHSE, the $w_{0}$
and $w_{\pi}$ could also capture the numbers of Floquet edge modes
at zero and $\pi$ quasienergies under the OBC, and thus are capable of describing
the bulk-edge correspondence of the related models. In the presence
of NHSE, we can retrieve the characterization of topological transitions
and bulk-edge correspondence in chiral symmetric, non-Hermitian Floquet
systems under the OBC through the open-bulk winding numbers, in analogy
with Eq.~(\ref{eq:W0StatOpen}). For the $\hat{U}_{1}$ and $\hat{U}_{2}$
in Eqs.~(\ref{eq:KickU1}) and (\ref{eq:KickU2}), we can define
a winding number for each of them under the OBC as \cite{ZhouNHFTP08}
\begin{equation}
	W_{\alpha}=-\frac{1}{L_{{\rm B}}}{\rm Tr}_{{\rm B}}({\cal S}{\cal Q}_{\alpha}[{\cal Q}_{\alpha},\hat{N}]),\qquad\alpha=1,2.\label{eq:W12}
\end{equation}
Here, the meanings of $L_{{\rm B}}$, ${\rm Tr}_{{\rm B}}$, ${\cal S}$
and $\hat{N}$ are the same as those in Eq.~(\ref{eq:W0StatOpen}).
The Floquet band projector in the time frame $\alpha$ is given by
\begin{equation}
	{\cal Q}_{\alpha}=\sum_{|\psi_{j}^{\alpha}\rangle\in{\rm bulk}}{\rm sgn}[{\rm Re}(E_{j})]|\psi_{j}^{\alpha}\rangle\langle\widetilde{\psi}_{j}^{\alpha}|,
\end{equation}
where $|\psi_{j}^{\alpha}\rangle$ is the $j$th bulk eigenstate of
$\hat{U}_{\alpha}$ ($\alpha=1,2$) with the quasienergy $E_{j}$
under the OBC. The linear combinations of $W_{1}$ and $W_{2}$ lead
to another pair of winding numbers \cite{ZhouNHFTP08}
\begin{equation}
	W_{0}=\frac{W_{1}+W_{2}}{2},\qquad W_{\pi}=\frac{W_{1}-W_{2}}{2}.\label{eq:W0P}
\end{equation}

In Subsec.~\ref{subsec:TI}, we will illustrate that with the help of
the winding numbers $(w_{0},w_{\pi})$ and $(W_{0},W_{\pi})$, a dual
topological characterization of the phase transitions, edge states,
and bulk-edge correspondence can be established for 1D, chiral symmetric
non-Hermitian Floquet systems under different boundary conditions,
regardless of whether the NHSE is present or not \cite{ZhouNHFTP08}.
Interestingly, the winding numbers $(w_{0},w_{\pi})$ may both become half-integer
quantized due to the presence of Floquet EP in the bulk, thus revealing the presence of Floquet exceptional topology. 
Meanwhile, the open-bulk winding numbers $(W_{0},W_{\pi})$ are always integer quantized.

\subsection{Dynamical indicators\label{subsec:DI}}

In this subsetion, we review two complementary dynamical probes in position
and momentum spaces. Both of them can be used to characterize the
topological properties of 1D non-Hermitian Floquet systems with chiral
symmetry. These indicators allow us to extract the topological winding
numbers of the system from its long-time stroboscopic dynamics \cite{ZhouNHFTP01,ZhouNHFTP02,ZhouNHFTP03}.
The measurement of these indicators could thus provide evidences for the existence of
non-Hermitian Floquet topological matter.

\subsubsection{Dynamic winding number (DWN)\label{subsec:DWN}}

The DWN \cite{DWN01,DWN02}, obtained from the long-time stroboscopic
average of spin textures, could provide us with information about
the bulk topological properties of non-Hermitian Floquet systems \cite{ZhouNHFTP03}.
Let us consider a 1D, chiral symmetric non-Hermitian Floquet system
with two quasienergy bands. Under the PBC, we can express its Floquet
operator in the symmetric time frame $\alpha$ ($=1,2$) as $\hat{U}_{\alpha}=\sum_{k\in{\rm BZ}}|k\rangle e^{-iH_{\alpha}(k)}\langle k|$.
Here $H_{\alpha}(k)$ is the effective Hamiltonian in time frame $\alpha$
and $k\in[-\pi,\pi)$ is the quasimomentum. $\hat{U}_{\alpha}$ and
$H_{\alpha}(k)$ share the same chiral symmetry ${\cal S}$, i.e.,
${\cal S}H_{\alpha}(k){\cal S}=-H_{\alpha}(k)$. For a non-Hermitian
system, the $\hat{U}_{\alpha}$ is generally not unitary and $H_{\alpha}(k)$
is also not Hermitian. The right and left biorthonormal eigenvectors
$\{|n_{\alpha}(k)\rangle\}$ and $\{|\widetilde{n}_{\alpha}(k)\rangle\}$
of $H_{\alpha}(k)$ satisfy the eigenvalue equations
\begin{equation}
	H_{\alpha}(k)|n_{\alpha}(k)\rangle=E_{n}(k)|n_{\alpha}(k)\rangle,
\end{equation}
\begin{equation}
	\langle\widetilde{n}_{\alpha}(k)|H_{\alpha}(k)=\langle\widetilde{n}_{\alpha}(k)|E_{n}(k).
\end{equation}
Here $n=\pm$ are the indices of the two Floquet bands with the quasienergies
$E_{\pm}(k)\equiv\pm E(k)$. The biorthonormal relationship requires
\begin{equation}
	\langle\widetilde{n}_{\alpha}(k)|n'_{\alpha}(k)\rangle=\delta_{nn'},
 \quad\sum_{n=\pm}|n_{\alpha}(k)\rangle\langle\widetilde{n}_{\alpha}(k)|=1,\quad\alpha=1,2.
\end{equation}
We consider the case in which the system is prepared in a general
initial state $|\psi_{\alpha}(k,0)\rangle=\sum_{n=\pm}c_{n}(k)|n_{\alpha}(k)\rangle$.
The corresponding initial state in the left Hilbert space reads $|\widetilde{\psi}_{\alpha}(k,0)\rangle=\sum_{n\pm}c_{n}(k)|\widetilde{n}_{\alpha}(k)\rangle$,
such that initially $\sum_{n=\pm}|c_{n}(k)|^{2}=1$ at each $k$.
After the stroboscopic evolution over a number of $\ell$ driving
periods, the right initial state becomes
\begin{equation}
	|\psi_{\alpha}(k,\ell)\rangle=\sum_{n=\pm}c_{n}(k)e^{-i\ell E_{n}(k)}|n_{\alpha}(k)\rangle.
\end{equation}
For the left initial state, we assume it to be evolved by a different
effective Hamiltonian $\widetilde{H}_{\alpha}(k)=\sum_{\pm}E_{n}(k)|\widetilde{n}_{\alpha}(k)\rangle\langle n_{\alpha}(k)|$,
so that after the evolution over $\ell$ driving periods it reaches
the state
\begin{equation}
	|\widetilde{\psi}_{\alpha}(k,\ell)\rangle=\sum_{n=\pm}c_{n}(k)e^{-i\ell E_{n}(k)}|\widetilde{n}_{\alpha}(k)\rangle.
\end{equation}
Note that the dynamical equation of $|\widetilde{\psi}_{\alpha}(k,0)\rangle$
we used here is different from that employed in our study of the APT in Subsec.~\ref{subsec:APT}.

The stroboscopic average of an observable $\hat{O}$ over $|\widetilde{\psi}_{\alpha}(k,t)\rangle$
at the time $t=\ell T$ is then given by
\begin{equation}
	\langle\hat{O}(k,\ell)\rangle_{\alpha}=\frac{\langle\widetilde{\psi}_{\alpha}(k,\ell)|\hat{O}|\psi_{\alpha}(k,\ell)\rangle}{\langle\widetilde{\psi}_{\alpha}(k,\ell)|\psi_{\alpha}(k,\ell)\rangle}.\label{eq:ST}
\end{equation}
Without the loss of generality, we can consider the chiral symmetric
$H_{\alpha}(k)$ to be in the form of
\begin{equation}
	H_{\alpha}(k)=h_{\alpha x}(k)\sigma_{x}+h_{\alpha y}(k)\sigma_{y}.\label{eq:H12DWN}
\end{equation}
Note that any two out of the three Pauli matrices ($\sigma_x, \sigma_y, \sigma_z$) can be chosen to
enter the $H_{\alpha}(k)$, and the rest Pauli matrix {[}e.g., $\sigma_{z}$
for the $H_{\alpha}(k)$ in Eq.~(\ref{eq:H12DWN}){]} plays the
role of the chiral symmetry operator ${\cal S}$. By diagonalizing
$H_{\alpha}(k)$, we obtain the biorthonormal eigenvectors as
\begin{equation}
	|n_{\alpha}(k)\rangle=\frac{1}{\sqrt{2}E_{n}(k)}\begin{bmatrix}h_{\alpha x}(k)-ih_{\alpha y}(k)\\
		E_{n}(k)
	\end{bmatrix},\qquad|\widetilde{n}_{\alpha}(k)\rangle=\frac{1}{\sqrt{2}E_{n}^{*}(k)}\begin{bmatrix}h_{\alpha x}^{*}(k)-ih_{\alpha y}^{*}(k)\\
		E_{n}^{*}(k)
	\end{bmatrix},
\end{equation}
where $E_{n}(k)=n\sqrt{h_{\alpha x}^{2}(k)+h_{\alpha y}^{2}(k)}$
for $n=\pm$. We can now compute the stroboscopic-averaged spin textures
in the long-time limit. For the $H_{\alpha}(k)$ in Eq.~(\ref{eq:H12DWN}),
this means that we need to find the multi-cycle averages of $\sigma_{x}$
and $\sigma_{y}$. According to Eq.~(\ref{eq:ST}), they are given by
\begin{equation}
	r_{j}^{\alpha}(k)\equiv\lim_{N\rightarrow\infty}\frac{1}{N}\sum_{\ell=1}^{N}\langle\sigma_{j}(k,\ell)\rangle_{\alpha},
\end{equation}
where $j=x,y$ and $\alpha=1,2$. $N$ counts the total number of
driving periods. From the averaged spin textures $[r_{x}^{\alpha}(k),r_{y}^{\alpha}(k)]$,
we can define the dynamic winding angle as
\begin{equation}
	\theta_{yx}^{\alpha}(k)\equiv\arctan\left[\frac{r_{y}^{\alpha}(k)}{r_{x}^{\alpha}(k)}\right].
\end{equation}
The net winding number of $\theta_{yx}^{\alpha}(k)$ over a cycle
in $k$-space defines the dynamic winding number (DWN) in the time frame
$\alpha$, i.e.,
\begin{equation}
	\nu_{\alpha}=\int_{-\pi}^{\pi}\frac{dk}{2\pi}\partial_{k}\theta_{yx}^{\alpha}(k),\qquad\alpha=1,2.\label{eq:DWN}
\end{equation}
In Ref.~\cite{ZhouNHFTP03}, it was proved with straightforward calculations
that in the limit $N\rightarrow\infty$, the $\nu_{\alpha}$ 
converges to the $w_{\alpha}$ in Eq.~(\ref{eq:w12}) if the initial
condition satisfies $c_{\pm}(k)\neq0$ at each $k$. Therefore, by
preparing the initial state at different $k$ under this condition
and measuring the averaged spin textures over a long stroboscopic
time, we can obtain the winding numbers $(w_{0},w_{\pi})$ of a chiral
symmetric 1D Floquet system (either Hermitian or non-Hermitian) through
the following combinations of $(\nu_{1},\nu_{2})$ in two symmetric
time frames \cite{ZhouNHFTP03}, i.e.,
\begin{equation}
	w_{0}=\frac{\nu_{1}+\nu_{2}}{2},\qquad w_{\pi}=\frac{\nu_{1}-\nu_{2}}{2}.
\end{equation}
In Sec.~\ref{sec:Res}, we will illustrate the application of
this dynamical-topological correspondence to non-Hermitian Floquet
topological insulators in one dimension. We will see that both integer
and half-integer quantized topological winding numbers can be extracted
from the DWN.

\subsubsection{Mean chiral displacement (MCD)\label{subsec:MCD}}

The MCD allows us to detect the winding numbers of a Floquet system
from the long-time averaged chiral displacement of an initially localized
wavepacket. It was first proposed as a means to probe the topological
invariants of chiral symmetric topological insulators in one dimension
\cite{MCD01,MCD02,MCD03}. Later, the MCD was used to obtain the winding
numbers of Floquet systems \cite{FTPLg04,FTPLg08} and also generalized
to 2D higher-order topological insulators \cite{FTPLg07}. Its applicability
was demonstrated both theoretically and experimentally \cite{MCD01,MCD03}.

For a chiral symmetric non-Hermitian system, the chiral displacement
in the symmetric time frame $\alpha$ ($=1,2$) can be defined as
\begin{equation}
	C_{\alpha}(t)={\rm Tr}[\rho_{0}(\hat{\widetilde{U}}_{\alpha}^{\dagger})^{t}(\hat{N}\otimes{\cal S})\hat{U}_{\alpha}^{t}].\label{eq:CD12}
\end{equation}
Here, $t=\ell T$ denotes the stroboscopic time, with $T$ being the
driving period. $\hat{N}$ is the unit cell position operator and
${\cal S}$ is the chiral symmetry operator. The initial state $\rho_{0}$
can be chosen as a state localized in the middle of the lattice. For
example, $\rho_{0}$ may take the form of $(|0\rangle\langle0|\otimes\sigma_{0})/2$
for a 1D bipartite lattice, where $|0\rangle$ is the eigenbasis of
the central unit cell and $\sigma_{0}$ is the identity operator acting on the
internal space of the two sublattices. Both the Floquet operator $\hat{U}_{\alpha}$
and its dual $\hat{\widetilde{U}}_{\alpha}$ respect the chiral symmetry
${\cal S}$. In the lattice representation, they can be expressed as
\begin{equation}
	\hat{U}_{\alpha}=\sum_{j=1}^{L}e^{-iE_{j}}|\psi_{j}^{\alpha}\rangle\langle\widetilde{\psi}_{j}^{\alpha}|,\qquad\hat{\widetilde{U}}_{\alpha}=\sum_{j=1}^{L}e^{-iE_{j}}|\widetilde{\psi}_{j}^{\alpha}\rangle\langle\psi_{j}^{\alpha}|.
\end{equation}
Here $L$ denotes the total number of degrees of freedom of the lattice.
$|\psi_{j}^{\alpha}\rangle$ and $\langle\widetilde{\psi}_{j}^{\alpha}|$
denote the right and left eigenvectors of $\hat{U}_{\alpha}$ with
the quasienergy $E_{j}$. They form a biorthonormal basis such that
\begin{equation}
\langle\widetilde{\psi}_{j}^{\alpha}|\psi_{m}^{\alpha}\rangle=\delta_{jm},\qquad\sum_{j=1}^{L}|\psi_{j}^{\alpha}\rangle\langle\widetilde{\psi}_{j}^{\alpha}|=1.
\end{equation}
Note that $\hat{\widetilde{U}}_{\alpha}$ is not the Hermitian conjugate of $\hat{U}_{\alpha}$ in general.

We now consider the stroboscopic long-time average of $C_{\alpha}(t)$
in the time frames $\alpha=1,2$ for a 1D non-Hermitian Floquet system
with chiral symmetry. Under the PBC, taking the initial state to be
$\rho_{0}=(|0\rangle\langle0|\otimes\sigma_{0})/2$ and performing
the Fourier transformation from position to momentum representations,
we find the $C_{\alpha}(t)$ in Eq.~(\ref{eq:CD12}) to be \cite{ZhouNHFTP02}
\begin{equation}
	C_{\alpha}(t)=\frac{1}{2}\int_{-\pi}^{\pi}\frac{dk}{2\pi}{\rm Tr}[\widetilde{U}_{\alpha}^{\dagger t}(k){\cal S}i\partial_{k}U_{\alpha}^{t}(k)].
\end{equation}
Here $k\in[-\pi,\pi)$ is the quasimomentum. $U_{\alpha}(k)=\langle k|\hat{U}_{\alpha}|k\rangle$
and $\widetilde{U}_{\alpha}(k)=\langle k|\hat{\widetilde{U}}_{\alpha}|k\rangle$
act on the internal degrees of freedom (spins and/or sublattices)
of the system at a fixed $k$. Taking the long-time stroboscopic average
and incorporating the normalization factor ${\rm Tr}[\rho_{0}(\hat{\widetilde{U}}_{\alpha}^{\dagger})^{t}\hat{U}_{\alpha}^{t}]$,
we find the expression of MCD as \cite{ZhouNHFTP02}
\begin{equation}
	\overline{C}_{\alpha}=\lim_{\ell\rightarrow\infty}\frac{1}{\ell T}\sum_{t=T}^{\ell T}\int_{-\pi}^{\pi}\frac{dk}{2\pi}\frac{{\rm Tr}\left[\widetilde{U}_{\alpha}^{\dagger t}(k){\cal S}i\partial_{k}U_{\alpha}^{t}(k)\right]}{{\rm Tr}\left[\widetilde{U}_{\alpha}^{\dagger t}(k)U_{\alpha}^{t}(k)\right]}.\label{eq:MCD0}
\end{equation}
For a given $H_{\alpha}(k)$ of the form in Eq.~(\ref{eq:H12DWN}),
we have ${\cal S}=\sigma_{z}$, $U_{\alpha}(k)=e^{-iH_{\alpha}(k)}$,
and $\widetilde{U}_{\alpha}(k)=e^{-iE(k)H_{\alpha}^{\dagger}(k)/E^{*}(k)}$,
where $E(k)=\frac{1}{2}{\rm Tr}[H_{\alpha}^{2}(k)]$. One can then
work out Eq.~(\ref{eq:MCD0}) explicitly and obtain \cite{ZhouNHFTP02}
\begin{equation}
	\overline{C}_{\alpha}=\frac{w_{\alpha}}{2},\qquad\alpha=1,2.\label{eq:MCD12}
\end{equation}
Here the $w_{\alpha}$ is nothing but the winding number of $\hat{U}_{\alpha}$
as defined in Eq.~(\ref{eq:w12}). Therefore, by measuring the
MCDs $(\overline{C}_{1},\overline{C}_{2})$ in two symmetric time
frames, we can determine the topological winding numbers $(w_{0},w_{\pi})$
of a 1D chiral symmetric Floquet system through the relations \cite{ZhouNHFTP02}
\begin{equation}
	w_{0}=C_{0}\equiv\overline{C}_{1}+\overline{C}_{2},\qquad w_{\pi}=C_{\pi}\equiv\overline{C}_{1}-\overline{C}_{2}.
\end{equation}
The MCD provides a real-space complementary to the DWN.
In Sec.~\ref{sec:Res}, we will demonstrate the usage of MCD to
dynamically probing the winding numbers of first- and second-order
non-Hermitian Floquet topological insulators in one \cite{ZhouNHFTP02}
and two \cite{ZhouNHFTP06} spatial dimensions.

\subsection{Localization transition and mobility edge\label{subsec:LT}}

In the last part of this subsection, we recap some tools that can be
used to characterize the real-to-complex spectral transitions, localization
transitions, and mobility edges in disordered non-Hermitian Floquet
systems \cite{ZhouNHFTP09,ZhouNHFTP11}. Considering the relevance
to this review, we focus on a 1D quadratic lattice model of the form
\begin{equation}
	\hat{H}(t)=\sum_{\langle n,n'\rangle}[J_{n}(t)e^{\gamma_{n}}\hat{c}_{n}^{\dagger}\hat{c}_{n'}+J_{n}^{*}(t)e^{-\gamma_{n}}\hat{c}_{n'}^{\dagger}\hat{c}_{n}]+\sum_{n}V_{n}(t)\hat{c}_{n}^{\dagger}\hat{c}_{n}.\label{eq:HQC}
\end{equation}
Here $\langle n,n'\rangle$ includes the lattice site indices $n$
and $n'$ with $n'>n$. $\hat{c}_{n}^{\dagger}$ ($\hat{c}_{n}$)
creates (annihilates) a particle on the site $n$ and the parameter
$\gamma\in\mathbb{R}$. The Hamiltonian $\hat{H}(t)$ is non-Hermitian
if $\gamma_{n}\neq0$ for some $n$ (nonreciprocal hopping) or $V_{n}(t)\neq V_{n}^{*}(t)$
(onsite gain and loss). $\hat{H}(t)$ is further time-periodic if
$J_{n}(t)=J_{n}(t+T)$ and $V_{n}(t)=V_{n}(t+T)$ for all $n$, with
$T$ being the driving period. The disorder terms may be included
within $V_{n}(t)$ (diagonal disorder) or $J_{n}(t)$ (off-diagonal
disorder). As an example, for a 1D non-Hermitian quasicrystal (NHQC)
with correlated onsite disorder, the $V_{n}(t)$ may take the form
of $V_{n}(t)=V(t)\cos(2\pi\alpha n+i\beta)$. Here $\alpha$ is irrational
and $\beta\in\mathbb{R}$, with $i\beta$ describing an imaginary
phase shift in the superlattice potential $V_{n}$.

The Floquet operator of the system described by the $\hat{H}(t)$
in Eq.~(\ref{eq:HQC}) takes the general form of $\hat{U}=\hat{\mathsf{T}}e^{-i\int_{0}^{T}\hat{H}(t)dt}=e^{-i\hat{H}_{{\rm eff}}T}$.
If $\hat{H}(t)$ respects the PT-symmetry, $\hat{H}_{{\rm eff}}$
could have the same symmetry and the quasienergy spectrum of $\hat{U}$
could be real. With the increase of the non-Hermitian parameter of
$\hat{H}(t)$, the Floquet spectrum of $\hat{U}$ may undergo a PT-breaking
transition, after which certain quasienergies of $\hat{U}$ (obtained
by solving $\hat{U}|\Psi\rangle=e^{-iE}|\Psi\rangle$) may acquire
nonvanishing imaginary parts. To take into account such a spectral
transition, we introduce the following quantities
\begin{equation}
	\max|{\rm Im}E|\equiv\max_{j\in\{1,...,L\}}(|{\rm Im}E_{j}|),\label{eq:EImax}
\end{equation}
\begin{equation}
	\rho\equiv N({\rm Im}E\neq0)/L.\label{eq:DOS}
\end{equation}
Here $E_{j}$ is the $j$th quasienergy eigenvalue of $\hat{U}$.
$L$ counts the Hilbert space dimension of $\hat{U}$, i.e., the total
number of degrees of freedom of the lattice. $N({\rm Im}E\neq0)$
denotes the number of quasienergy eigenvalues whose imaginary parts
are nonzero. $\rho$ thus describes the density of states with complex
quasienergies in the system. In the PT-invariant phase, we would have
$\max|{\rm Im}E|=\rho=0$, implying that all the quasienergies of
$\hat{U}$ are real. In the PT-broken phase, we would have $\max|{\rm Im}E|>0$
and $\rho\in(0,1]$, meaning that there is a finite number of eigenstates
of $\hat{U}$ whose quasienergies are complex. Specially, we would
have $\rho\simeq1$ if almost all the Floquet eigenstates of $\hat{U}$
possess complex quasienergies. Therefore, by locating the positions
where both the $\max|{\rm Im}E|$ and $\rho$ start to deviate from
zero in the parameter space, we can identify the PT transition points
for the Floquet spectrum of $\hat{U}$. We will see that due to the
interplay between periodic drivings and non-Hermitian effects, both
PT-symmetry breaking and restoring transitions could be induced by
tuning a single parameter in the Floquet NHQC.

The localization nature of the Floquet eigenstates of $\hat{U}$ can
be characterized by the statistics of its quasienergy levels and the
inverse participation ratios (IPRs). Let us denote the normalized
right eigenvectors of $\hat{U}$ and their corresponding quasienergies
as $\{|\psi_{j}\rangle|j=1,...,L\}$ and $\{E_{j}|j=1,...,L\}$. Along
the real axis, the spacing between the $j$th and the $(j-1)$th quasienergies
of $\hat{U}$ is given by $\epsilon_{j}={\rm Re}E_{j}-{\rm Re}E_{j-1}$,
from which we obtain the ratio between two adjacent spacings of quasienergy
levels as
\begin{equation}
	g_{j}=\frac{\min(\epsilon_{j},\epsilon_{j+1})}{\max(\epsilon_{j},\epsilon_{j+1})},\qquad j=2,...,L-1.
\end{equation}
Here the $\max(\epsilon_{j},\epsilon_{j+1})$ and $\min(\epsilon_{j},\epsilon_{j+1})$
are the maximum and minimum between the $\epsilon_{j}$ and $\epsilon_{j+1}$,
respectively. The statistical property of adjacent gap ratios can
then be obtained by averaging over all $g_{j}$ in the thermodynamic
limit, i.e.,
\begin{equation}
	\overline{g}=\lim_{L\rightarrow\infty}\frac{1}{L}\sum_{j}g_{j}.\label{eq:AGR}
\end{equation}
We have $\overline{g}\rightarrow0$ if all the bulk Floquet eigenstates
of $\hat{U}$ are extended. Comparatively, we expect $\overline{g}$
to approach a constant $\overline{g}_{\max}>0$ if all the bulk eigenstates
of $\hat{U}$ are localized. If we find $\overline{g}\in(0,\overline{g}_{\max})$,
extended and localized eigenstates of $\hat{U}$ should coexist and
be separated by some mobility edges in the quasienergy spectrum. The
$\overline{g}$ can thus be utilized to distinguish between phases with different
localization nature in 1D non-Hermitian Floquet systems from the perspective
of level statistics. In the lattice representation, we can expand
the right eigenvector $|\psi_{j}\rangle$ as $|\psi_{j}\rangle=\sum_{n=1}^{L}\psi_{n}^{j}|n\rangle$,
where $\psi_{n}^{j}=\langle n|\psi_{j}\rangle$ and $\sum_{n}|\psi_{n}^{j}|^{2}=1$.
The inverse and normalized participation ratios of $|\psi_{j}\rangle$
in the real space can then be defined as
\begin{equation}
	{\rm IPR}_{j}=\sum_{n=1}^{L}|\psi_{n}^{j}|^{4},\qquad{\rm NPR}_{j}=\frac{1}{L}{\rm IPR}_{j}^{-1}.
\end{equation}
For a localized Floquet eigenstate $|\psi_{j}\rangle$, we have
${\rm IPR}_{j}\rightarrow\lambda_{j}$ and ${\rm NPR}_{j}\rightarrow0$
in the thermodynamic limit, where the Lyapunov exponent $\lambda_{j}$
of $|\psi_{j}\rangle$ could be a function of its quasienergy $E_{j}$.
If $|\psi_{j}\rangle$ represents an extended state, we 
have ${\rm IPR}_{j}\rightarrow0$ and ${\rm NPR}_{j}\rightarrow1$.
The global localization features of the Floquet system described by
$\hat{U}$ can then be extracted from the combined information of
$\{{\rm IPR}_{j}|j=1,...,L\}$ and $\{{\rm NPR}_{j}|j=1,...,L\}$.
For ease of usage, we introduce the following localization indicators
\begin{equation}
	{\rm IPR}_{\max}=\max_{j\in\{1,...,L\}}({\rm IPR}_{j}),
\end{equation}
\begin{equation}
	{\rm IPR}_{\min}=\min_{j\in\{1,...,L\}}({\rm IPR}_{j}),
\end{equation}
\begin{equation}
	\zeta=\log_{10}({\rm IPR}_{{\rm ave}}\cdot{\rm NPR}_{{\rm ave}}),
\end{equation}
where the ${\rm IPR}_{{\rm ave}}=\frac{1}{L}\sum_{j=1}^{L}{\rm IPR}_{j}$
and ${\rm NPR}_{{\rm ave}}=\frac{1}{L}\sum_{j=1}^{L}{\rm NPR}_{j}$
are the averages of ${\rm IPR}_{j}$ and ${\rm NPR}_{j}$ over all
Floquet eigenstates. It is not hard to see that in the metallic
phase where all Floquet eigenstates are extended, we have ${\rm IPR}_{\max}\rightarrow0$,
${\rm IPR}_{\min}\rightarrow0$ and $\zeta\sim-\log_{10}(L)\rightarrow-\infty$
in the thermodynamic limit $L\rightarrow\infty$. Instead, in the insulating
phase with all Floquet eigenstates being localized, we
have ${\rm IPR}_{\max}>0$, ${\rm IPR}_{\min}>0$ and $\zeta\rightarrow-\infty$.
In the critical phase where extended and localized Floquet eigenstates
are coexistent, we have ${\rm IPR}_{\max}>0$, ${\rm IPR}_{\min}\rightarrow0$
together with a finite $\zeta$. Assembling the information obtained
from ${\rm IPR}_{\max}$, ${\rm IPR}_{\min}$, and $\zeta$ thus allows
us to distinguish the extended, localized, and critical mobility edge
phases of a non-Hermitian Floquet system with disorder. These tools
will be applied to our study of the Floquet NHQC in Subsec.~\ref{subsec:TQC}.

We can also probe the transport nature of distinct non-Hermitian Floquet
phases from the wave packet dynamics. Let us consider a generic and
normalized initial state $|\Psi(0)\rangle$ in the lattice representation.
After the stroboscopic evolution over a number of $\ell$ driving
periods by $\hat{U}$, the final state turns out to be $|\Psi'(t=\ell T)\rangle=\hat{U}^{\ell}|\Psi(0)\rangle$.
Since the Floquet operator $\hat{U}$ is not unitary for a non-Hermitian
$\hat{H}(t)$ in general, the norm of $|\Psi(0)\rangle$ cannot be
preserved during the evolution. We can express the normalized state
at $t=\ell T$ as $|\Psi(t)\rangle=|\Psi'(t)\rangle/\sqrt{\langle\Psi'(t)|\Psi'(t)\rangle}$.
The real space expansion $|\Psi(t)\rangle=\sum_{n=1}^{L}\psi_{n}(t)|n\rangle$
then provides us with the probability amplitude $\psi_{n}(t)=\langle n|\Psi(t)\rangle$
of the normalized final state $|\Psi(t)\rangle$ on the lattice site
$n$ at time $t$. Using the collection of amplitudes $\{\psi_{n}(t)|n=1,...,L\}$,
we can define the following dynamical quantities
\begin{equation}
	\langle x(t)\rangle=\sum_{n=1}^{L}n|\psi_{n}(t)|^{2},\qquad\langle x^{2}(t)\rangle=\sum_{n=1}^{L}n^{2}|\psi_{n}(t)|^{2},\label{eq:Avexx2}
\end{equation}
\begin{equation}
	\Delta x(t)=\sqrt{\langle x^{2}(t)\rangle-\langle x(t)\rangle^{2}},\qquad v(t)=\frac{1}{t}\sqrt{\langle x^{2}(t)\rangle},\label{eq:DelxVt}
\end{equation}
where the stroboscopic time $t=\ell T$ and $\ell\in\mathbb{Z}$.
It is clear that the $\langle x(t)\rangle$, $\Delta x(t)$, and $v(t)$
describe the center, standard deviation, and spreading speed of the
wavepacket $|\Psi(t)\rangle$ in the lattice representation, respectively.
For simplicity, we usually choose the initial state $|\Psi(0)\rangle$
to be exponentially localized at a single site that is deep inside
the bulk of the lattice. If the Floquet system described by $\hat{U}$
resides in a localized phase, we expect the $\langle x(t)\rangle$
and $\Delta x(t)$ to stay around their initial values, which means
that the wavepacket almost does not move and spread. In this case,
the speed $v(t)$ should also tend to zero for a long-time evolution
($\ell\gg1$). If the system stays in an extended phase, we expect
the increasing of $\Delta x(t)$ with time due to the spreading of
the initial wavepacket. Meanwhile, the $\langle x(t)\rangle$ may
or may not change with time, depending on whether the hopping amplitudes
are symmetric. For a nearest-neighbor, nonreciprocal hopping as in
Eq.~(\ref{eq:HQC}), we may have $\langle x(t)\rangle\propto t$ and
$\Delta x(t)\propto\sqrt{t}$ for a metallic phase of the system.
The $v(t)$ should take a maximal possible value $v_{\max}(t)$ in
this case. If the system is prepared in a critical mobility edge phase,
we expect both the $\langle x(t)\rangle$ and $\Delta x(t)$ to show
intervening behaviors, while the averaged spreading speed $v(t)$
should satisfy $0<v(t)<v_{\max}(t)$. Therefore, we can exploit the
dynamical quantities $\langle x(t)\rangle$, $\Delta x(t)$, and $v(t)$
to discriminate phases with different localization properties in disordered
non-Hermitian Floquet systems. We will illustrate such an application
for Floquet NHQC in Subsec.~\ref{subsec:TQC}.

\section{Non-Hermitian Floquet phases of matter\label{sec:Res}}

This section collects and treats typical examples of non-Hermitian
Floquet phases discovered in our previous work \cite{ZhouNHFTP01,ZhouNHFTP02,ZhouNHFTP03,ZhouNHFTP04,ZhouNHFTP05,ZhouNHFTP06,ZhouNHFTP07,ZhouNHFTP08,ZhouNHFTP09,ZhouNHFTP10,ZhouNHFTP11}.
We will see that the interplay between periodic driving fields
and gain/loss or nonreciprocal effects could induce rich phases and
transitions in non-Hermitian Floquet systems.

\subsection{Non-Hermitian Floquet exceptional topology\label{subsec:FET}}

Let us start with a simple model, whose Floquet effective Hamiltonian
and quasienergy dispersion are given by Eqs.~(\ref{eq:HeffEP})
and (\ref{eq:QuasiE}). To be explicit, we choose 
\begin{equation}
	h_{x}=\sin k,\qquad h_{z}=\mu+\cos k,\label{eq:hxz}
\end{equation}
where $k\in(-\pi,\pi]$ denotes the quasimomentum in the first Brillouin
zone, and the system parameter $\mu\in\mathbb{R}$. The $H_{{\rm eff}}$
in Eq.~(\ref{eq:HeffEP}) thus describes the Bloch Hamiltonian of
a 1D two-band lattice model under the PBC in the rotating frame, where $\mu$
corresponds to the amplitude of onsite potential and the nearest-neighbor
hopping amplitude has been chosen to be the unit of energy. From
Eqs.~(\ref{eq:QuasiE}) and (\ref{eq:hxz}), we find that the two
quasienergy bands of $H_{{\rm eff}}$ could meet with each other at
the quasienergy $\omega/2$ when 
\begin{equation}
	\mu=\begin{cases}
		\frac{\omega}{2}\pm\frac{\gamma}{2}-1, & {\rm for}\quad k=0,\\
		\frac{\omega}{2}\pm\frac{\gamma}{2}+1, & {\rm for}\quad k=\pi.
	\end{cases}\label{eq:muPBC}
\end{equation}

If $\mu$ satisfies one of the above two equalities, there will be
a second-order Floquet EP at $k=0$ or $k=\pi$ in the conventional
Brillouin zone. Note that both the quasienergy of this Floquet EP
and its location in the parameter space depend on the frequency $\omega$,
which highlights the impact of the harmonic driving field on phase
transitions in the system. Moreover, the appearance of an EP in $k$-space
usually implies the breakdown of bulk-edge correspondence in conventional
topological phases. This issue might be overcome by incorporating
the formalism of generalized Brillouin zone (GBZ) and non-Bloch band
theory \cite{NHSE05,NHSE07}. Following the standard recipe, we first
make the substitutions $e^{ik}\rightarrow\beta$ and $e^{-ik}\rightarrow\beta^{-1}$,
where the complex number $\beta\in{\rm GBZ}$. The effective Hamiltonian
in Eq.~(\ref{eq:HeffEP}) now takes the form of 
\begin{equation}
	H_{{\rm eff}}(\beta)=\frac{\omega}{2}\sigma_{0}+\left(\frac{\beta-\beta^{-1}}{2i}+i\frac{\gamma}{2}\right)\sigma_{x}+\left(\mu+\frac{\beta+\beta^{-1}}{2}-\frac{\omega}{2}\right)\sigma_{z},
\end{equation}
with the quasienergy bands 
\begin{equation}
	\varepsilon_{\pm}(\beta)=\frac{\omega}{2}\pm\sqrt{\left(\frac{\beta-\beta^{-1}}{2i}+\frac{i\gamma}{2}\right)^{2}+\left(\mu-\frac{\omega}{2}+\frac{\beta+\beta^{-1}}{2}\right)^{2}}.
\end{equation}
We next focus on the non-constant part of $\varepsilon_{\pm}(\beta)$,
whose square takes the form of
\begin{equation}
	\epsilon_{\pm}^{2}(\beta)=\frac{a\beta^{2}+b\beta+c}{\beta},\label{eq:epmbeta2}
\end{equation}
where
\begin{equation}
	a=\mu-\frac{\omega}{2}+\frac{\gamma}{2},\qquad b=1+\left(\mu-\frac{\omega}{2}\right)^{2}-\frac{\gamma^{2}}{4},\qquad c=\mu-\frac{\omega}{2}-\frac{\gamma}{2}.
\end{equation}
The $\varepsilon_{\pm}^{2}(\beta)$ is a Laurent polynomial of $\beta$.
According to Ref.~\cite{NHSE07} (see also Ref.~\cite{NHCI01}), $\beta\in{\rm GBZ}$
for the $H_{{\rm eff}}(\beta)$ if $\epsilon_{\pm}^{2}(\beta)=\epsilon_{\pm}^{2}(\beta e^{i\theta})$,
with $\theta$ being a phase factor. For Eq.~(\ref{eq:epmbeta2}),
this means that $\beta^{2}=ce^{-i\theta}/a$. The GBZ is thus a circle
of the radius
\begin{equation}
	r=|\beta|=\left|\frac{\mu-\omega/2-\gamma/2}{\mu-\omega/2+\gamma/2}\right|^{1/2}.\label{eq:r}
\end{equation}
It is clear that the GBZ radius is controlled by both the driving
field (through $\omega$) and the non-Hermitian effect (through $\gamma$)
. In the Hermitian limit ($\gamma\rightarrow0$), we have $r\rightarrow1$
and the GBZ is reduced to the conventional BZ, as expected. The GBZ
becomes ill defined if $\mu-\omega/2=\pm\gamma/2$, where the radius
$r$ is zero or infinity. Finally, we can use the GBZ to determine
the gap closing (phase transition) points of the system under the
OBC. Setting the $\epsilon_{\pm}^{2}(\beta)=0$ in Eq.~(\ref{eq:epmbeta2}),
we find
\begin{equation}
	\mu=\frac{\omega}{2}\pm\begin{cases}
		\sqrt{\gamma^{2}/4+1}, & |\gamma|\leq2,\\
		\sqrt{\gamma^{2}/4\pm1}, & |\gamma|>2.
	\end{cases}\label{eq:muOBC}
\end{equation}
These bulk gap-closing conditions are clearly different from those
found in conventional BZ under the PBC in Eq.~(\ref{eq:muPBC}).
Using the transfer matrix method \cite{NHSE09}, we can further obtain
the parameter regions in which degenerate Floquet edge states appear
at the quasienergy $\omega/2$ (anomalous Floquet $\pi$ edge modes),
i.e.,
\begin{equation}
	\mu-\frac{\omega}{2}\in\begin{cases}
		\left(-\sqrt{\gamma^{2}/4+1},\sqrt{\gamma^{2}/4+1}\right), & |\gamma|\leq2,\\
		\left(-\sqrt{\gamma^{2}/4+1},-\sqrt{\gamma^{2}/4-1}\right)\cup\left(\sqrt{\gamma^{2}/4-1},\sqrt{\gamma^{2}/4+1}\right), & |\gamma|>2.
	\end{cases}\label{eq:muEdge}
\end{equation}
Finally, we can characterize the different topological phases of the
system by a non-Bloch winding number \cite{NHSE10}
\begin{equation}
	W=\int_{{\rm GBZ}}\frac{d\beta}{4\pi i}{\rm Tr}[\sigma_{y}{\cal Q}(\beta)\partial_{\beta}{\cal Q}(\beta)],\label{eq:WGBZ}
\end{equation}
where $\sigma_{y}$ is the chiral symmetry operator of the shifted
effective Hamiltonian $H_{{\rm eff}}(\beta)-\frac{\omega}{2}\sigma_{0}$.
The Q matrix is defined as ${\cal Q}(\beta)=|\psi_{+}(\beta)\rangle\langle\widetilde{\psi}_{+}(\beta)|-|\psi_{-}(\beta)\rangle\langle\widetilde{\psi}_{-}(\beta)|$.
$|\psi_{\pm}(\beta)\rangle$ are the right eigenvectors of $H_{{\rm eff}}(\beta)$
with the quasienergies $\varepsilon_{\pm}(\beta)$. $\langle\widetilde{\psi}_{\pm}(\beta)|$
are the corresponding left eigenvectors. For our model, their explicit
expressions are given by
\begin{equation}
	|\psi_{\pm}(\beta)\rangle=\frac{1}{\sqrt{2(\varepsilon_{\pm}-\frac{\omega}{2})(\varepsilon_{\pm}-h_{z})}}\begin{pmatrix}h_{x}+i\gamma/2\\
		\varepsilon_{\pm}-h_{z}
	\end{pmatrix},
\end{equation}
\begin{equation}
	\langle\widetilde{\psi}_{\pm}(\beta)|=\frac{1}{\sqrt{2(\varepsilon_{\pm}-\frac{\omega}{2})(\varepsilon_{\pm}-h_{z})}}\begin{pmatrix}h_{x}+i\gamma/2 & \varepsilon_{\pm}-h_{z}\end{pmatrix}.
\end{equation}

\begin{figure}
	\begin{centering}
		\includegraphics[scale=0.6]{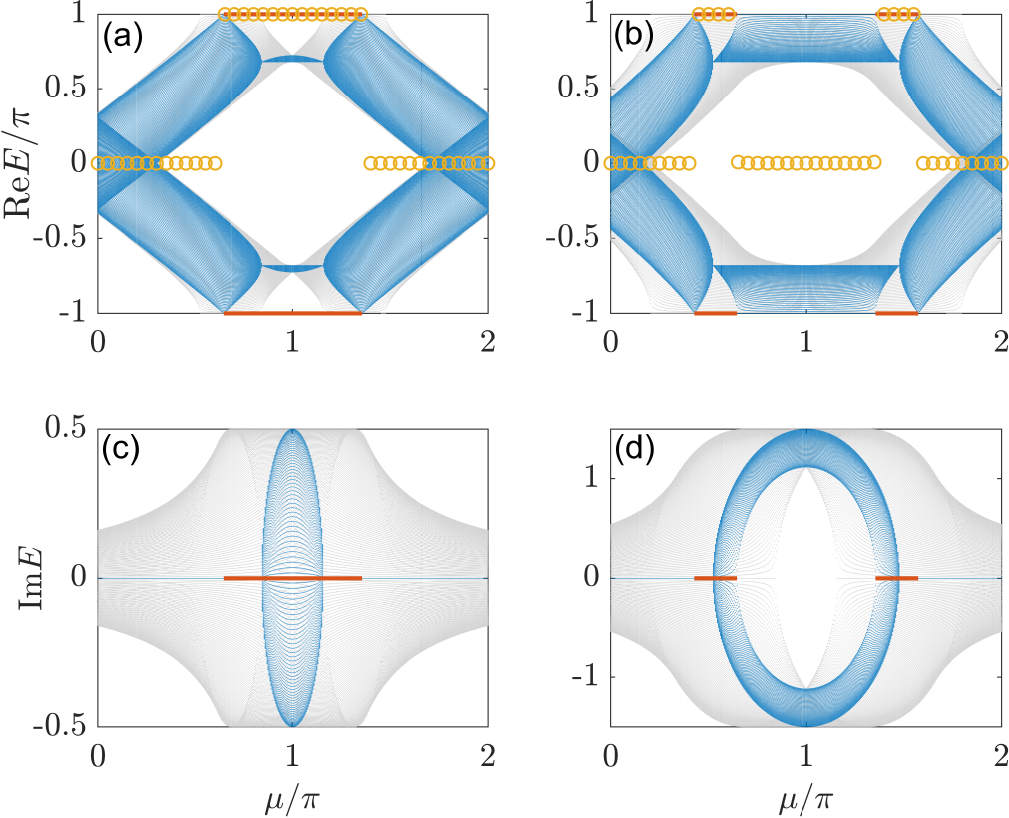}
		\par\end{centering}
	\caption{Floquet spectra and winding numbers of the harmonically driven 1D
		non-Hermitian lattice model. The gray dots at each $\mu$ denote the
		real {[}in (a) and (b){]} and imaginary {[}in (c) and (d){]} parts
		of quasienergies obtained under the PBC in the conventional BZ. The
		blue dots at each $\mu$ denote the real {[}in (a) and (b){]} and
		imaginary {[}in (c) and (d){]} parts of quasienergies obtained under
		the OBC in the GBZ. The yellow circles denote the winding number $W$,
		which is equal to $1$ ($0$) in the topological (trivial) phases with
		(without) anomalous Floquet edge modes at the quasienergies $\pm\pi$,
		which are highlighted by the red solid lines. Other system parameters
		are chosen as $(\omega,\gamma)=(2\pi,1)$ {[}$(\omega,\gamma)=(2\pi,3)${]}
		for the panels (a) and (c) {[}(b) and (d){]}. \label{fig:FET}}
\end{figure}

In Fig.~\ref{fig:FET}, we show the Floquet spectra of $H_{{\rm eff}}$
under different boundary conditions and the non-Bloch winding numbers
of $H_{{\rm eff}}$ versus $\mu$ for some typical cases. We see that
the Floquet spectra and the gap closing points at the quasienergies
$E=\pm\pi$ could indeed be very different under the PBC and the OBC.
Some high-order EPs can be identified from the bulk Floquet spectra
under the OBC, which clearly signify the emergence of Floquet exceptional
topology. Besides, the non-Bloch winding number $W$ can correctly
discriminate different topological phases and characterize topological
phase transitions accompanied by quasienergy-band touchings under
the OBC at $E=\pi$. We find $W=1$ ($W=0$) in the topologically
nontrivial (trivial) phases with (without) anomalous Floquet $\pi$
edge modes. The bulk-edge correspondence is then recovered. The gap-closing
points and the parameter regions with Floquet $\pi$ modes are found
to be perfectly coincident with the predictions of Eqs.~(\ref{eq:muOBC})
and (\ref{eq:muEdge}). These observations demonstrate the applicability
of non-Bloch band theories to non-Hermitian Floquet systems. It is
also straightforward to check that the system possesses Floquet NHSEs
in a broad parameter regime under the OBC. The simple model introduced
in this subsection thus allows us to get a bird's-eye view on the
nontrivial topological phenomena that could be brought about by the
interplay between Floquet driving fields and non-Hermitian effects.
Further examples with more striking features will be reviewed in the
following subsections.

\subsection{Non-Hermitian Floquet topological insulators\label{subsec:TI}}

In this subsection, we review three types of non-Hermitian Floquet
topological insulators in one and two spatial dimensions. For all
the cases, we uncover that the collaboration between driving and gain/loss
or nonreciprocal effects could induce topological insulating states
unique to non-Hermitian Floquet systems. They are featured by large
topological invariants, many topological edge or corner modes, and
separated by rich topological phase transitions. The bulk-edge (or
bulk-corner) correspondence will also be established for each class of
systems considered in this subsection.

\subsubsection{First-order topological phase\label{subsec:1TI}}

\begin{figure}
	\begin{centering}
		\includegraphics[scale=0.34]{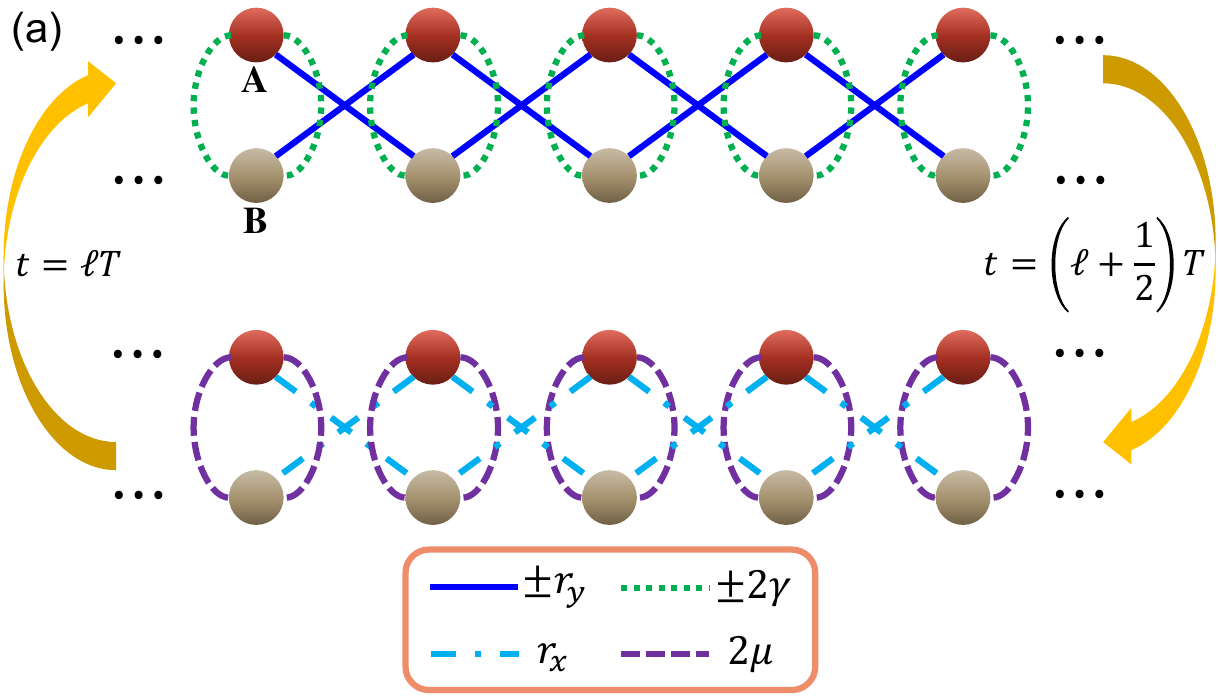}\includegraphics[scale=0.315]{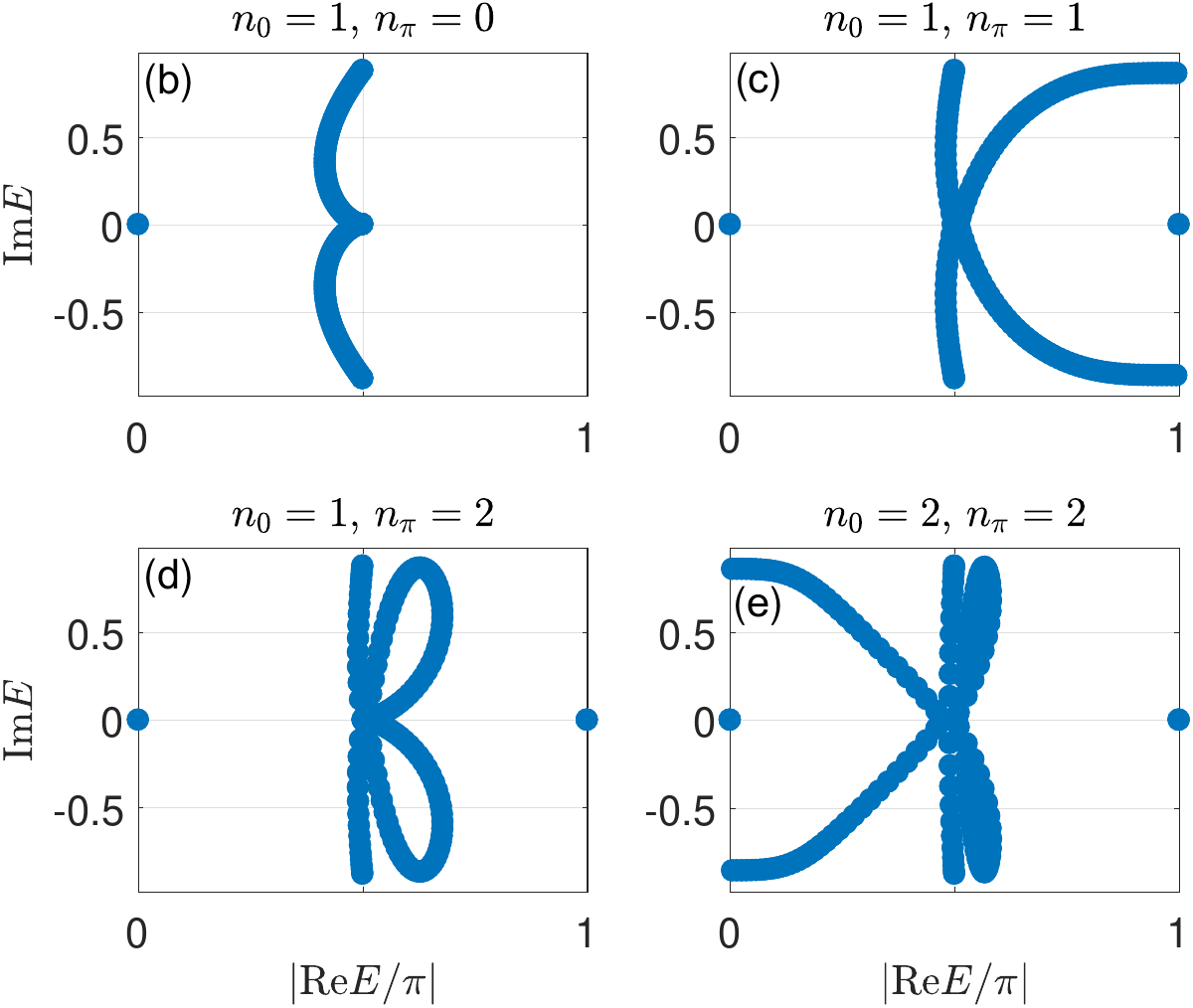}
		\par\end{centering}
	\medskip{}
	
	\begin{centering}
		\includegraphics[scale=0.3]{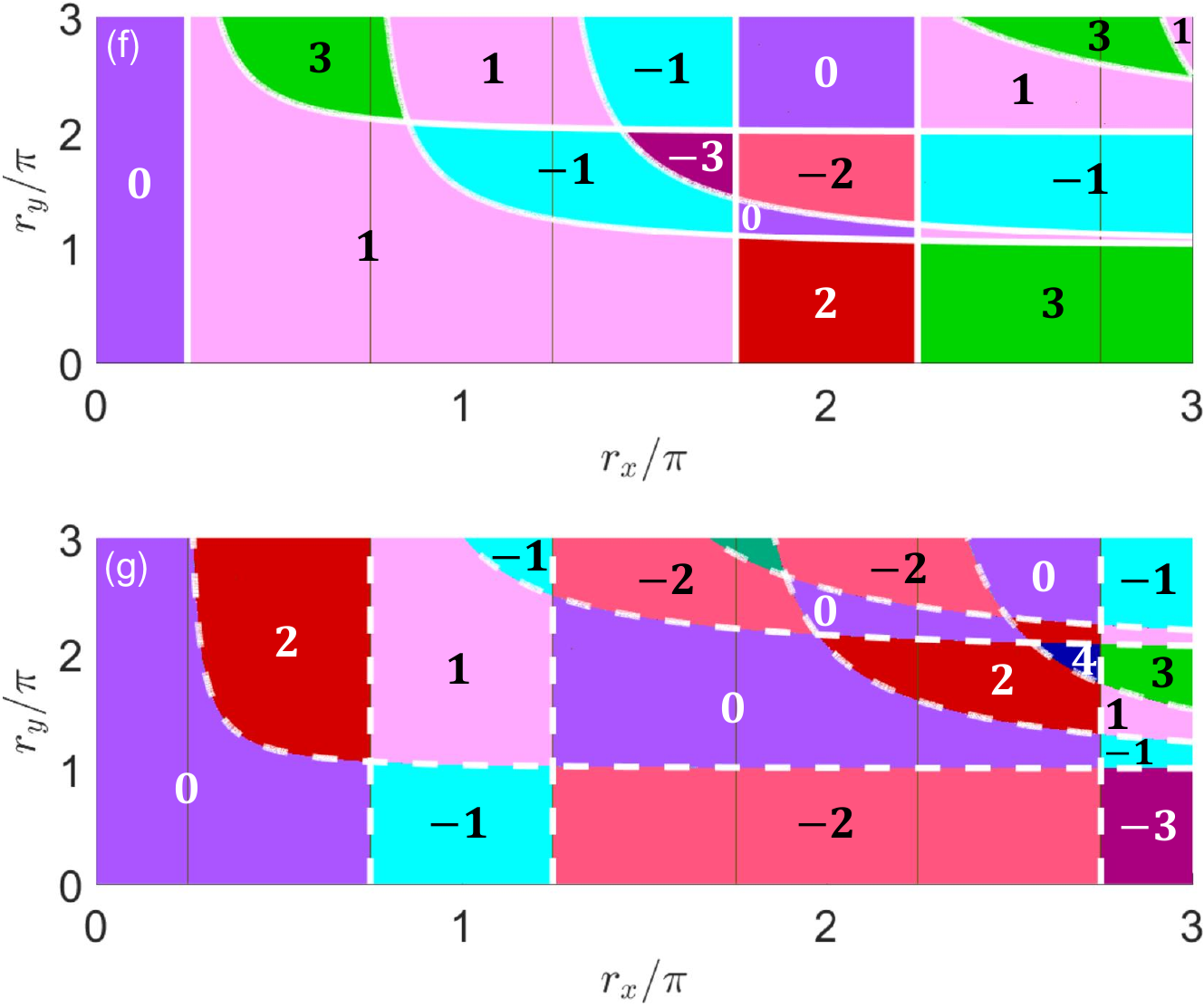}$\,\,$\includegraphics[scale=0.3]{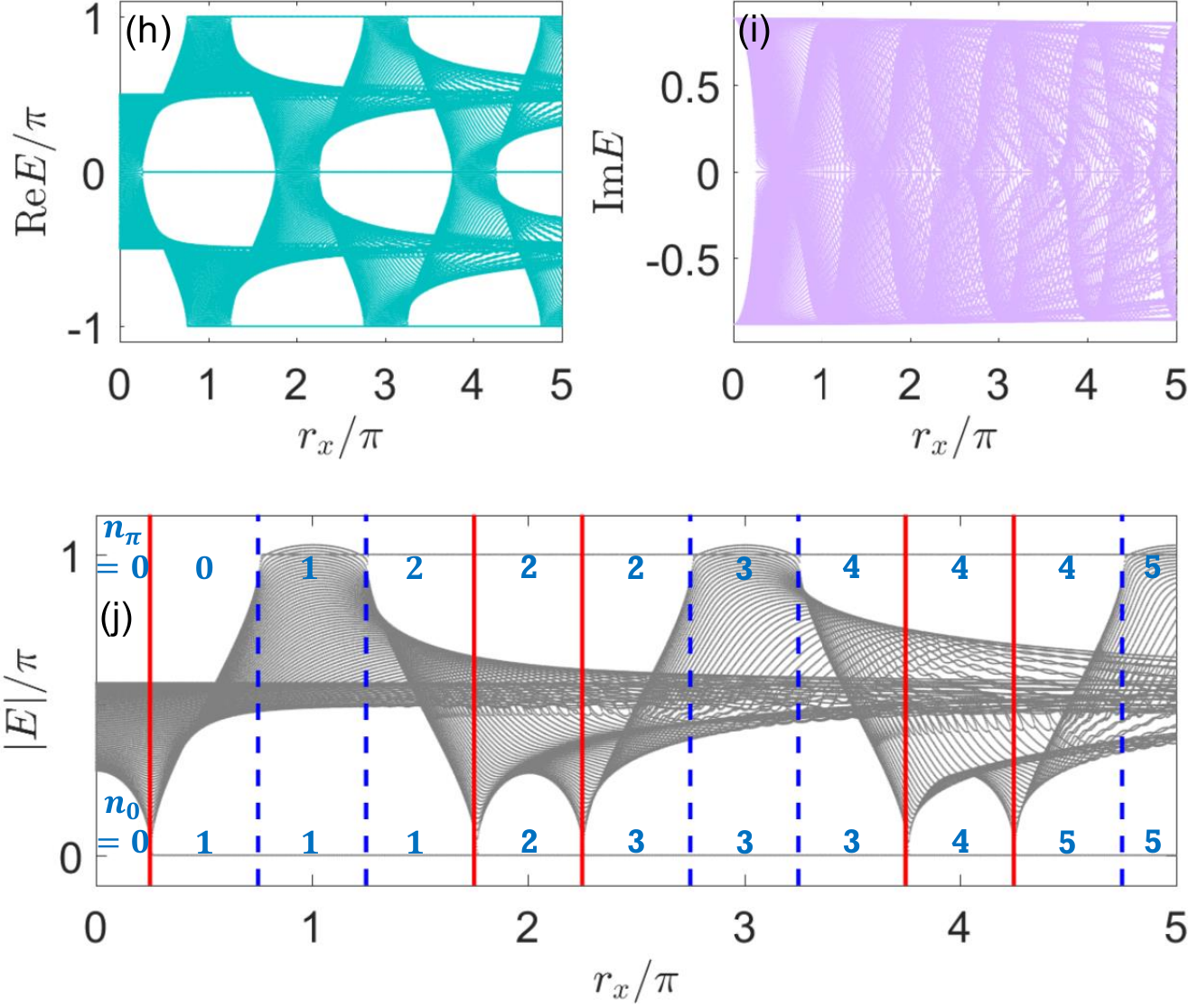}
		\par\end{centering}
	\caption{\textbf{Non-Hermitian Floquet topological insulator}: (a) a schematic
		diagram of the lattice model, (b)--(e) quasienergy spectra on the
		complex plane under the OBC, (f)--(g) topological phase diagrams
		under the PBC, and (h)--(j) Floquet spectra and edge states under
		the OBC \cite{ZhouNHFTP01}. \label{fig:NHFTI1}}
\end{figure}

We start with the characterization of 1D non-Hermitian Floquet topological
insulators. One typical model that incorporates their rich topological
properties is the following periodically quenched dimerized tight-binding
lattice, whose time-dependent Hamiltonian reads \cite{ZhouNHFTP01}
\begin{equation}
	\hat{H}(t)=\begin{cases}
		\hat{H}_{1}, & t\in[\ell T,\ell T+T/2),\\
		\hat{H}_{2}, & t\in[\ell T+T/2,\ell T+T),
	\end{cases}
\end{equation}
where
\begin{equation}
	\hat{H}_{1}=\sum_{n}(ir_{y}\hat{c}_{n+1}^{\dagger}\hat{c}_{n}+{\rm H.c.}+2i\gamma\hat{c}_{n}^{\dagger}\hat{c}_{n})\otimes\sigma_{y},
\end{equation}
\begin{equation}
	\hat{H}_{2}=\sum_{n}(r_{x}\hat{c}_{n}^{\dagger}\hat{c}_{n+1}+\mu\hat{c}_{n}^{\dagger}\hat{c}_{n}+{\rm H.c.})\otimes\sigma_{x}.
\end{equation}
Here $T$ is the driving period and we will set $\hbar=T=1$ in our
calculations. $\hat{c}_{n}^{\dagger}$ $(\hat{c}_{n})$ creates (annihilates)
a particle in the unit cell $n$ of the lattice. $\sigma_\alpha$, $\alpha={x,y,z}$,
are Pauli matrices acting on the two sublattice degrees of freedom
A and B within each unit cell. The system parameters $r_{x},r_{y},\mu,\gamma$
are all real. The non-Hermitian effect is introduced by the nonreciprocal
intracell coupling term $2i\gamma\sigma_{y}$ applied over the first
half of each driving period {[}see Fig.~\ref{fig:NHFTI1}(a) for an
illustration of the model{]}. Experimentally, such a term might be
realized by coupled-resonator optical waveguide with asymmetric internal
scattering \cite{MalzardPRL2015}.

Considering the one-period evolution of the system from $t=\ell+0^{-}$
to $t=\ell+1+0^{-}$, the Floquet operator of $\hat{H}(t)$ takes
the form of $\hat{U}=e^{-\frac{i}{2}\hat{H}_{2}}e^{-\frac{i}{2}\hat{H}_{1}}$.
Its quasienergy spectrum and Floquet eigenstates are obtained by solving
the eigenvalue equation $\hat{U}|\psi\rangle=e^{-iE}|\psi\rangle$.
Under the OBC, we show the Floquet spectrum of $\hat{U}$ on the complex
quasienergy plane ${\rm Re}E-{\rm Im}E$ for some typical cases in
Figs.~\ref{fig:NHFTI1}(b)--(e) {[}with $(\mu,r_{y},\gamma)=(0,\pi/2,{\rm arccosh}(\sqrt{2}))$,
$r_{x}=\pi/2,\pi,3\pi/2,2\pi$, and the number of unit cells $N=150${]}.
The numbers of degenerate Floquet edge modes at the quasienergies
$E=0$ and $E=\pi$ are denoted by $n_{0}$ and $n_{\pi}$ in the
corresponding figure captions. We observe one or multiple pairs of
edge modes at both the center ($E=0$) and boundary ($E=\pi$) of
the first quasienergy Brillouin zone. Interestingly, when the numbers
of these edge modes are the same, i.e., $n_{0}=n_{\pi}$, we find
them to appear in different types of quasienergy gaps {[}see 
Figs.~\ref{fig:NHFTI1}(c) and \ref{fig:NHFTI1}(e){]}. For example, we
find a pair of Floquet zero modes in the line gap at $E=0$, while
another pair of Floquet $\pi$ modes are found in the point gap at
$E=\pi$. This is rather different from the situation in Hermitian
Floquet systems, where the quasienergy is real and there is no distinction
between point and line gaps. It is also different from the cases
encountered in non-Hermitian static systems with two bands, where
topological edge modes can only appear in either a point gap or a
line gap at $E=0$. The Floquet spectra with hybrid (point plus line)
quasienergy gaps are thus unique to non-Hermitian Floquet systems.

To understand the origin of the non-Hermitian Floquet edge modes at
$E=0$ and $E=\pi$, we study the bulk topological properties of the
system. Under the PBC, the Floquet operator of the system reads
\begin{equation}
	U(k)=e^{-i(\mu+r_{x}\cos k)\sigma_{x}}e^{-i(r_{y}\sin k+i\gamma)\sigma_{y}},
\end{equation}
where $k\in[-\pi,\pi)$ is the quasimomentum. The quasienergy spectrum
of $U(k)$ is given by 
\begin{equation}
	\pm E(k)=\pm\arccos[\cos(\mu+r_{x}\cos k)\cos(r_{y}\sin k+i\gamma)],
\end{equation}
and the gap-closing conditions can be obtained analytically by setting
$\cos[E(k)]=\pm1$ \cite{ZhouNHFTP01}. Transforming $U(k)$ to symmetric
time frames, we obtain
\begin{equation}
	U_{1}(k)=e^{-\frac{i}{2}(\mu+r_{x}\cos k)\sigma_{x}}e^{-i(r_{y}\sin k+i\gamma)\sigma_{y}}e^{-\frac{i}{2}(\mu+r_{x}\cos k)\sigma_{x}},
\end{equation}
\begin{equation}
	U_{2}(k)=e^{-\frac{i}{2}(r_{y}\sin k+i\gamma)\sigma_{y}}e^{-i(\mu+r_{x}\cos k)\sigma_{x}}e^{-\frac{i}{2}(r_{y}\sin k+i\gamma)\sigma_{y}}.
\end{equation}
It is clear that both $U_{1}(k)$ and $U_{2}(k)$ respect the chiral
(sublattice) symmetry ${\cal S}=\sigma_{z}$. We can thus characterize
the non-Hermitian Floquet topological phases of $U(k)$ by the winding
numbers $(w_{0},w_{\pi})$ according to Eq.~(\ref{eq:w0p}). In
Figs.~\ref{fig:NHFTI1}(f) and \ref{fig:NHFTI1}(g), we present the
$w_{0}$ and $w_{\pi}$ of $U(k)$ versus $(r_{x},r_{y})$ respectively,
yielding the topological phase diagram of the system. We find various
non-Hermitian Floquet insulating phases. They are characterized by
large integer winding numbers and separated by a series of topological
phase transitions with quasienergy level crossings at $E=0$ {[}white
solid lines in Fig.~\ref{fig:NHFTI1}(f){]} and $E=\pi$ {[}white
dashed lines in Fig.~\ref{fig:NHFTI1}(g){]}. The winding numbers
$(w_{0},w_{\pi})$ could become arbitrarily large with the increase
of system parameters (e.g., the intercell hopping amplitude $r_{x}$),
which implies that Floquet phases with large topological invariants
could indeed survive even with finite non-Hermitian effects {[}$\mu=0$
and $\gamma={\rm arccosh}(\sqrt{2})$ in Figs.~\ref{fig:NHFTI1}(f)--(g){]}.
Moreover, these winding numbers correctly count the numbers of degenerate
Floquet edge modes $n_{0}$ and $n_{\pi}$ at $E=0$ and $E=\pi$
under the OBC. That is, we have the bulk-edge correspondence for our
1D chiral symmetric non-Hermitian Floquet topological insulator as
\cite{ZhouNHFTP01}
\begin{equation}
	n_{0}=2|w_{0}|,\qquad n_{\pi}=2|w_{\pi}|.\label{eq:NHFTIBBC}
\end{equation}
Representative examples of the Floquet spectra under the OBC are shown
in Figs.~\ref{fig:NHFTI1}(h)--(j) {[}with $\mu=0$, $r_{y}=\pi/2$,
$\gamma={\rm arccosh}(\sqrt{2})$ and the number of unit cells $N=150${]},
which provide verifications for the found bulk-edge correspondence
in Eq.~(\ref{eq:NHFTIBBC}). Note in passing that the system does
not show NHSE under the OBC for $\mu=0$, even though the Floquet
spectra possess point quasienergy gaps under the PBC. Meanwhile, many
pairs of degenerate edge modes at the quasienergies $E=0$ and $E=\pi$
are found in the topological phases with large winding numbers. As
these topological edge modes could survive in the presence of non-Hermitian
effects, they may provide more resources for the realization of robust
quantum state transfer and quantum computing schemes in open systems.

\begin{figure}
	\begin{centering}
		\includegraphics[scale=0.4]{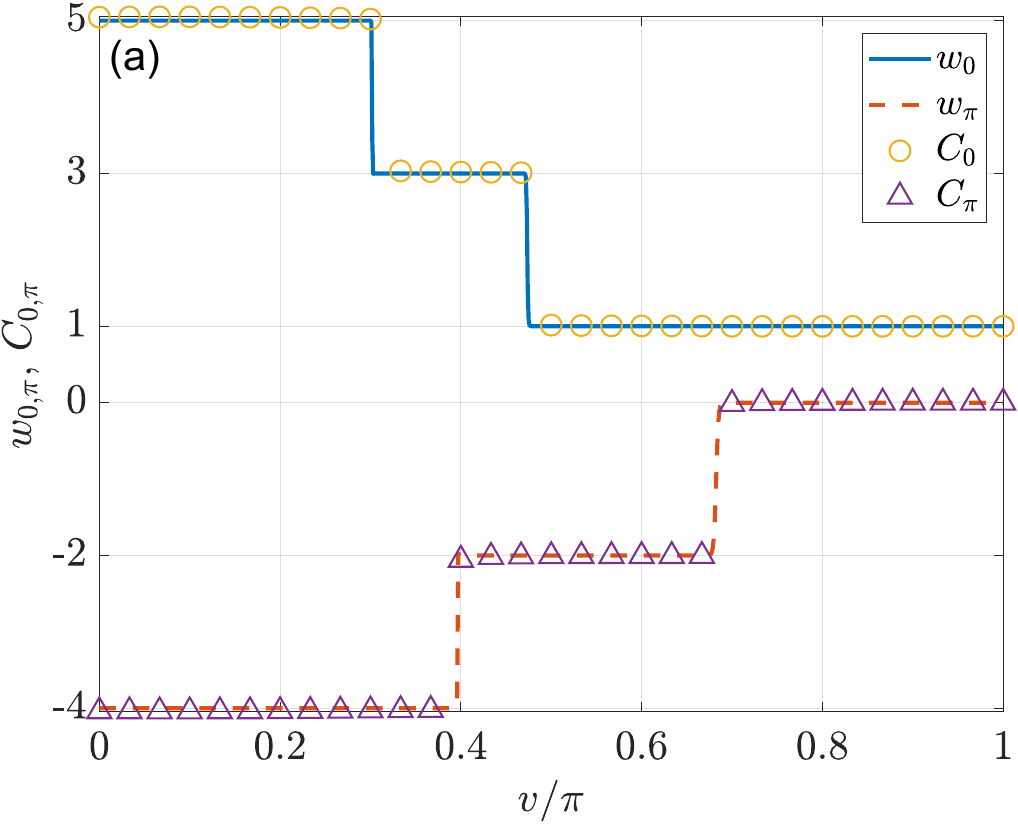}$\quad$\includegraphics[scale=0.4]{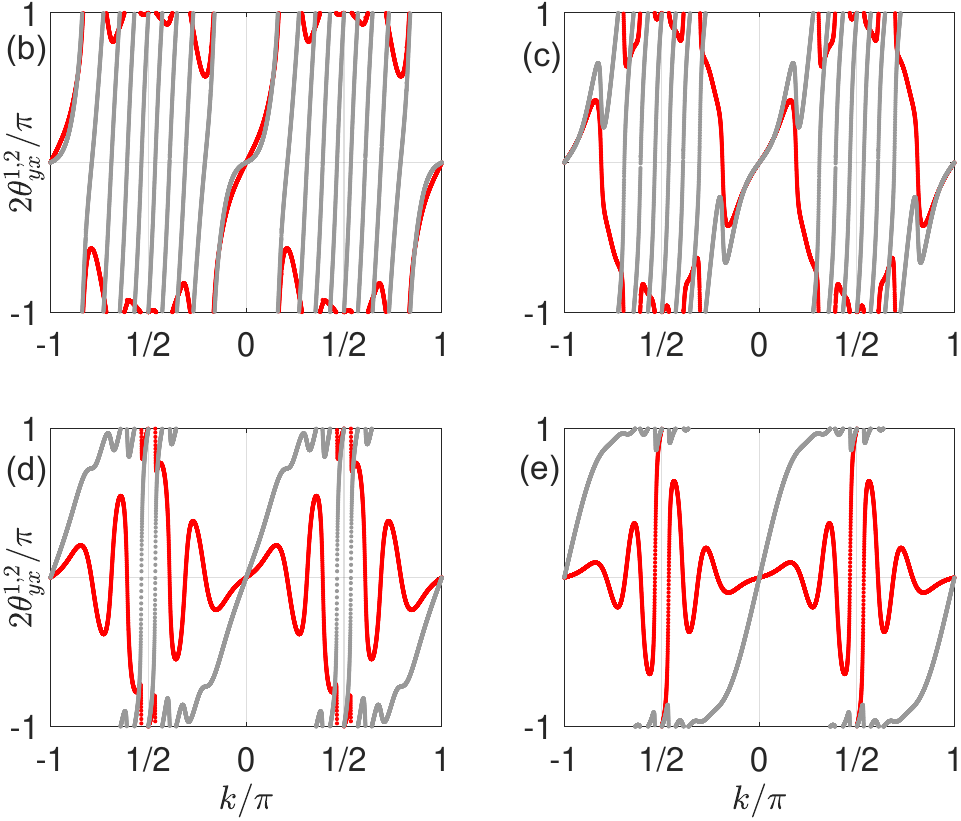}
		\par\end{centering}
	\caption{\textbf{Dynamical characterization of non-Hermitian Floquet topological
			phases}: (a) winding numbers ($w_{0},w_{\pi}$), MCDs ($C_{0},C_{\pi}$),
		and (b)--(e) dynamic winding numbers from stroboscopic time-averaged
		spin textures \cite{ZhouNHFTP03}. \label{fig:NHFTI2}}
\end{figure}

The winding numbers of non-Hermitian Floquet topological insulators
can be experimentally probed by measuring the stroboscopic time-averaged
spin textures \cite{ZhouNHFTP01} or the MCDs \cite{ZhouNHFTP02},
as introduced in Subsecs.~\ref{subsec:DWN} and \ref{subsec:MCD}.
Following Ref.~\cite{ZhouNHFTP03}, we consider a periodically quenched
non-Hermitian lattice model, whose Floquet operator in the momentum
space reads $U(k)=e^{-iJ_{2}\sin k\sigma_{y}}e^{-iJ_{1}\cos k\sigma_{x}}$.
Here $J_{1}=u_{1}+iv_{1}$, $J_{2}=u_{2}+iv_{2}$, and $u_{1},u_{2},v_{1},v_{2}\in\mathbb{R}$.
The winding numbers $(w_{0},w_{\pi})$ of $U(k)$ can be obtained
using the approach of symmetric time frames {[}see Eqs.~(\ref{eq:KickU1})--(\ref{eq:w0p}){]}.
Following the recipes outlined in Subsecs.~\ref{subsec:DWN} and \ref{subsec:MCD},
we obtain the MCDs in Fig. \ref{fig:NHFTI2}(a) {[}with $v=v_{1}=v_{2}$
and $(u_{1},u_{2})=(4.5\pi,0.5\pi)${]} and the dynamic winding angles
of time-averaged spin textures in Figs. \ref{fig:NHFTI2}(b)--(e)
{[}with $(u_{1},u_{2})=(4.5\pi,0.5\pi)$ and $v=v_{1}=v_{2}=0.2\pi,0.35\pi,0.6\pi,0.9\pi${]}.
In all the cases, the obtained MCDs and dynamic winding numbers are
coincident with the winding numbers $(w_{0},w_{\pi})$ of Floquet
operator $U(k)$ even when there are finite non-Hermitian effects
($v\neq0$). The dynamic winding numbers and the MCDs thus provide
us with two complementary approaches to dynamically characterize 1D
non-Hermitian Floquet topological phases with chiral (sublattice)
symmetry in momentum and in real spaces, respectively. They can be
applied to detect non-Hermitian Floquet matter in different types
of experimental setups \cite{NHFTP06,NHFTP15}.

\begin{figure}
	\begin{centering}
		\includegraphics[scale=0.45]{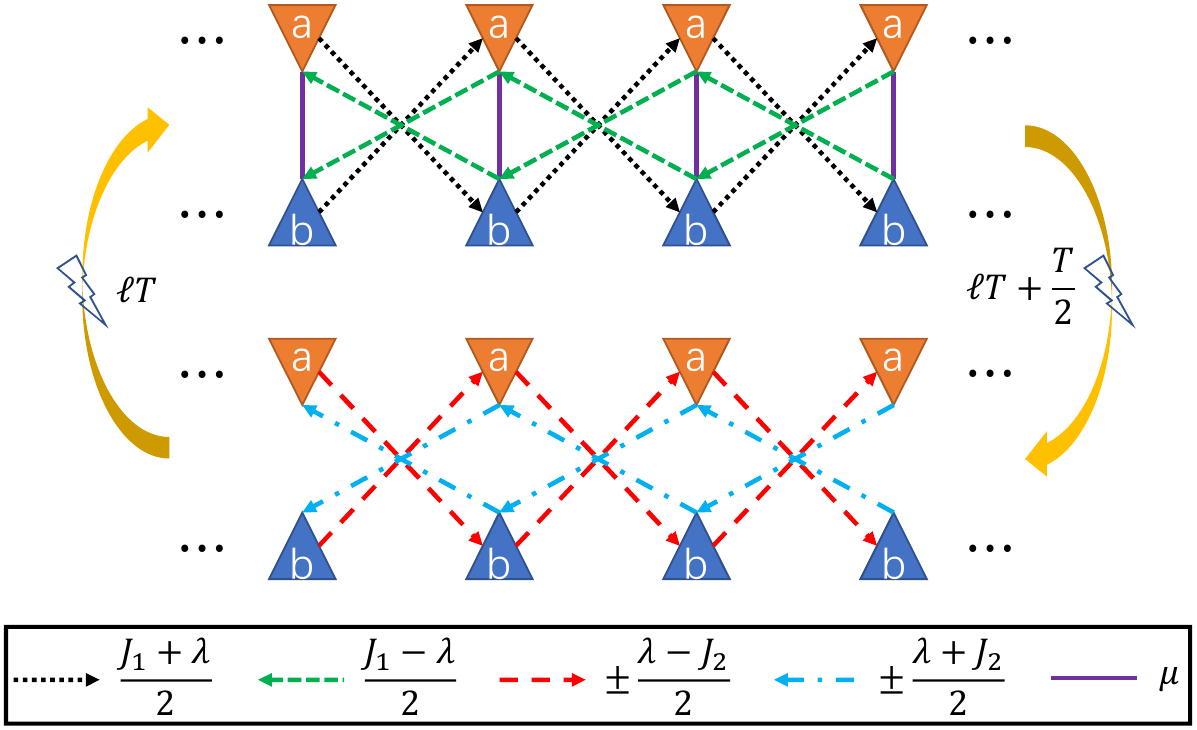}
		\par\end{centering}
	\medskip{}
	
	\begin{centering}
		\includegraphics[scale=0.415]{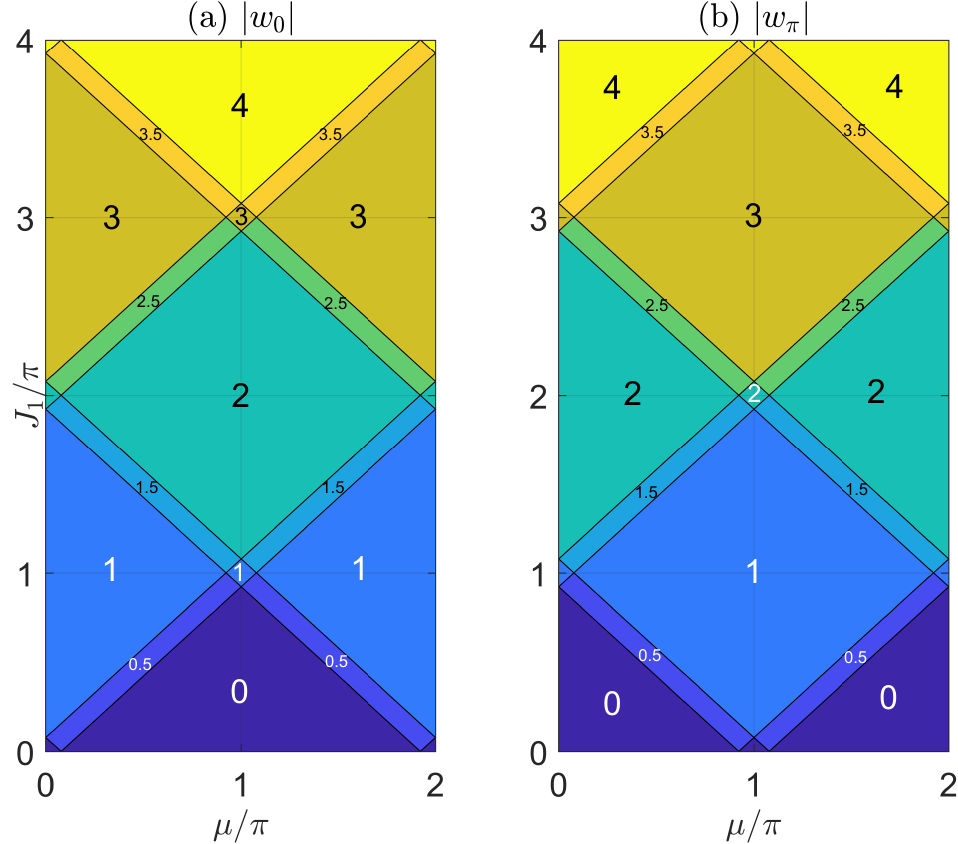}$\,$\includegraphics[scale=0.415]{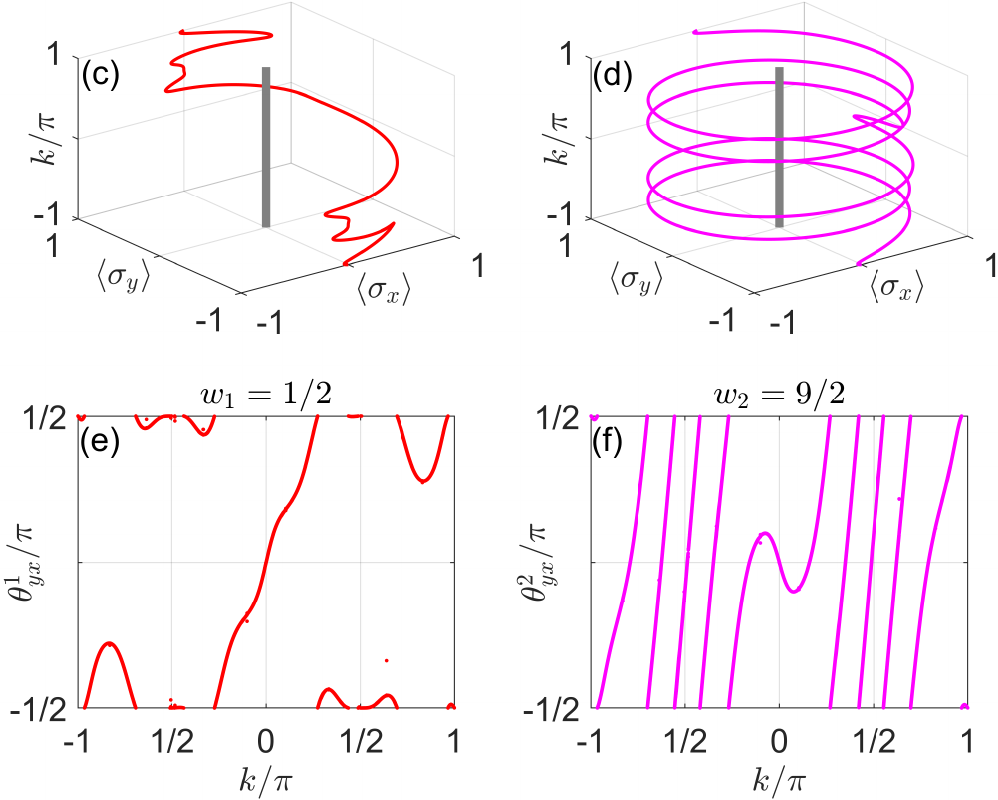}
		\par\end{centering}
	\medskip{}
	
	\begin{centering}
		\includegraphics[scale=0.39]{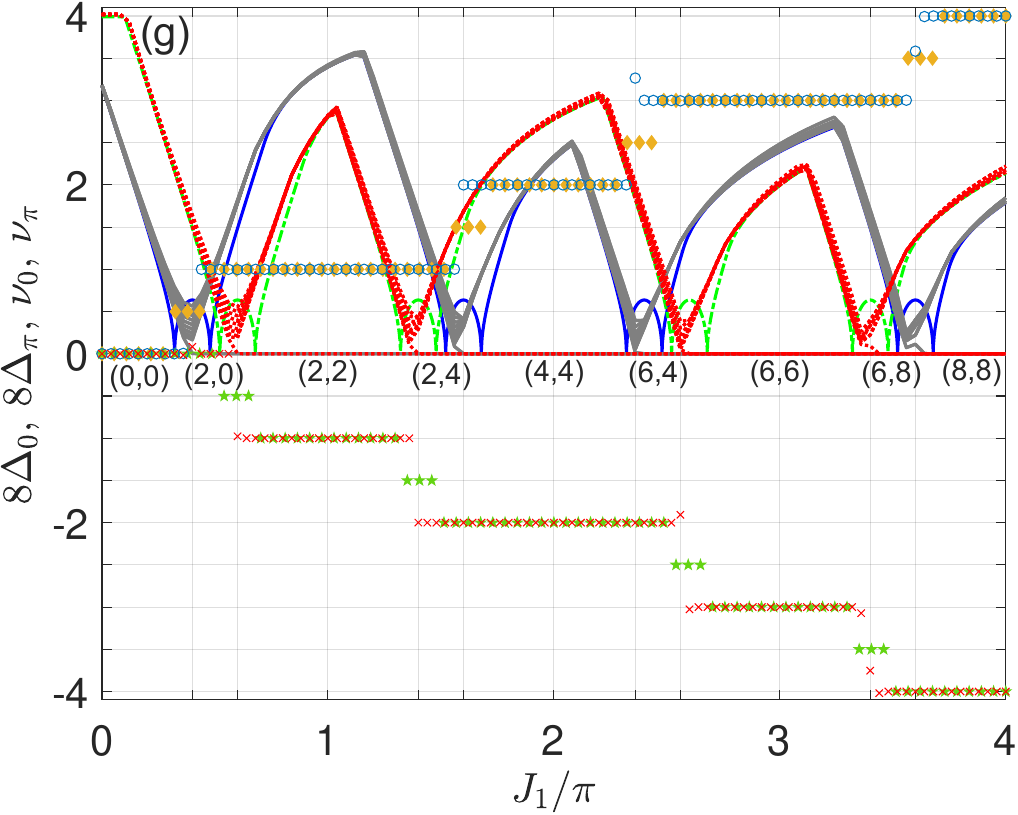}$\,$\includegraphics[scale=0.35]{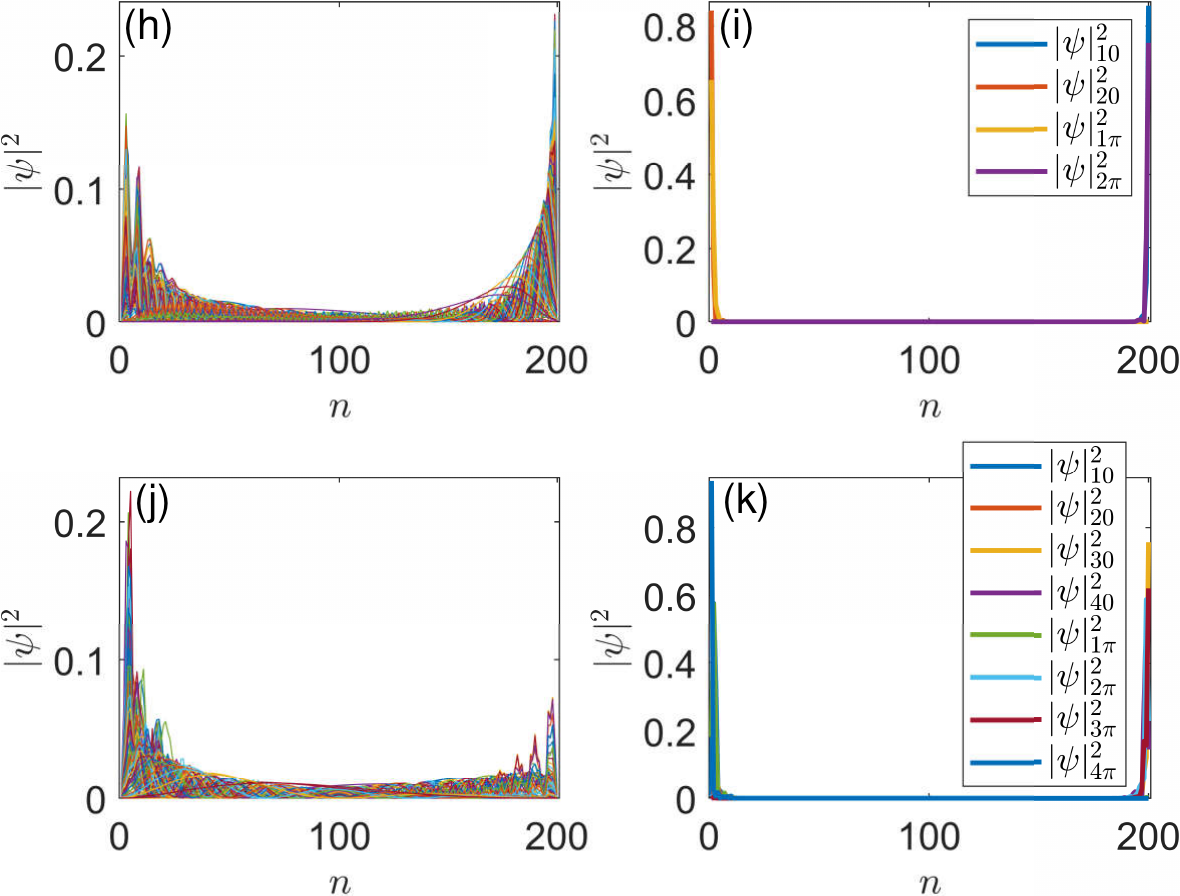}
		\par\end{centering}
	\caption{\textbf{Non-Hermitian Floquet topological insulator} \textbf{with
			NHSE}: a schematic diagram of the lattice model, (a)--(b) topological
		phase diagrams, (c)--(f) static and dynamic winding angles, (g) quasienergy
		gap functions under the OBC, and (h)--(k) Floquet topological corner
		modes at zero, $\pi$ quasienergies and skin-localized bulk modes.
		Under the OBC, the numbers of Floquet zero and $\pi$ edge modes in
		each topological phase are shown below the horizontal axis of (g).
		System parameters are $(J_{1},J_{2},\mu,\lambda)=(\pi,0.5\pi,0.4\pi,0.25)$
		and $(J_{1},J_{2},\mu,\lambda)=(2\pi,0.5\pi,0.4\pi,0.25)$ for (h,
		i) and (j, k) \cite{ZhouNHFTP08}. \label{fig:NHFTI3}}
\end{figure}

The models we considered above in this subsection do not possess the
NHSE. It remains unclear whether the rich non-Hermitian Floquet topology
could coexist with NHSEs, and how to characterize the topological
bulk-edge correspondence in the presence of Floquet NHSE. To address
this issue, we consider a periodically quenched non-Hermitian SSH
model \cite{ZhouNHFTP08}, whose $k$-space Hamiltonian under the
PBC takes the form of
\begin{equation}
	H(k,t)=\begin{cases}
		(\mu+J_{1}\cos k+i\lambda\sin k)\sigma_{x}, & t\in[\ell T,\ell T+T/2),\\
		(J_{2}\sin k+i\lambda\cos k)\sigma_{y}, & t\in[\ell T+T/2,\ell T+T).
	\end{cases}
\end{equation}
Here $\mu$ and $J_{1}/2+J_{2}/2$
are the intracell and intercell hopping amplitudes. The non-Hermiticity
is introduced by the asymmetric parts of hopping amplitudes $\pm i\lambda/2$
between the two sublattices. A sketch of the model is shown in 
Fig.~\ref{fig:NHFTI3}. The Floquet operator associated with $H(k,t)$
(for $\hbar=1$ and $T=2$) reads 
\begin{equation}
	U(k)=e^{-i(J_{2}\sin k+i\lambda\cos k)\sigma_{y}}e^{-i(\mu+J_{1}\cos k+i\lambda\sin k)\sigma_{x}}.
\end{equation}
In symmetric time frames, it respects the chiral symmetry ${\cal S}=\sigma_{z}$
\cite{ZhouNHFTP08}. We may thus use the winding numbers $(w_{0},w_{\pi})$
to characterize the Floquet topological phases of $U(k)$. The calculation
of such winding numbers following Eqs.~(\ref{eq:KickU1})--(\ref{eq:w0p})
leads to the topological phase diagrams in Fig.~\ref{fig:NHFTI3}(a)--(b).
Interestingly, we find that despite Floquet topological insulators
with large integer winding numbers, there are also various phases
with half-integer quantized topological invariants. These unique phases
are absent in the Hermitian limit of the model. Their appearance is
due to the Floquet exceptional topology induced by non-Hermitian effects
in the system. The parameter regions which accommodate half-integer
topological phases may further show Floquet NHSEs under the OBC. In
Figs.~\ref{fig:NHFTI3}(c)--(f), we compare the dynamic winding angles
of $U(k)$ with the winding patterns of its Floquet Hamiltonian vectors
in different time frames {[}with $(J_{1},J_{2},\mu,\lambda)=(2.4\pi,0.5\pi,0.4\pi,0.25)${]}
\cite{ZhouNHFTP08}. Their coincidence implies that we can use the
dynamic winding numbers introduced in Subsec.~\ref{subsec:DWN} to
probe the half-integer quantized topological invariants of 1D non-Hermitian
Floquet topological insulators with chiral symmetry.

Under the OBC, the bulk Floquet spectra and the gap-closing points
of quasienergies are found to be different from those under the PBC,
as shown in Fig.~\ref{fig:NHFTI3}(g) {[}with $(J_{2},\mu,\lambda)=(0.5\pi,0.4\pi,0.25)${]}
\cite{ZhouNHFTP08}. This means that the bulk winding numbers $w_{0}$
(diamonds) and $w_{\pi}$ (pentagrams) of $U(k)$ cannot be used to
characterize the edge states and topological phase transitions under
the OBC. To overcome this issue, we switch to the OBC winding numbers
$(\nu_{0},\nu_{\pi})$ {[}defined as the $(W_{0},W_{\pi})$ in 
Eq.~(\ref{eq:W0P}){]}. These winding numbers are found to be integer
quantized. They change their values only when the Floquet spectrum
of the system closes its gap at $E=0$ or $E=\pi$ under the OBC. Furthermore,
in each gapped phase, the winding numbers $(\nu_{0},\nu_{\pi})$ {[}circles
and crosses in Fig.~\ref{fig:NHFTI3}(g){]} are related to the numbers
of zero and $\pi$ Floquet edge modes $(n_{0},n_{\pi})$ through the
bulk-edge correspondence $(n_{0},n_{\pi})=2(|\nu_{0}|,|\nu_{\pi}|)$.
This relation holds even with NHSEs in our system \cite{ZhouNHFTP08}.
Therefore, our work provides a dual topological characterization of
non-Hermitian Floquet topological insulators under different boundary
conditions. In Figs.~\ref{fig:NHFTI3}(h)--(k), we present examples
of the skin-localized bulk states {[}in Figs.~\ref{fig:NHFTI3}(h) and \ref{fig:NHFTI3}(j){]}
and two types of Floquet edge states {[}in Figs.~\ref{fig:NHFTI3}(i) and
\ref{fig:NHFTI3}(k){]} in our system, which are also consistent with
the theoretical predictions \cite{ZhouNHFTP08}.

Putting together, we found different types of 1D non-Hermitian Floquet
topological insulators in a series of studies \cite{ZhouNHFTP01,ZhouNHFTP02,ZhouNHFTP03,ZhouNHFTP05,ZhouNHFTP08}.
All the models considered in these studies exhibit rich non-Hermitian
Floquet phases with large topological invariants, many topological
edge states, and unique topological transitions induced by the interplay
between non-Hermitian effects and time-periodic driving fields. The
bulk-edge correspondences and dynamic topological characterizations
of these intriguing new phases are also established, leading to an
explicit and all-round physical picture of 1D non-Hermitian Floquet
topological matter. In the following subsections, we uncover the essential
role of Floquet engineering in other types of non-Hermitian topological
matter.

\subsubsection{Second-order topological phase\label{subsec:2TI}}

We next consider the example of a non-Hermitian Floquet second-order
topological insulator (SOTI) in two dimensions. An $n$th-order topological
insulator in $d$ ($d\geq n$) spatial dimensions possesses localized
eigenmodes along its $(d-n)$-dimensional boundaries, which are topologically
nontrivial and protected by the symmetries of the $d$-dimensional
bulk. For example, an SOTI in two dimensions usually owns localized
topological states around its zero-dimensional geometric corners.
Assisted by time-periodic drivings, degenerate corner modes can further
appear at different quasienergies in Floquet second-order topological
phases \cite{FTPLg07,ZhouFSOTSC}.

\begin{figure}
	\begin{centering}
		\includegraphics[scale=0.32]{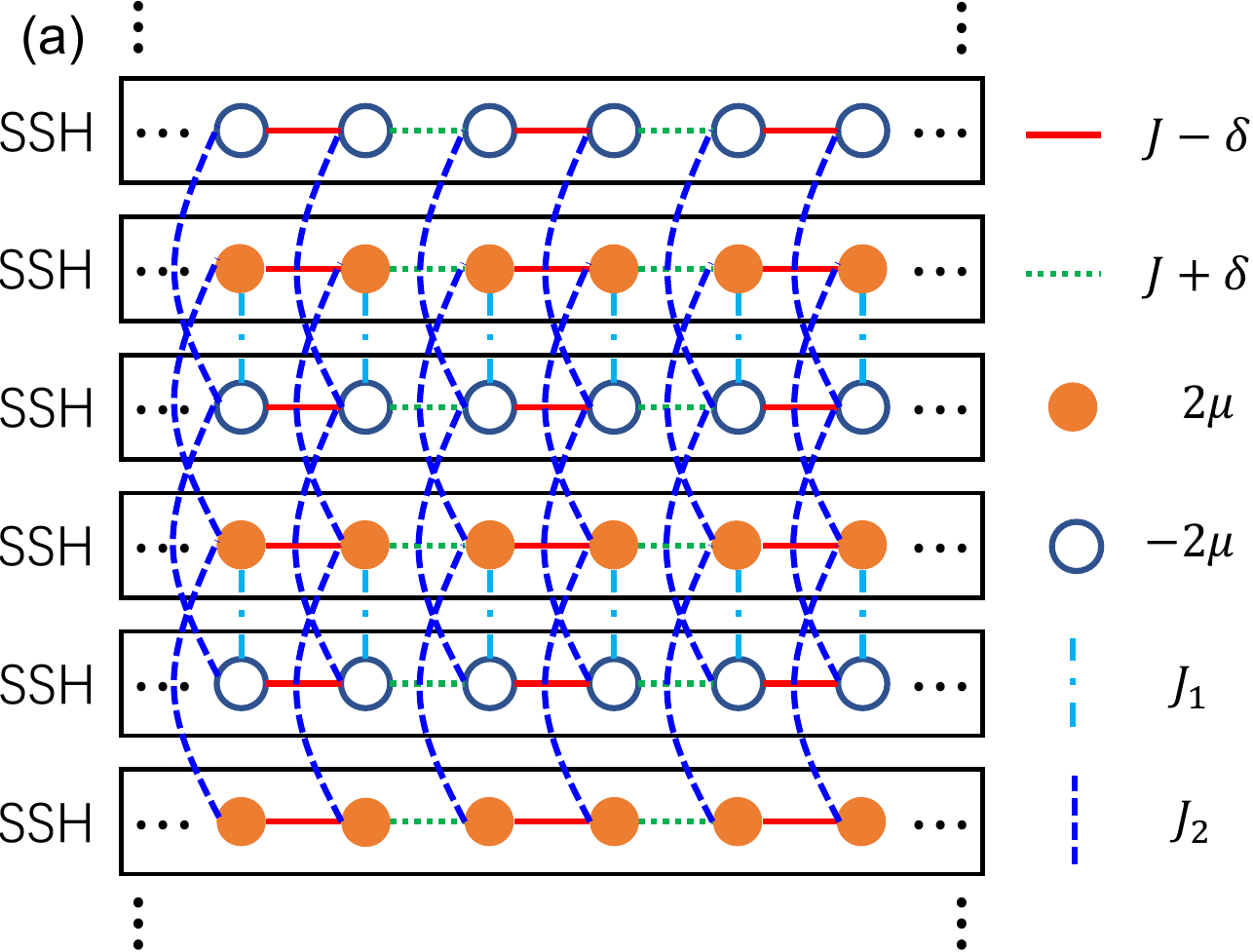}\includegraphics[scale=0.38]{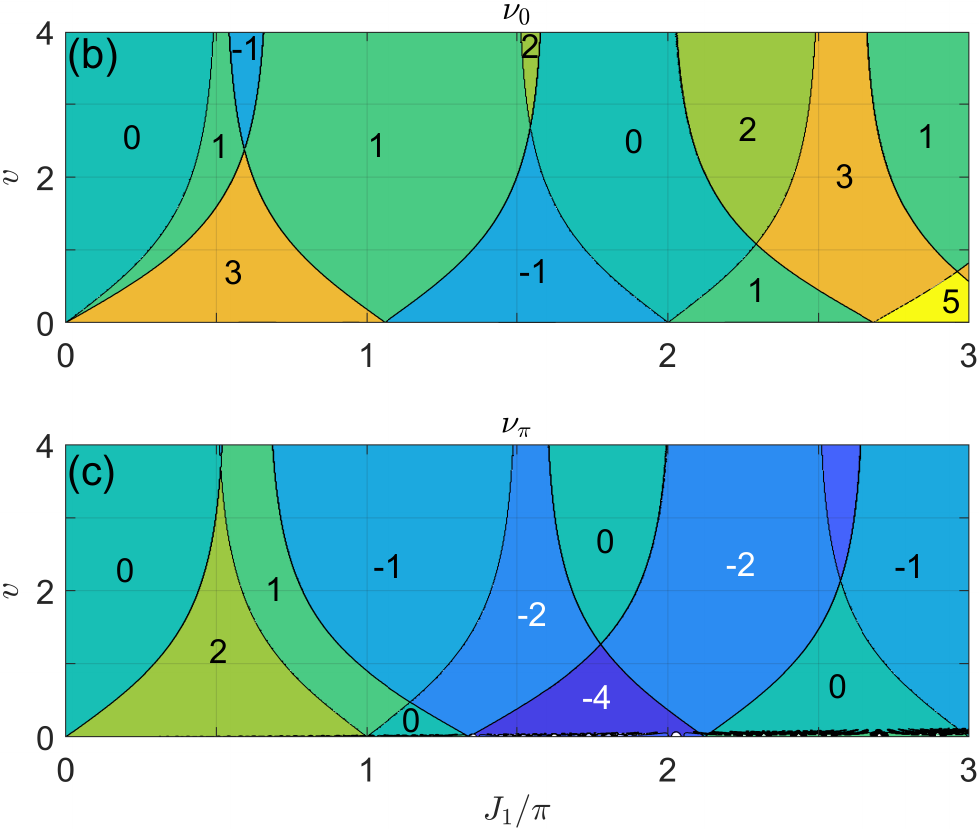}
		\par\end{centering}
	\medskip{}
	
	\begin{centering}
		\includegraphics[scale=0.38]{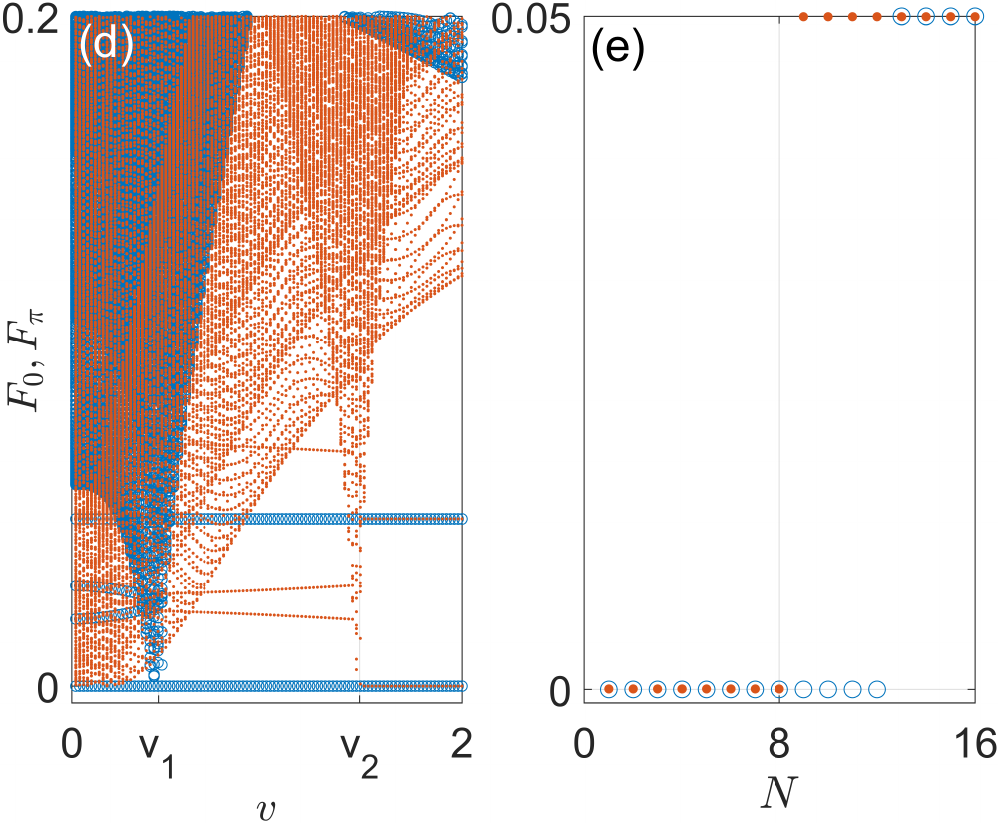}$\qquad$\includegraphics[scale=0.385]{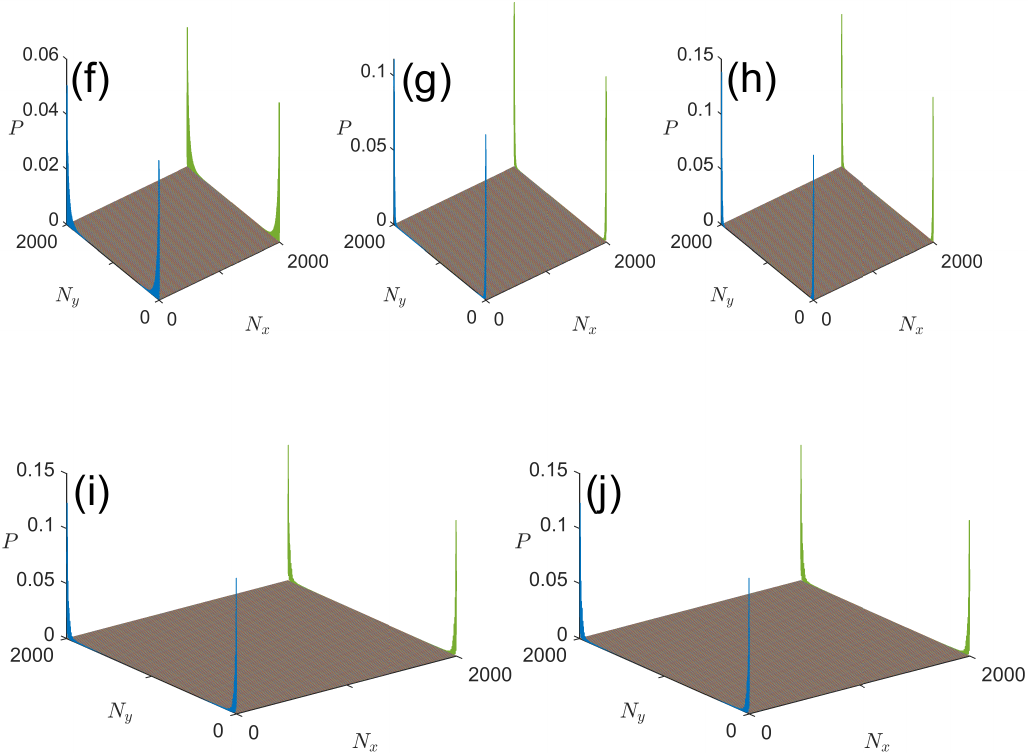}
		\par\end{centering}
	\caption{\textbf{Non-Hermitian Floquet SOTI}: (a) a schematic diagram of the
		static lattice model, (b)--(c) topological phase diagrams, (d)--(e)
		quasienergy gap functions under the OBC, (f)--(h) Floquet topological
		corner modes at the quasienergy zero, and (i)--(j) Floquet topological
		corner modes at the quasienergy $\pi$ \cite{ZhouNHFTP06}. \label{fig:NHFSOTI}}
\end{figure}

In this subsection, we focus on one typical model of non-Hermitian
Floquet SOTI, which is obtained following the recipe of coupled-wire
construction \cite{FTPLg07}. A schematic diagram of the static lattice
model is shown in Fig.~\ref{fig:NHFSOTI}(a). The system is formed
by a stack of SSH chains along the $y$-direction, with dimerized
couplings $(J_{1},J_{2})$ between adjacent chains. We consider a
Floquet variant of the model by applying time-periodic quenches to
the coupling and onsite parameters $(J_{1},J_{2},\mu)$ along the
$y$-direction. Under the PBC along both $x$ and $y$ directions,
the time-dependent Bloch Hamiltonian of the Floquet model takes the
form of
\begin{equation}
	H(k_{x},k_{y})=H_{x}(k_{x})\otimes\tau_{0}+\sigma_{0}\otimes H_{y}(k_{y},t),
\end{equation}
where $(k_{x},k_{y})\in[-\pi,\pi)\times[-\pi,\pi)$ are the quasimomenta.
$\sigma_{0}$ and $\tau_{0}$ are two by two identity matrices. The
Hamiltonians of the 1D subsystems $H_{x}(k_{x})$ and $H_{y}(k_{y},t)$
are explicitly given by
\begin{equation}
	H_{x}(k_{x})=[(J-\delta)+(J+\delta)\cos k_{x}]\sigma_{x}+(J+\delta)\sin k_{x}\sigma_{y},\label{eq:Hxkx}
\end{equation}
\begin{equation}
	H_{y}(k_{y},t)=\begin{cases}
		2J_{1}\cos k_{y}\tau_{x} & t\in[\ell T,\ell T+T/2)\\
		2(\mu+J_{2}\sin k_{y})\tau_{z} & t\in[\ell T+T/2,\ell T+T)
	\end{cases}.\label{eq:Hyky}
\end{equation}
Here $T$ is the driving period and $\ell\in\mathbb{Z}$. $\sigma_{x,y,z}$
and $\tau_{x,y,z}$ are Pauli matrices acting on the sublattice degrees
of freedom in $x$ and $y$ directions, respectively. Setting $\hbar=T=1$,
$J=\delta=\Delta/2$, and choosing $\mu=u+iv$ ($u,v\in\mathbb{R}$),
the Floquet operator of the system in ${\bf k}$-space is found to
be \cite{ZhouNHFTP06}
\begin{equation}
	U(k_{x},k_{y})=e^{-i\Delta(\cos k_{x}\sigma_{x}+\sin k_{x}\sigma_{y})}\otimes e^{-ih_{z}(k_{y})\tau_{z}}e^{-ih_{x}(k_{y})\tau_{x}},
\end{equation}
where
\begin{equation}
	h_{x}(k_{y})=J_{1}\cos k_{y},\qquad h_{z}(k_{y})=u+iv+J_{2}\sin k_{y}.
\end{equation}
The non-Hermitian effect is brought about by the onsite gain and loss
encoded in the term $iv\tau_{z}$.

In symmetric time frames, the system has the chiral (sublattice) symmetry
${\cal S}=\sigma_{z}\otimes\tau_{y}$ \cite{ZhouNHFTP06}. One can
thus characterize its Floquet second-order topological phases by integer
winding numbers \cite{ZhouNHFTP06}
\begin{equation}
	\nu_{0}=\frac{\nu_{1}+\nu_{2}}{2},\qquad\nu_{\pi}=\frac{\nu_{1}-\nu_{2}}{2}.
\end{equation}
Here $\nu_{\alpha}=ww_{\alpha}$ for $\alpha=1,2$. $w$ is the winding
number of the static Hamiltonian $\Delta(\cos k_{x}\sigma_{x}+\sin k_{x}\sigma_{y})$,
which is always equal to one for the topological flat-band limit of
the SSH model. $(w_{1},w_{2})$ are the winding numbers of the subsystem
Floquet operator $e^{-ih_{z}(k_{y})\tau_{z}}e^{-ih_{x}(k_{y})\tau_{x}}$,
which can be defined via the recipe outlined in Subsec.~\ref{subsec:Sym}.
They both take integer-quantized values even with the non-Hermitian
effects considered in our system \cite{ZhouNHFTP06}.

In Fig.~\ref{fig:NHFSOTI}(b), we show the topological phase diagrams
of $U(k_{x},k_{y})$ in the parameter space $J_{1}-v$ {[}with $(\Delta,J_{2},u)=(\pi/20,3\pi,0)${]}.
We find that both the winding numbers $(\nu_{0},\nu_{\pi})$ can take
large values, which indicates the presence of non-Hermitian Floquet
SOTI phases with large topological invariants in our setting. Furthermore,
with the increase of the gain and loss strength $v$, the system can
undergo topological phase transitions and even enter non-Hermitian
Floquet SOTI phases with larger winding numbers. These non-Hermiticity
boosted topological signatures are also not expected in non-driven
systems. Their appearance is thus due to the corporation between Floquet
driving fields and non-Hermitian gain and loss. 

Under the OBC along both $x$ and $y$ directions, localized Floquet
eigenmodes with the quasienergies zero and $\pi$ appear at the four
corners of the square lattice. Some spatial profiles of these corner
modes are shown in Figs.~\ref{fig:NHFSOTI}(f)--(j) {[}with $(\Delta,J_{1},J_{2},u,v)=(\pi/20,2.5\pi,3\pi,0,2)${]}.
There are twelve (eight) Floquet corner modes at the quasienergy zero
($\pi$). More generally, their numbers $(n_{0},n_{\pi})$ are related
to the winding numbers $(\nu_{0},\nu_{\pi})$ through the algebraic
relations \cite{ZhouNHFTP06}
\begin{equation}
	(n_{0},n_{\pi})=4(|\nu_{0}|,|\nu_{\pi}|).
\end{equation}
We thus established the bulk-corner correspondence for a class of
non-Hermitian Floquet SOTI with chiral symmetry. With the growth of
the non-Hermitian parameter $v$, topological phase transitions accompanied
by the increase of bulk winding numbers $(\nu_{0},\nu_{\pi})$ and
corner mode numbers $(n_{0},n_{\pi})$ are observed from the gap functions
$(F_{0},F_{\pi})\equiv(\frac{1}{\pi}|\varepsilon|,\frac{1}{\pi}\sqrt{(|{\rm Re}\varepsilon|-\pi)^{2}+({\rm Im}\varepsilon)^{2}})$
under the OBC (here $\varepsilon$ denotes the quasienergy), as showcased
in Figs.~\ref{fig:NHFSOTI}(d)--(e). The topological phase transition
points $v_{1}$ and $v_{2}$ under the OBC are precisely consistent
with the gap closing points of the 2D bulk under the PBC, which can
be found analytically \cite{ZhouNHFTP06}. Finally, even with non-Hermitian
effects, the bulk winding numbers $(\nu_{0},\nu_{\pi})$ can be dynamically
probed by 2D generalizations of the MCD as introduced in 
Subsec.~\ref{subsec:MCD} (see Ref.~\cite{ZhouNHFTP06} for more details).

Overall, we discovered rich non-Hermitian Floquet SOTI phases in a
periodically quenched 2D lattice with balanced gain and loss. Different
from the static case, each SOTI phase is now depicted by two topological
invariants $(\nu_{0},\nu_{\pi})$. Many non-Hermitian Floquet SOTIs
with large topological invariants and gain- or loss-induced topological
phase transitions were identified. Under the OBC, the winding numbers
$(\nu_{0},\nu_{\pi})$ determine the numbers of symmetry-protected
Floquet corner modes at zero and $\pi$ quasienergies. The interplay
between driving and dissipation thus results in a series of non-Hermitian
Floquet SOTI phases with multiple zero and $\pi$ corner modes, which
may find applications in topological state preparations, information scrambling and quantum
computation. The work of Ref.~\cite{ZhouNHFTP06} offers one of the
earliest findings of rich non-Hermitian Floquet topological phases
beyond one spatial dimension. In the meantime, it introduces a generic
scheme to construct non-Hermitian Floquet higher-order topological
phases across different physical dimensions, which is expected to be applicable
across insulating, superconducting and semi-metallic systems.
Note in passing that
some latter studies also considered the NHSEs and anomalous $\pi$
modes in Floquet higher-order topological phases with somewhat different
focuses \cite{NHFTP28,NHFTP31,NHFTP38}.

\subsubsection{qth-root topological phase\label{subsec:qrTI}}

We now consider an exotic class of non-Hermitian Floquet topological
phase, which could have symmetry-protected boundary states at the
quasienergies $p\pi/q$, with $p,q\in\mathbb{Z}^{\pm}$ and $p/q\neq0,1$.
A systematic recipe of obtaining these new phases is to take the nontrivial
$q$th-root of a propagator $\hat{U}$ that describes Floquet topological
matter. The general procedure is developed in Ref.~\cite{ZhouNHFTP10},
which generalizes the previous schemes of taking $2^{n}$th and $3^{n}$th
roots for static Hamiltonians \cite{RootTP01,RootTP02,RootTP03,RootTP04,RootTP05,RootTP06,RootTP07,RootTP08,RootTP09,RootTP10}
to Floquet systems. The basic idea is first outlined in Ref.~\cite{RootRaditya},
and then expanded by utilizing $\mathbb{Z}_{q}$ generalizations of
Pauli matrices as ancillary degrees of freedom before taking the $q$th-root
of a Floquet operator in an enlarged Hilbert space. It is in parallel
with Dirac's original idea of adding internal degrees of freedom for
electrons before taking the square-root to get their relativistic
wave equation \cite{RootDirac}.

\begin{figure}
	\begin{centering}
		\includegraphics[scale=0.275]{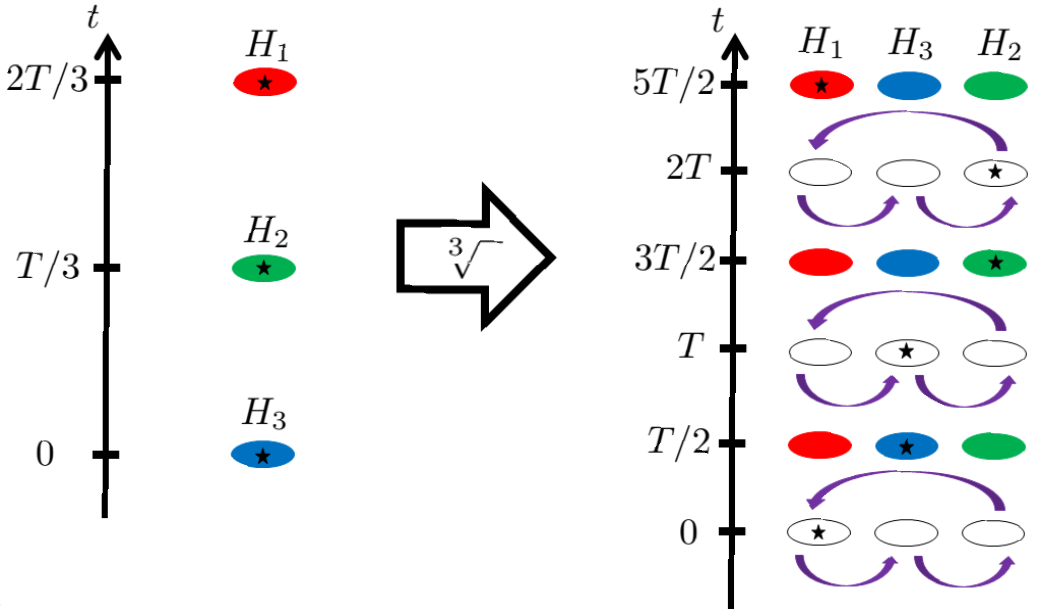}$\,\,$\includegraphics[scale=0.256]{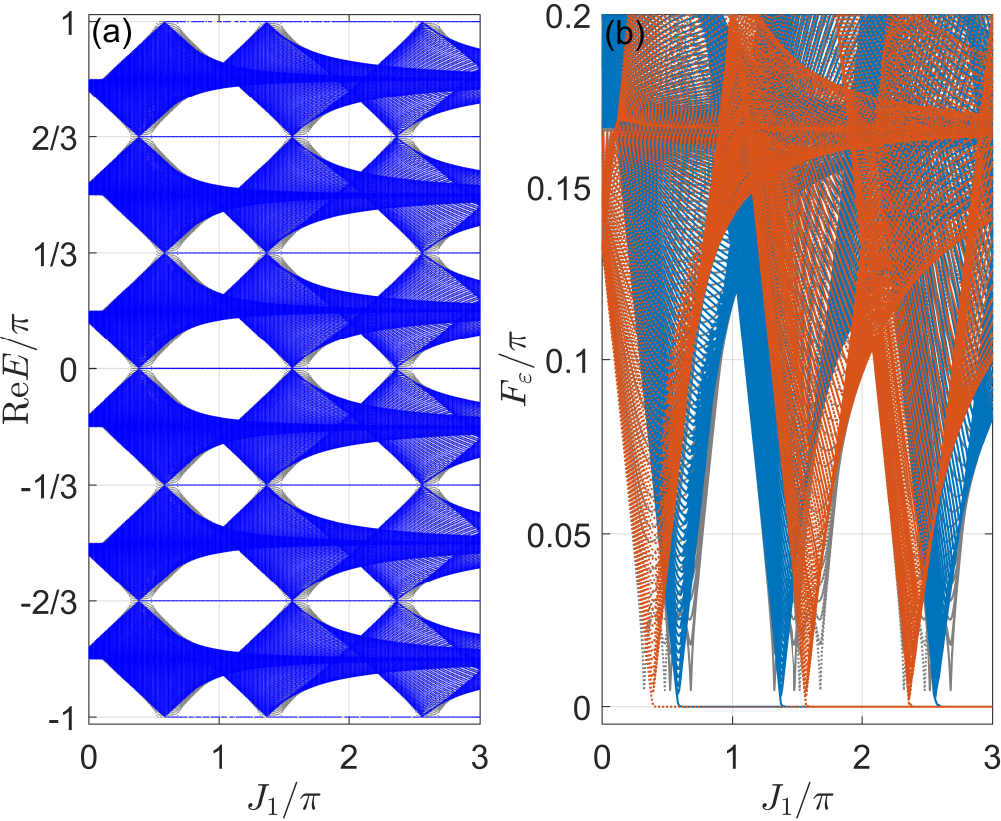}$\,\,$\includegraphics[scale=0.256]{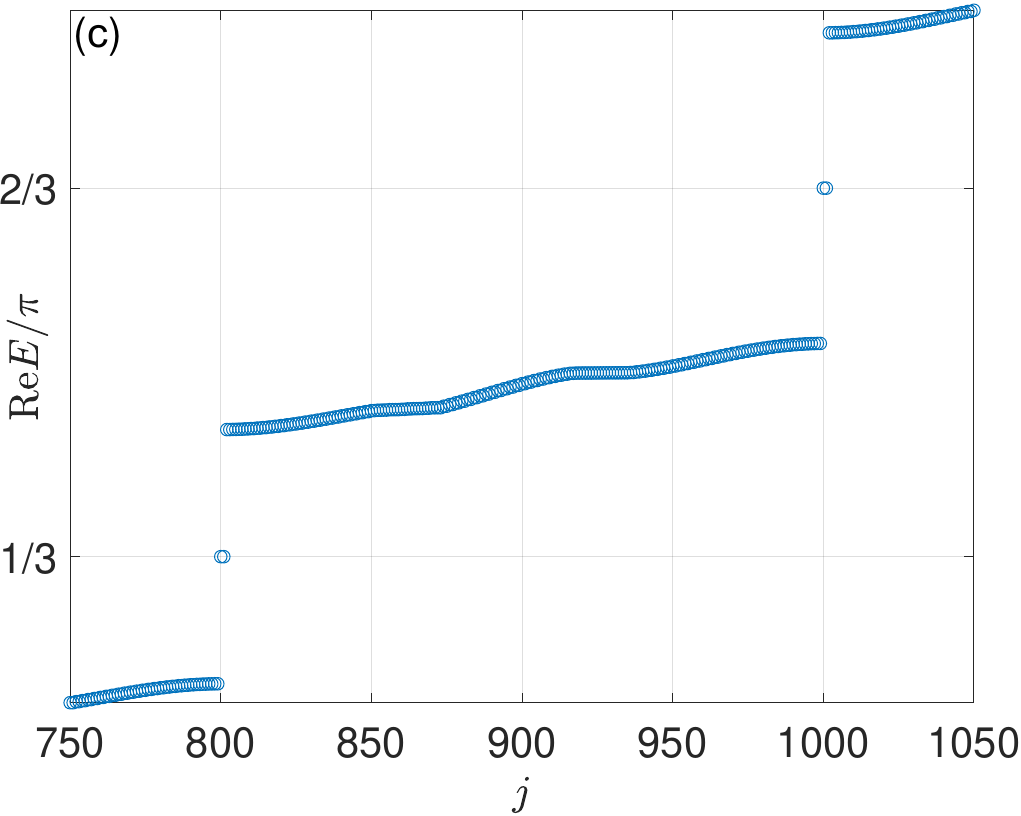}
		\par\end{centering}
	\medskip{}
	
	\begin{centering}
		\includegraphics[scale=0.258]{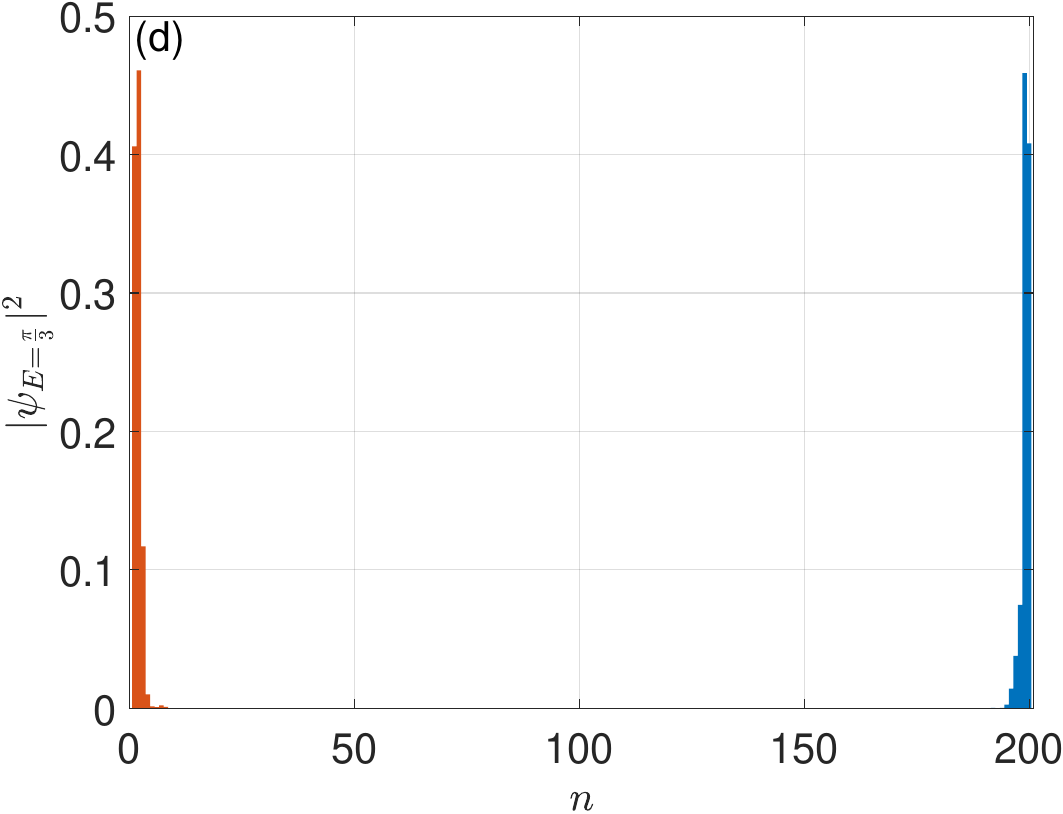}\includegraphics[scale=0.258]{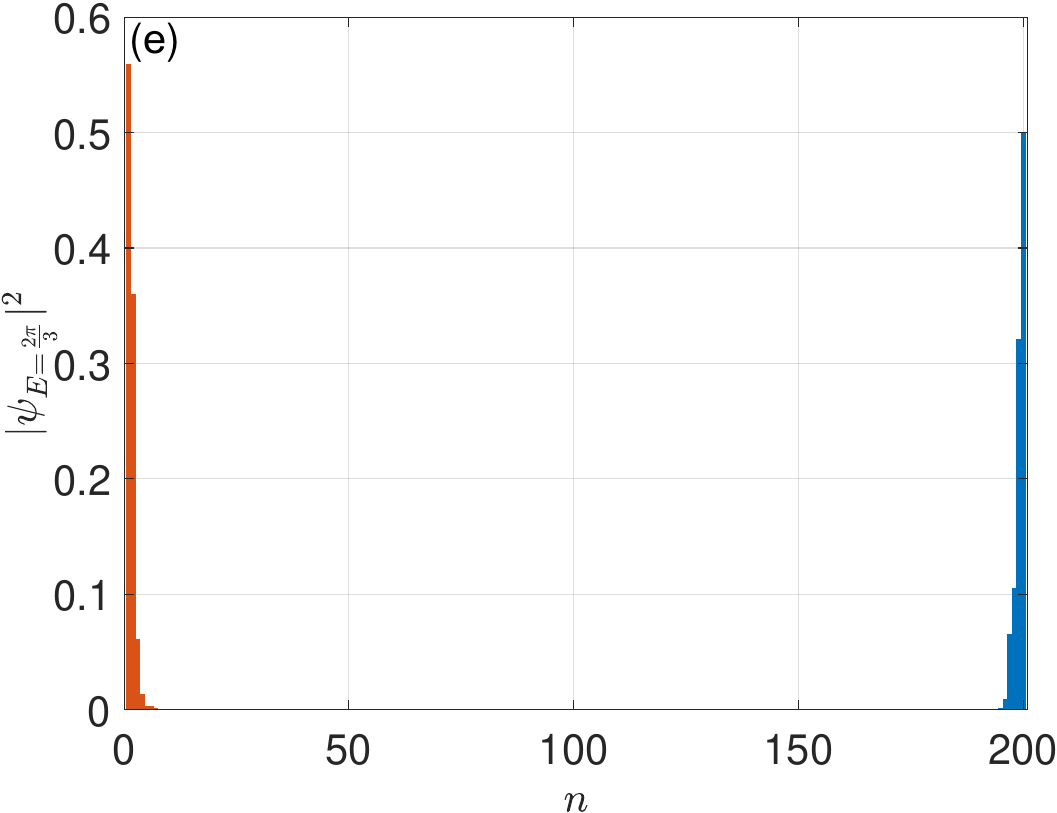}\includegraphics[scale=0.258]{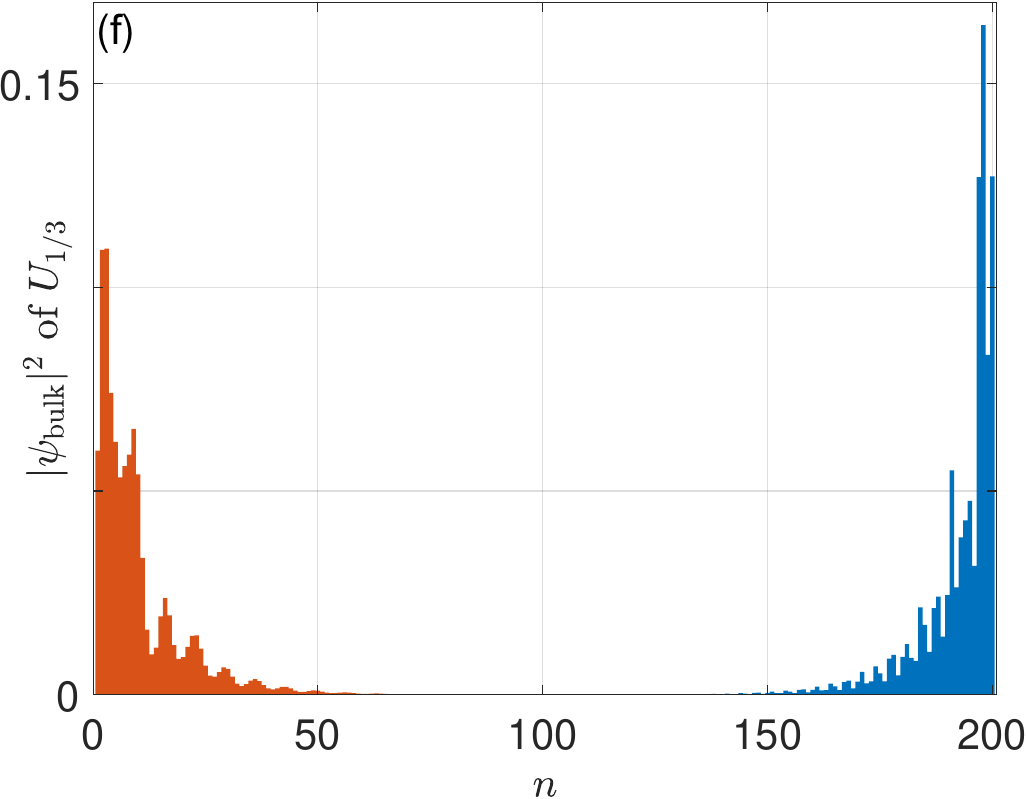}
		\par\end{centering}
	\caption{\textbf{Cubic-root non-Hermitian Floquet topological insulator}: a
		schematic diagram of taking the cubit-root for a Floquet operator,
		(a) quasienergy spectrum under the PBC (in grey) and OBC (in blue),
		(b) gap functions under the PBC (in grey) and OBC (in red and blue),
		(c) Floquet topological edge modes with $E=\pi/3,2\pi/3$, and (d)--(f)
		spatial profiles of Floquet edge modes and bulk skin modes \cite{ZhouNHFTP10}.
		\label{fig:QRNHFTI}}
\end{figure}

To be explicit, we consider the construction of a cubic-root non-Hermitian
Floquet topological insulator. An illustration of the scheme is given
in Fig.~\ref{fig:QRNHFTI}. Let us consider a three-step periodically
quenched parent system, whose time-periodic Hamiltonian takes the
form of \cite{ZhouNHFTP10}
\begin{equation}
	\hat{H}(t)=\begin{cases}
		\hat{H}_{3}, & t\in[\ell,\ell+1/3),\\
		\hat{H}_{2}, & t\in[\ell+1/3,\ell+2/3),\\
		\hat{H}_{1}, & t\in[\ell+2/3,\ell+1),
	\end{cases}
\end{equation}
where $\ell\in\mathbb{Z}$ and we have assumed $\hbar=T=1$. The Floquet
operator of the system is then given by $\hat{U}=e^{-\frac{i}{3}\hat{H}_{1}}e^{-\frac{i}{3}\hat{H}_{2}}e^{-\frac{i}{3}\hat{H}_{3}}$.
Following the methodology of Ref.~\cite{ZhouNHFTP10}, the nontrivial cubic-root
$\hat{U}_{1/3}$ of $\hat{U}$ is found to be
\begin{equation}
	\hat{U}_{1/3}=\begin{pmatrix}0 & e^{-i\frac{\hat{H}_{1}}{3}} & 0\\
		0 & 0 & e^{-i\frac{\hat{H}_{2}}{3}}\\
		e^{-i\frac{\hat{H}_{3}}{3}} & 0 & 0
	\end{pmatrix}.\label{eq:U1ov3}
\end{equation}
The cube of $\hat{U}_{1/3}$ then gives 
\begin{equation}
	\hat{U}_{1/3}^{3}=\begin{pmatrix}e^{-\frac{i}{3}\hat{H}_{1}}e^{-\frac{i}{3}\hat{H}_{2}}e^{-\frac{i}{3}\hat{H}_{3}} & 0 & 0\\
		0 & e^{-\frac{i}{3}\hat{H}_{2}}e^{-\frac{i}{3}\hat{H}_{3}}e^{-\frac{i}{3}\hat{H}_{1}} & 0\\
		0 & 0 & e^{-\frac{i}{3}\hat{H}_{3}}e^{-\frac{i}{3}\hat{H}_{1}}e^{-\frac{i}{3}\hat{H}_{2}}
	\end{pmatrix}.\label{eq:U1ov3CB}
\end{equation}
It is clear that $\hat{U}_{1/3}^{3}$ contains three identical copies
of $\hat{U}$, in the sense that its three diagonal blocks are all
related by similarity transformations and thus sharing the same spectrum
and topological properties with $\hat{U}$ \cite{ZhouNHFTP10}. 

As an example, we choose the Floquet model introduced in 
Ref.~\cite{ZhouNHFTP08} as the parent system, whose Hamiltonian can now
be expressed as
\begin{equation}
	\hat{H}(t)=\begin{cases}
		\frac{3}{4}\hat{H}_{1} & t\in[\ell,\ell+1/3)\\
		\frac{3}{2}\hat{H}_{2} & t\in[\ell+1/3,\ell+2/3)\\
		\frac{3}{4}\hat{H}_{1} & t\in[\ell+2/3,\ell+1)
	\end{cases},
\end{equation}
where
\begin{equation}
	\hat{H}_{1}=J_{2}\sum_{n}(i\hat{c}_{n+1}^{\dagger}\hat{c}_{n}+{\rm H.c.})\otimes\sigma_{y}+i\lambda\sum_{n}(\hat{c}_{n}^{\dagger}\hat{c}_{n+1}+{\rm H.c.})\otimes\sigma_{y},
\end{equation}
\begin{equation}
	\hat{H}_{2}=\sum_{n}(\mu\hat{c}_{n}^{\dagger}\hat{c}_{n}+J_{1}\hat{c}_{n}^{\dagger}\hat{c}_{n+1}+{\rm H.c.})\otimes\sigma_{x}+i\lambda\sum_{n}(i\hat{c}_{n+1}^{\dagger}\hat{c}_{n}+{\rm H.c.})\otimes\sigma_{x}.
\end{equation}
The cubit-root Floquet operator $\hat{U}_{1/3}$ of the system then
takes the form of Eq.~(\ref{eq:U1ov3}). The non-Hermitian effect
is introduced by the asymmetric hopping amplitude $i\lambda$. The
Floquet quasienergy spectrum and gap functions of $\hat{U}_{1/3}$
versus $J_{1}$ under the PBC (grey dots) and OBC (blue dots, red
and blue lines) are shown in Figs.~\ref{fig:QRNHFTI}(a)--(b) {[}with
$(J_{2},\mu,\lambda)=(0.5\pi,0.4\pi,0.25)$ and the length of lattice
$L=400${]}. Here the gap function 
\begin{equation}
	F_{\varepsilon}\equiv\sqrt{({\rm Re}E-\varepsilon)^{2}+({\rm Im}E)^{2}},
\end{equation}
and the blue (red) lines in the Fig.~\ref{fig:QRNHFTI}(b) denote
$F_{\pi/3}$ ($F_{2\pi/3}$). We observe a series of phase transitions
accompanied by quasienergy-gap closings at $E=p\pi/q$ for $p=0,1,2,3$
and $q=3$. After these transitions, more and more degenerate Floquet
edge modes emerge at the fractional quasienergies $E=\pi/3$ and $2\pi/3$,
which are impossible in conventional Floquet topological insulators
with two quasienergy bands in one dimension. These fractional-quasienergy
edge modes are thus unique to cubit-root Floquet topological insulators,
either Hermitian or non-Hermitian. A zoom in of their quasienergies
for one exemplary case is shown in Fig.~\ref{fig:QRNHFTI}(c) {[}with
$(J_{1},J_{2},\mu,\lambda)=(\pi,0.5\pi,0.4\pi,0.25)$ and the length
of lattice $L=400${]}, and their spatial profiles are shown in 
Figs.~\ref{fig:QRNHFTI}(d) {[}for the $E=\pi/3$ eigenmodes{]} and \ref{fig:QRNHFTI}(e)
{[}for the $E=2\pi/3$ eigenmodes{]}. Moreover, the numbers of these
fractional-quasienergy edge modes $(n_{\pi/3},n_{2\pi/3})$ are determined
by the open-bulk winding numbers $(\nu_{0},\nu_{\pi})$ \cite{ZhouNHFTP08}
of the parent system $\hat{U}=e^{-i\frac{\hat{H}_{1}}{4}}e^{-i\frac{\hat{H}_{2}}{2}}e^{-i\frac{\hat{H}_{1}}{4}}$.
More precisely, we have the following bulk-edge correspondence for
our cubic-root non-Hermitian Floquet topological insulator $\hat{U}_{1/3}$,
i.e., \cite{ZhouNHFTP10}
\begin{equation}
	n_{0}=n_{2\pi/3}=2|\nu_{0}|,\qquad n_{\pi/3}=n_{\pi}=2|\nu_{\pi}|,
\end{equation}
where $n_{0}$ and $n_{\pi}$ are the numbers of zero and $\pi$ degenerate
Floquet edge modes. Finally, we notice that the system also possesses
NHSE under the OBC, as reflected by the profiles of its Floquet bulk
states in Fig.~\ref{fig:QRNHFTI}(f) {[}with $(J_{1},J_{2},\mu,\lambda)=(\pi,0.5\pi,0.4\pi,0.25)$
and the length of lattice $L=400${]}. In Figs.~\ref{fig:QRNHFTI}(a)
and \ref{fig:QRNHFTI}(b), we also notice the discrepancy between
the gap-closing points of the system under the PBC and OBC. Nevertheless,
the open-bulk winding numbers $(\nu_{0},\nu_{\pi})$ correctly capture
the bulk-edge correspondence of the system under the OBC even with
NHSEs. Therefore, the work conducted in Ref.~\cite{ZhouNHFTP10} also
established a dual topological characterization for $q$th-root non-Hermitian
Floquet topological insulators. The formalism developed in 
Ref.~\cite{ZhouNHFTP10} is equally applicable to the construction of $q$th-root
Floquet topological insulators, superconductors, and semimetals in
other (non-)Hermitian systems and across different physical dimensions.
The fractional quasienergy edge modes at $E=p\pi/q$ might also be
employed to generate boundary time crystals with different temporal
periodicity and topological properties \cite{FTPExp13}. They may also
find applications in the Floquet quantum computing.

\subsection{Non-Hermitian Floquet topological superconductors\label{subsec:TSC}}

Similar to the static case, non-Hermitian Floquet topological phases
could also appear in superconducting systems. We review one such example
in this subsection, which admits many Floquet Majorana zero and $\pi$
edge modes even with non-Hermitian effects.

The model we are going to consider describes a Floquet Kitaev chain,
whose superconducting pairing terms are subject to time-periodic kicks.
The Hamiltonian of the model takes the form of
\begin{equation}
	\hat{H}(t)=\frac{1}{2}\sum_{n}[J(\hat{c}_{n}^{\dagger}\hat{c}_{n+1}+{\rm H.c.})+\Delta\delta_{T}(t)(\hat{c}_{n}\hat{c}_{n+1}+{\rm H.c.})+\mu(2\hat{c}_{n}^{\dagger}\hat{c}_{n}-1)],\label{eq:NHPKKC}
\end{equation}
where $J$ is the nearest-neighbor hopping amplitude, $\Delta$ is
the p-wave superconducting pairing amplitude, and $\mu$ is the chemical
potential. $\delta_{T}(t)\equiv\sum_{\ell\in\mathbb{Z}}\delta(t/T-\ell)$
implements delta kickings periodically with the period $T$. The system
is made non-Hermitian by setting $J=J_{r}+iJ_{i}$ or $\mu=\mu_{r}+i\mu_{i}$
with $J_{i}\neq0$ or $\mu_{i}\neq0$, respectively. Possible experimental
realizations of the delta kickings and non-Hermitian terms in this model
are discussed in Ref.~\cite{ZhouNHFTP04}.

\begin{figure}
	\begin{centering}
		\includegraphics[scale=0.41]{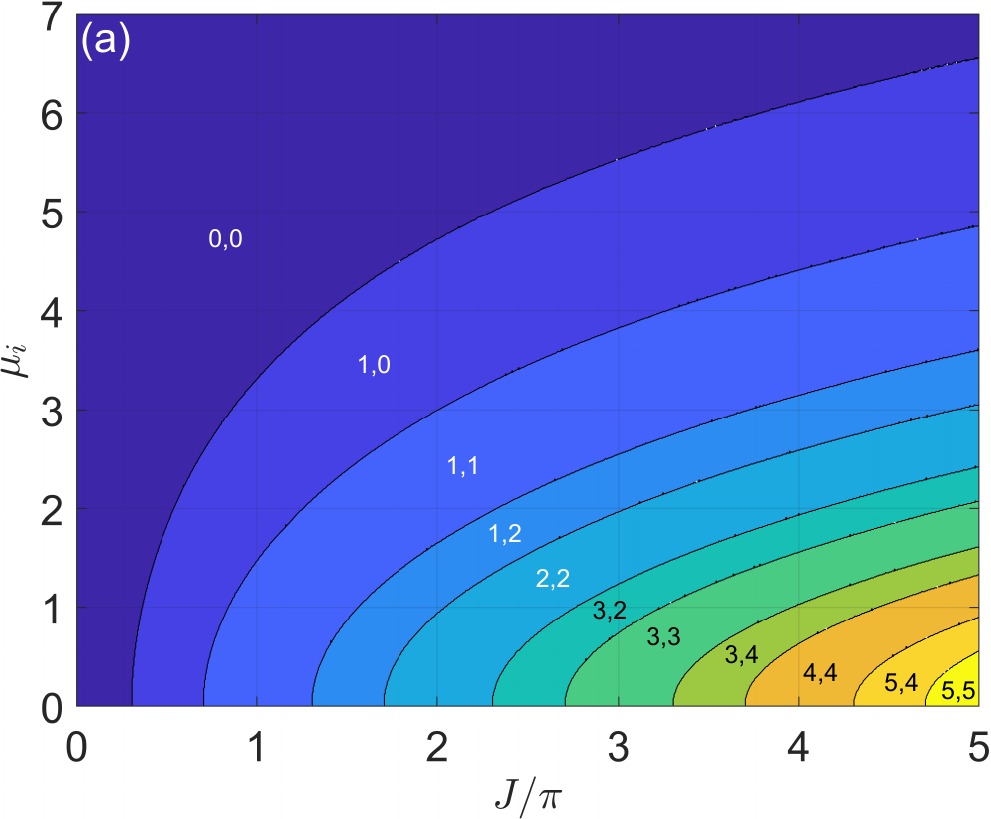}\includegraphics[scale=0.41]{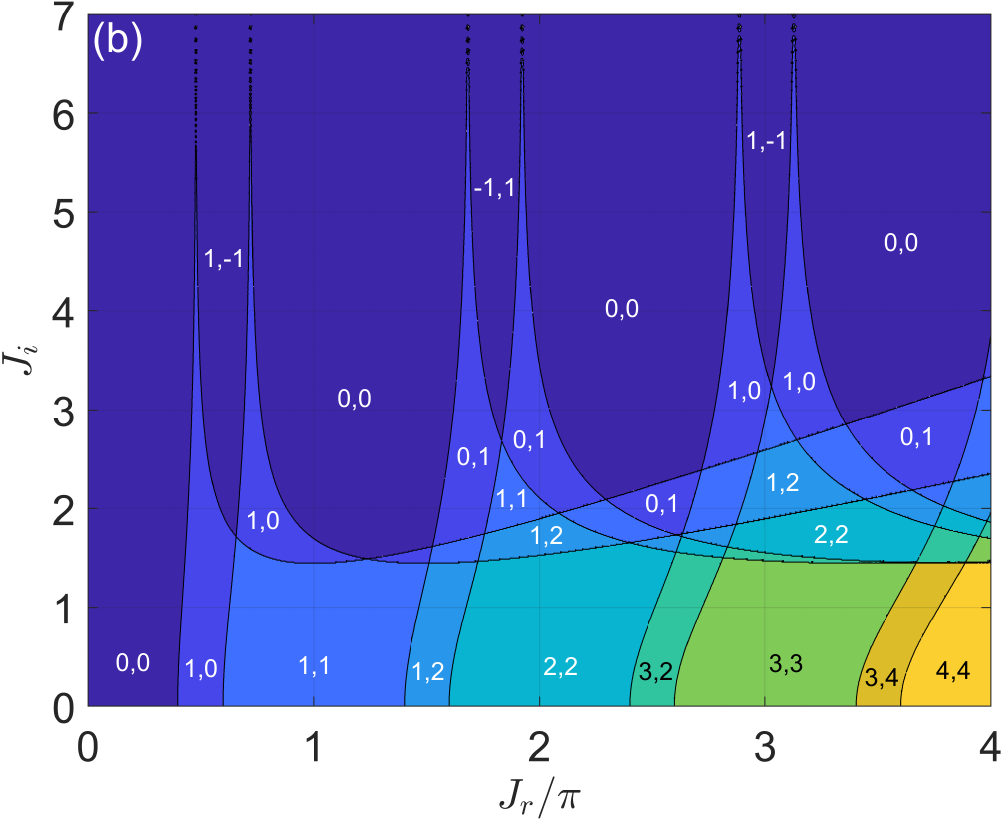}
		\par\end{centering}
	\medskip{}
	
	\begin{centering}
		\includegraphics[scale=0.41]{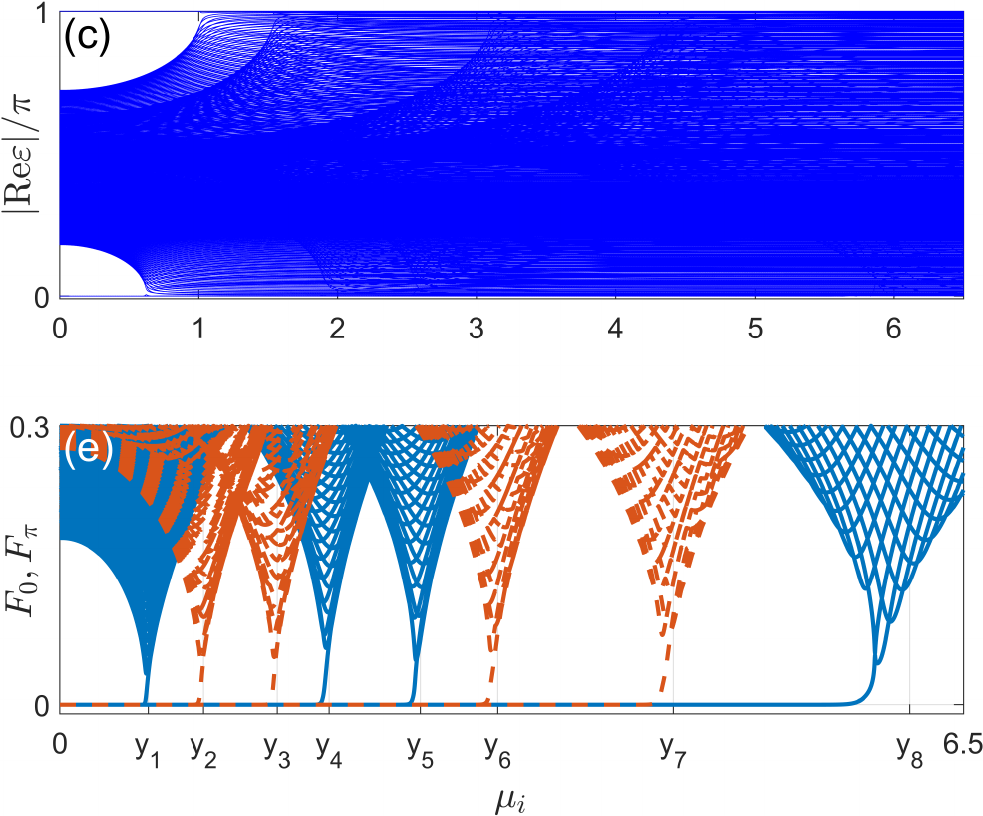}$\,\,\,$\includegraphics[scale=0.41]{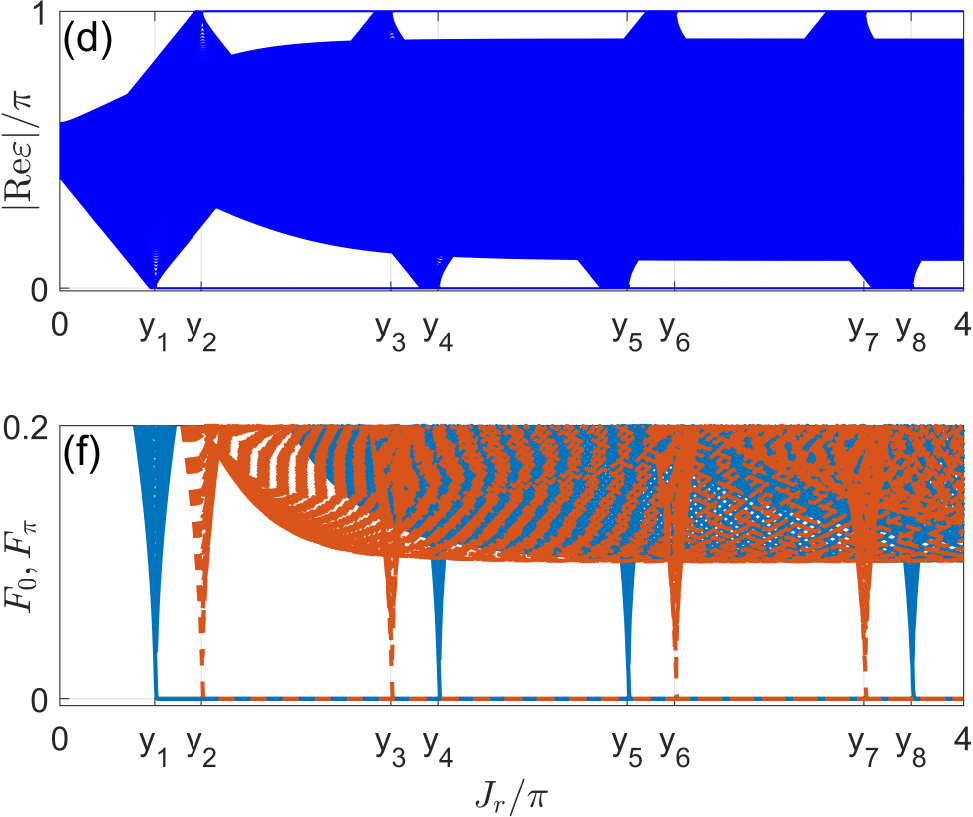}
		\par\end{centering}
	\caption{\textbf{Periodically kicked non-Hermitian Kitaev chain}: topological
		phase diagrams under the PBC {[}in (a) and (b){]}, Floquet spectra
		under the OBC {[}in (c) and (d){]} and quasienergy gap functions under
		the OBC {[}in (e) and (f){]}. The length of lattice is $L=1000$ for
		(c)--(f). In (e) and (f), the solid and dashed lines at $F_{0}=0$
		and $F_{\pi}=0$ denote non-Hermitian Floquet Majorana zero and $\pi$
		edge modes \cite{ZhouNHFTP04}. \label{fig:PKNHKC}}
\end{figure}

Under the PBC and in symmetric time frames, the Floquet operator of
$H(t)$ respects the chiral (sublattice) symmetry ${\cal S}=\sigma_{x}$.
Its bulk topological phases can then be characterized by the winding
numbers $(w_{0},w_{\pi})$ [see Eqs.~(\ref{eq:KickU1})--(\ref{eq:w0p})],
for which we choose
\begin{equation}
	\hat{H}_{0}=\frac{1}{2}\sum_{n}[J(\hat{c}_{n}^{\dagger}\hat{c}_{n+1}+{\rm H.c.})+\mu(2\hat{c}_{n}^{\dagger}\hat{c}_{n}-1)],
\end{equation}
\begin{equation}
	\hat{H}_{1}=\frac{1}{2}\sum_{n}\Delta(\hat{c}_{n}\hat{c}_{n+1}+{\rm H.c.}),
\end{equation}
and assume $\hbar=T=1$. Typical topological phase diagrams of the
system are shown in Figs.~\ref{fig:PKNHKC}(a) {[}for $J\in\mathbb{R}$,
$\mu\in\mathbb{C}$, $(\mu_{r},\Delta)=(0.3\pi,0.5\pi)${]} and \ref{fig:PKNHKC}(b)
{[}for $\mu\in\mathbb{R}$, $J\in\mathbb{C}$, $(\mu,\Delta)=(0.4\pi,0.9\pi)${]}.
Different topological phases correspond to different uniformly colored
regions and the winding numbers $(w_{0},w_{\pi})$ of each phase
are labeled therein. The boundaries between different phases can be
analytically determined from the gap-closing conditions of the system
(see Ref.~\cite{ZhouNHFTP04}). Two notable features are brought
about by the interplay between periodic drivings and non-Hermitian
effects. First, the system could undergo rich topological phase transitions
and enter Floquet superconducting phases with arbitrarily
large topological winding numbers in principle. These phases could support arbitrarily
many zero and $\pi$ Majorana edge modes in the thermodynamic limit,
which might be employed to engineer boundary time crystals and realize
Floquet quantum computing protocols. Second, Floquet superconducting
phases with larger winding numbers and therefore stronger topological
signatures may appear with the increase of the non-Hermitian parameter.
Therefore, new topological states could be induced and stabilized
solely by non-Hermitian effects in Floquet superconducting systems.
This is not expected in the non-driven counterparts of our model,
where nontrivial topological properties are usually destroyed
with the growth of non-Hermitian effects.

\begin{table}
	\begin{centering}
		\begin{tabular}{|c|c|c|c|c|c|}
			\hline 
			$\mu_{i}$ {[}in Fig.~\ref{fig:PKNHKC}(e){]} & $(w_{0},w_{\pi})$ & $(n_{0},n_{\pi})$ & $J_{r}$ {[}in Fig.~\ref{fig:PKNHKC}(f){]} & $(w_{0},w_{\pi})$ & $(n_{0},n_{\pi})$\tabularnewline
			\hline 
			\hline 
			$(0,y_{1})$ & $(4,4)$ & $(8,8)$ & $(0,y_{1})$ & $(0,0)$ & $(0,0)$\tabularnewline
			\hline 
			$(y_{1},y_{2})$ & $(3,4)$ & $(6,8)$ & $(y_{1},y_{2})$ & $(1,0)$ & $(2,0)$\tabularnewline
			\hline 
			$(y_{2},y_{3})$ & $(3,3)$ & $(6,6)$ & $(y_{2},y_{3})$ & $(1,1)$ & $(2,2)$\tabularnewline
			\hline 
			$(y_{3},y_{4})$ & $(3,2)$ & $(6,4)$ & $(y_{3},y_{4})$ & $(1,2)$ & $(2,4)$\tabularnewline
			\hline 
			$(y_{4},y_{5})$ & $(2,2)$ & $(4,4)$ & $(y_{4},y_{5})$ & $(2,2)$ & $(4,4)$\tabularnewline
			\hline 
			$(y_{5},y_{6})$ & $(1,2)$ & $(2,4)$ & $(y_{5},y_{6})$ & $(3,2)$ & $(6,4)$\tabularnewline
			\hline 
			$(y_{6},y_{7})$ & $(1,1)$ & $(2,2)$ & $(y_{6},y_{7})$ & $(3,3)$ & $(6,6)$\tabularnewline
			\hline 
			$(y_{7},y_{8})$ & $(1,0)$ & $(2,0)$ & $(y_{7},y_{8})$ & $(3,4)$ & $(6,8)$\tabularnewline
			\hline 
			$(y_{8},6.5)$ & $(0,0)$ & $(0,0)$ & $(y_{8},4\pi)$ & $(4,4)$ & $(8,8)$\tabularnewline
			\hline 
		\end{tabular}
		\par\end{centering}
	\caption{\textbf{Periodically kicked non-Hermitian Kitaev chain}: bulk topological
		winding numbers $(w_{0},w_{\pi})$ and numbers of Floquet Majorana
		zero and $\pi$ edge modes $(n_{0},n_{\pi})$ in different topological
		phases. The numerical values of phase transition points $y_{1}\sim y_{8}$
		for $\mu_{i}$ can be analytically found and are approximately given
		by 0.64, 1.03, 1.57, 1.94, 2.60, 3.15, 4.41, and 6.11. The numerical
		values of phase transition points $y_{1}\sim y_{8}$ for $J_{r}/\pi$
		can also be found analytically and are given approximately by 0.42,
		0.63, 1.47, 1.68, 2.51, 2.72, 3.56, and 3.77 \cite{ZhouNHFTP04}.
		\label{tab:PKNHKC}}
\end{table}

Under the OBC, the quasienergy spectrum and edge states of the system
can be obtained by diagonalizing the Floquet operator $\hat{U}=e^{-i\hat{H}_{0}}e^{-i\hat{H}_{1}}$
in Majorana or BdG bases \cite{ZhouNHFTP04}. As examples, the real
parts of Floquet spectra and gap functions $(F_{0},F_{\pi})$ versus
$\mu_{i}$ {[}with $(J,\Delta,\mu_{r})=(4\pi,0.5\pi,0.3\pi)${]} and
$J_{r}$ {[}with $(J_{i},\Delta,\mu)=(1,0.4\pi,0.9\pi)${]} are shown in
Figs.~\ref{fig:PKNHKC}(c), \ref{fig:PKNHKC}(e), and Figs.~\ref{fig:PKNHKC}(d),
\ref{fig:PKNHKC}(f), respectively. $y_{1}\sim y_{8}$ refer to gap
closing points of the Floquet spectrum under the PBC (see Table \ref{tab:PKNHKC}).
The gap functions are defined as
\begin{equation}
	F_{0}=\frac{1}{\pi}\sqrt{({\rm Re}\varepsilon)^{2}+({\rm Im}\varepsilon)^{2}},\qquad F_{\pi}=\frac{1}{\pi}\sqrt{(|{\rm Re}\varepsilon|-\pi)^{2}+({\rm Im}\varepsilon)^{2}},\label{eq:F0P}
\end{equation}
where $\varepsilon$ includes all the quasienergies of $\hat{U}$
under the OBC. The eigenmodes with $F_{0}=0$ and $F_{\pi}=0$ thus
have the quasienergies zero and $\pi$, respectively. In Table \ref{tab:PKNHKC},
we list the numbers of Floquet Majorana edge modes $(n_{0},n_{\pi})$
with the quasienergies $(0,\pi)$ and the bulk topological winding
numbers $(w_{0},w_{\pi})$. A simple bulk-edge correspondence can
be further found between these numbers, i.e.,
\begin{equation}
	n_{0}=2|w_{0}|,\qquad n_{\pi}=2|w_{\pi}|.\label{eq:PKNHKC-BBC}
\end{equation}
These relations hold in other parameter regions of our model as well. We conclude
that our non-Hermitian Floquet Kitaev chain could indeed possess many
Majorana zero and $\pi$ edge modes due to its large winding numbers.
The localization nature of these Majorana modes was also demonstrated
in Ref.~\cite{ZhouNHFTP04}.

Therefore, our study in Ref.~\cite{ZhouNHFTP04} established the topological
characterization and bulk-edge correspondence of 1D non-Hermitian
Floquet topological superconductors that belong to an extended
BDI symmetry class. The collaboration between time-periodic driving
fields and non-Hermitian effects was found to produce rich Floquet
superconducting phases with large topological winding numbers and
many Majorana edge modes at two distinct quasienergies zero and $\pi$.
These non-Hermitian Floquet Majorana modes might allow Floquet quantum
computing schemes to be more robust to environmental-induced nonreciprocity,
dissipation, and quasiparticle poisoning effects. The existence of
many pairs of Floquet Majorana modes may also create stronger transport
signals at the ends of the chain, making it easier
to experimentally detect their topological properties in open-system settings
\cite{NHFTP19}. In future studies, topics like non-Hermitian Floquet topological
superconductors under different driving protocols, in other symmetry
classes, with more complicated lattice effects (e.g., sublattice structures,
long-range hoppings or disorder), in higher spatial dimensions and
with NHSEs, deserve to be explored. Possible changes in topological
classifications due to many-body effects in non-Hermitian Floquet
superconductors are also awaited to be revealed.

\subsection{Non-Hermitian Floquet quasicrystals\label{subsec:TQC}}

In this subsection, we go beyond the clean limit of driven non-Hermitian
lattices and showcase that the interplay among correlated disorder,
temporal driving, and non-Hermitian effects could yield Floquet quasicrystals
with rich PT-symmetry breaking transitions, localization transitions,
and topological phase transitions. We will consider systems under
both high-frequency and near-resonant driving fields.

\begin{figure}
	\begin{centering}
		\includegraphics[scale=0.65]{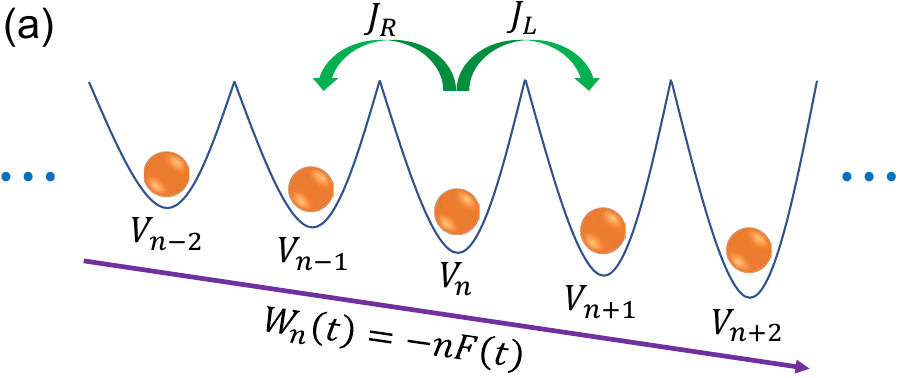}
		\par\end{centering}
	\medskip{}
	
	\begin{centering}
		\includegraphics[scale=0.6]{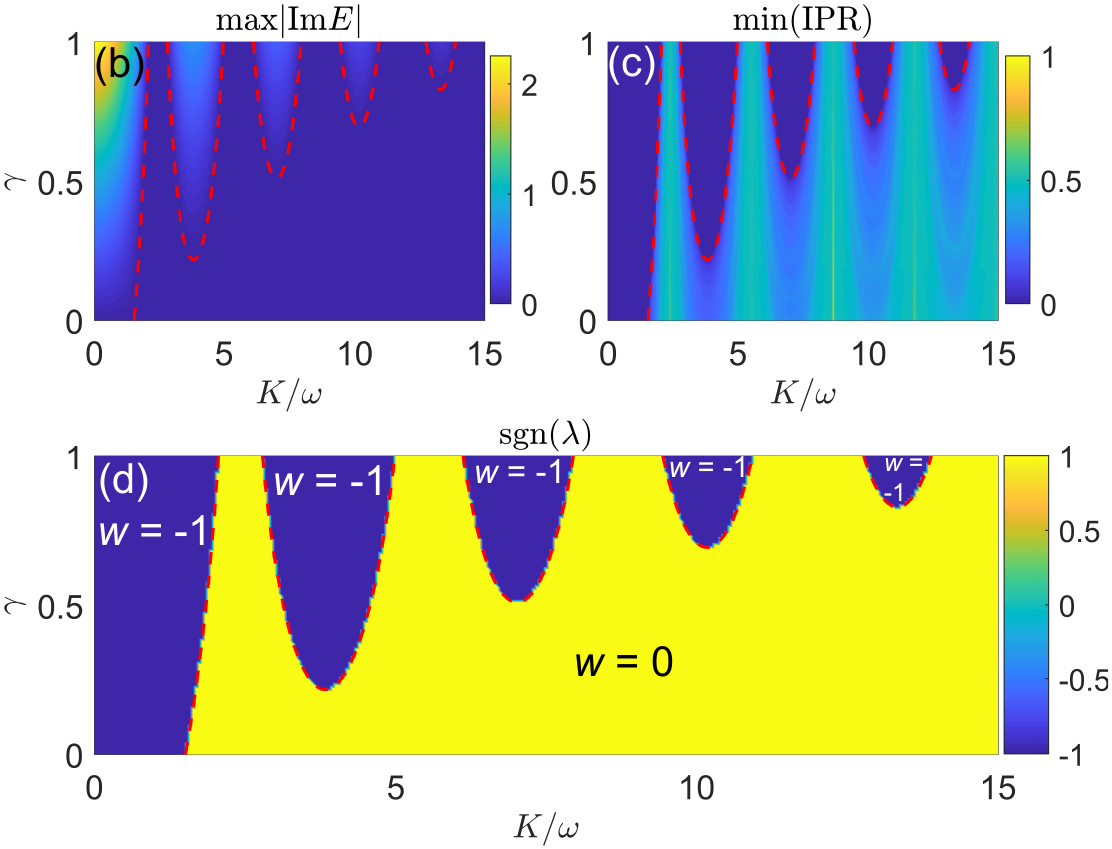}
		\par\end{centering}
	\caption{\textbf{Floquet NHQC under high-frequency driving forces}: (a) a schematic
		diagram of the lattice model, (b) PT transitions of the Floquet spectrum,
		(c) localization transitions, and (d) topological transitions \cite{ZhouNHFTP09}.
		\label{fig:FNHQC1}}
\end{figure}

We start with the example of a 1D non-Hermitian quasicrystal under
high-frequency harmonic driving forces. The time-periodic Hamiltonian
of the model takes the form of \cite{ZhouNHFTP09}
\begin{equation}
	\hat{H}(t)=\sum_{n}\{J(e^{-\gamma}\hat{c}_{n}^{\dagger}\hat{c}_{n+1}+e^{\gamma}\hat{c}_{n+1}^{\dagger}\hat{c}_{n})+[V\cos(2\pi\alpha n)-nK\cos(\omega t)]\hat{c}_{n}^{\dagger}\hat{c}_{n}\}.\label{eq:HtNHQC1}
\end{equation}
An illustration of the model is given in Fig.~\ref{fig:FNHQC1}(a).
Here $\gamma\in\mathbb{R}$ and $Je^{-\gamma}$ ($Je^{\gamma}$) describes
the right-to-left (left-to-right) nearest-neighbor hopping amplitude.
The hopping is nonreciprocal and thus $\hat{H}(t)\neq\hat{H}^{\dagger}(t)$
if $\gamma\neq0$. $\hat{c}_{n}^{\dagger}$ ($\hat{c}_{n}$) creates
(annihilates) a particle on the lattice site $n$. $V\in\mathbb{R}$
is the amplitude of an onsite potential, and $\alpha$ is chosen to
be irrational in order for the potential to be spatially quasiperiodic.
$K\in\mathbb{R}$ is the driving amplitude and $\omega$ is the driving
frequency. This model can be viewed as a Floquet and quasicrystal
variant of the Hatano-Nelson model \cite{NHDsod01}. In the absence
of the driving force and under the PBC, all the eigenstates of the
system are extended (localized) with complex (real) eigenvalues if
$|V|<|2J|e^{|\gamma|}$ ($|V|>|2J|e^{|\gamma|}$) \cite{NHQC02}.
The system could thus undergo a PT transition, a localization-delocalization
transition, and also a topological transition accompanied by the quantized
change of its spectral winding number {[}Eq.~(\ref{eq:EWN}){]} at
$|V|=|2J|e^{|\gamma|}$ \cite{NHQC02}.

Transforming the Hamiltonian $\hat{H}(t)$ in Eq.~(\ref{eq:HtNHQC1})
to the rotating frame and applying the method discussed in 
Subsec.~\ref{subsec:HFA}, we can obtain the Floquet effective Hamiltonian
of our system in the high-frequency limit \cite{ZhouNHFTP09}, i.e.,
\begin{equation}
	\hat{H}_{{\rm F}}=\sum_{n}\left[{\cal J}_{0}\left(\frac{K}{\omega}\right)J(e^{-\gamma}\hat{c}_{n}^{\dagger}\hat{c}_{n+1}+e^{\gamma}\hat{c}_{n+1}^{\dagger}\hat{c}_{n})+V\cos(2\pi\alpha n)\hat{c}_{n}^{\dagger}\hat{c}_{n}\right].\label{eq:HeffNHQC1}
\end{equation}
Here ${\cal J}_{0}(K/\omega)$ is the Bessel function of first kind,
which is a non-monotonous function of $K/\omega$. Under the PBC,
the Floquet NHQC described by $\hat{H}_{{\rm F}}$ could thus undergo
multiple and reentrant PT, localization and topological transitions
with the change of the ratio $K/\omega$ between the driving amplitude
and driving frequency whenever \cite{ZhouNHFTP09}
\begin{equation}
	|V|=|2J{\cal J}_{0}(K/\omega)|e^{|\gamma|}.\label{eq:PbNHQC1}
\end{equation}
This is indeed the case, as demonstrated by the maximum of the imaginary
parts of quasienergies, the minimum of IPRs, and the winding numbers
in Figs.~\ref{fig:FNHQC1}(b)--(d) {[}for $(J,V,\alpha)=(1,1,\frac{\sqrt{5}-1}{2})$
and the length of lattice $L=610${]}. In the metallic phase, we have
$\max|{\rm Im}E|>0$, $\min({\rm IPR})\rightarrow0$, and the winding
number $w=-1$, which means that all the Floquet eigenstates of $\hat{H}_{{\rm F}}$
are extended with complex quasienergies. In the insulating phase,
we have $\max|{\rm Im}E|=0$, $\min({\rm IPR})>0$, and the winding
number $w=0$, implying that all the Floquet eigenstates of $\hat{H}_{{\rm F}}$
are localized with real quasienergies. The boundaries between these
phases {[}red dashed lines in Figs.~\ref{fig:FNHQC1}(b)--(d){]}
are precisely depicted by Eq.~(\ref{eq:PbNHQC1}). The Lyapunov
exponent $\lambda=\ln|(Ve^{-|\gamma|})/[2J{\cal J}_{0}(K/\omega)]|$
is quasienergy-independent for all the Floquet eigenstates, which
is positive (negative and thus ill-defined) in the localized (extended)
phases \cite{ZhouNHFTP09}. Therefore, it is clear that the phases
and transitions in the quasicrystal Hatano-Nelson model can be significantly
modified by Floquet driving fields even in the high-frequency limit.
The periodic driving also provides us with a flexible knob to control
and engineer different types of phase transitions in NHQCs, with further
examples discussed in Ref.~\cite{ZhouNHFTP09}.

\begin{figure}
	\begin{centering}
		\includegraphics[scale=0.6]{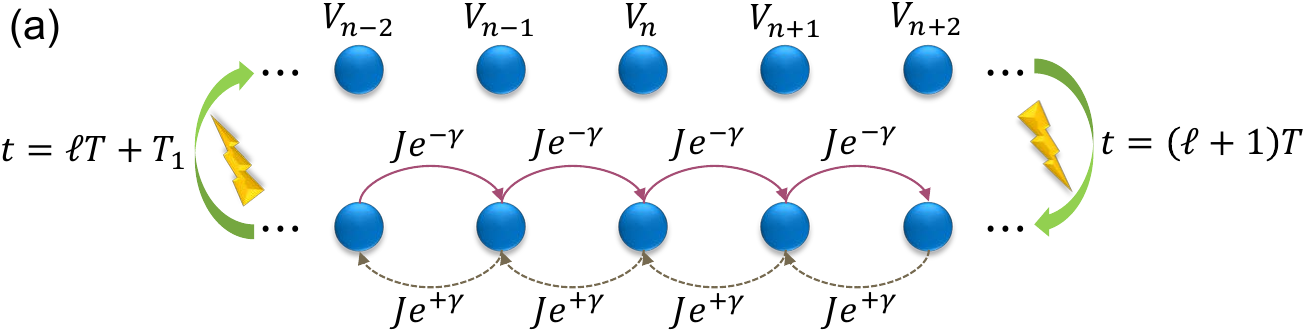}
		\par\end{centering}
	\medskip{}
	
	\begin{centering}
		\includegraphics[scale=0.38]{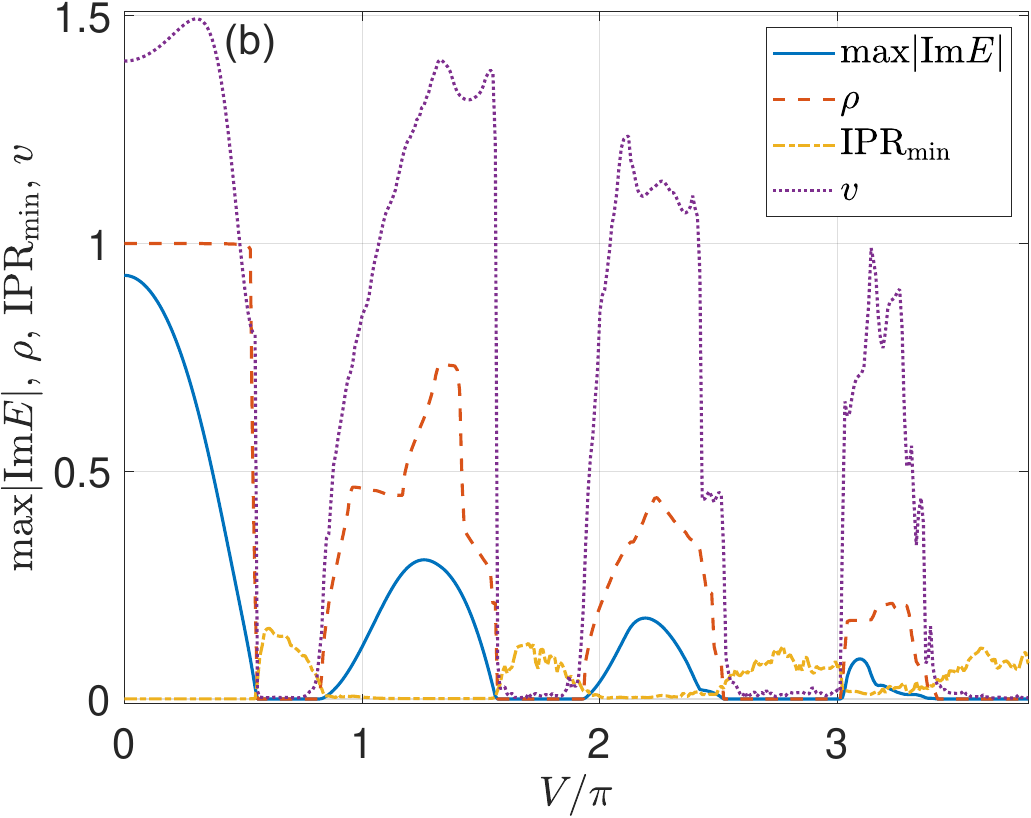}$\,\,$\includegraphics[scale=0.395]{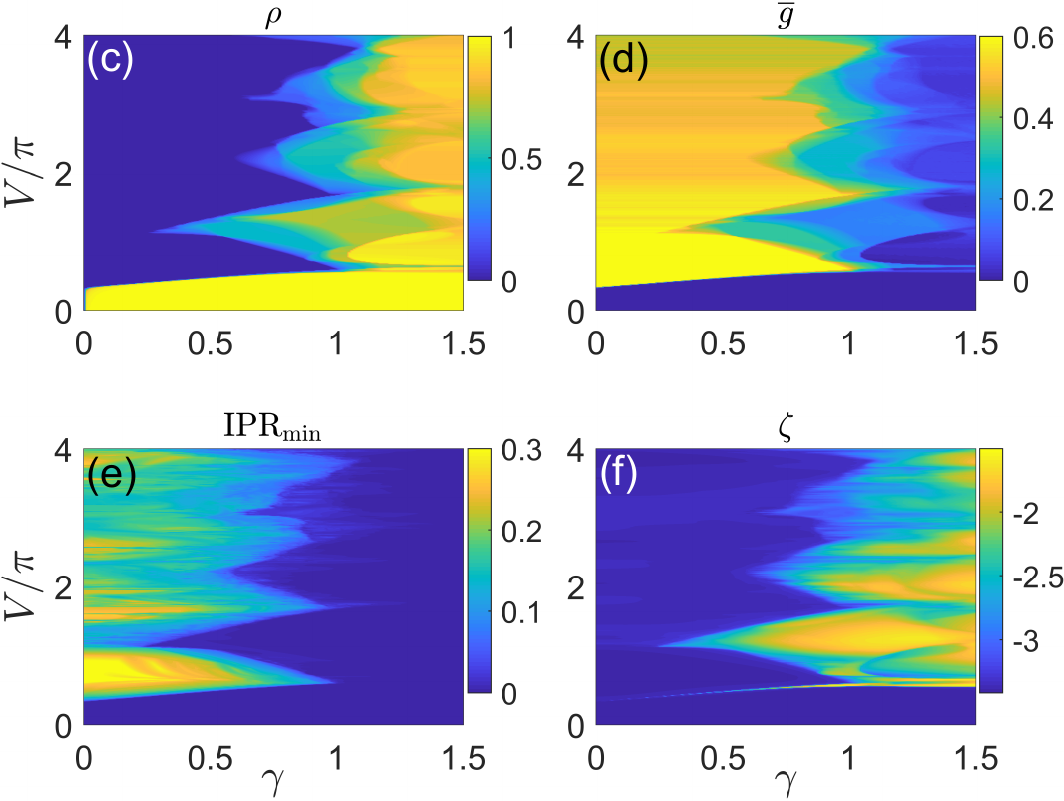}
		\par\end{centering}
	\caption{\textbf{Floquet NHQC under time-periodic kickings}: (a) a schematic
		diagram of the lattice model, (b)--(f) PT transitions of the Floquet
		spectrum, localization transitions of Floquet states and critical
		phases with mobility edges of the system under the PBC. The initial
		state $|\psi(0)\rangle=\sum_{n}\delta_{n0}|0\rangle$ for the calculation
		of spreading speed $v=v(t)$ is chosen to be localized at the center of the lattice,
		and the time average in Eq.~(\ref{eq:DelxVt}) is taken over $t=1000$
		driving periods \cite{ZhouNHFTP11}. \label{fig:FNHQC2}}
\end{figure}

We move on to the example of an NHQC under near-resonant driving fields.
In this case, the interplay between driving and non-Hermitian effects
not only induces multiple and reentrant localization transitions but
also generates critical mobility edge phases that are absent in non-driven
limits. A schematic diagram of our model is shown in Fig.~\ref{fig:FNHQC2}(a).
Its Hamiltonian takes the form of $\hat{H}(t)=\hat{K}$ for $t\in[\ell T,\ell T+T_{1})$
and $\hat{H}(t)=\hat{V}$ for $t\in[\ell T+T_{1},\ell T+T_{1}+T_{2})$,
where $\ell\in\mathbb{Z}$ and the driving period $T=T_{1}+T_{2}$.
The system is piecewisely quenched between $\hat{K}=J\sum(e^{\gamma}\hat{c}_{n}^{\dagger}\hat{c}_{n+1}+e^{-\gamma}\hat{c}_{n+1}^{\dagger}\hat{c}_{n})$
and $\hat{V}=V\sum_{n}\cos(2\pi\alpha n)\hat{c}_{n}^{\dagger}\hat{c}_{n}$
over each driving period. For an irrational $\alpha$, we thus arrive
at a periodically quenched, spatially quasiperiodic variant of the
Hatano-Nelson model, whose Floquet operator is given by \cite{ZhouNHFTP11}
\begin{equation}
	\hat{U}=e^{-i\mathsf{V}\sum_{n}\cos(2\pi\alpha n)\hat{c}_{n}^{\dagger}\hat{c}_{n}}e^{-i\mathsf{J}\sum(e^{\gamma}\hat{c}_{n}^{\dagger}\hat{c}_{n+1}+e^{-\gamma}\hat{c}_{n+1}^{\dagger}\hat{c}_{n})},
\end{equation}
where $\mathsf{V}=VT_{2}/\hbar$ and $\mathsf{J}=JT_{1}/\hbar$. Solving
the eigenvalue equation $\hat{U}|\psi\rangle=e^{-iE}|\psi\rangle$
and using the tools introduced in Subsec.~\ref{subsec:LT}, we could
obtain the maximal imaginary parts of quasienergies $\max|{\rm Im}E|$,
the density of states with complex eigenvalues $\rho$, the minimum
of IPRs and the spreading velocity of an initially localized wavepacket
$v$. A collection of these quantities versus the strength of quasiperiodic
potential $V$ is shown in Fig.~\ref{fig:FNHQC2}(b) for a typical
case {[}with $(J,\gamma,\alpha)=(\pi/6,0.8,\frac{\sqrt{5}-1}{2})$
and the length of lattice $L=4181${]}. We observe that with the increase
of $V$ from $V=0$, the system first undergoes a complex-to-real
PT transition in its quasienergy spectrum, which is also accompanied
by a localization transition of all its Floquet eigenstates from spatially
extended to localized. Interestingly, with the further enhancement
of the quasiperiodic potential $V$ (thus with stronger correlated
disorder), some localized Floquet states can again become extended
with real eigenvalues. The system then enters a critical phase, in
which extended and localized eigenstates coexist and are separated
by mobility edges on the complex quasienergy plane (see also the
Fig.~3 of Ref.~\cite{ZhouNHFTP11}). The further increase of $V$ leads
to reentrant transitions between localized and critical mobility edge
phases in the system. This is also reflected by the two-parameter
phase diagrams in Figs.~\ref{fig:FNHQC2}(c)--(f) {[}see 
Subsec.~\ref{subsec:HFA} for the definitions of $\rho$, $\overline{g}$,
${\rm IPR}_{\min}$, and $\zeta${]}. Note that both the critical phases
and the reentrant localization transitions are absent in the non-driven
Hatano-Nelson quasicrystal. They are brought about by the nearest-resonant
Floquet driving field. It induces long-range spatial couplings and
quasienergy windings in the system, thus yielding the observed phenomena
(see Ref.~\cite{ZhouNHFTP11} for more detailed discussions).

In experiments, Floquet NHQCs might be realized by ultracold atoms
in driven and quasiperiodic optical superlattices with particle losses
\cite{SkinExp06,NHFTP16}. Static NHQCs have also been realized by
non-unitary photonic quantum walks \cite{NHQC04,NHQC05}. Signatures
of PT breaking transitions, localization transitions, mobility edges,
and topological properties related to NHSEs can be extracted from
dynamical observables of initially localized wavepackets \cite{NHQC04,NHQC05}.
Since quantum walk models are intrinsically dynamical, they provide
natural platforms to realize and detect the Floquet NHQCs reviewed
in this subsection. Meanwhile, as the discrete-time quantum walk carries
a synthetic spin-half degree of freedom, it may also be utilized
to simulate other types of NHQCs, such as those with lattice dimerizations
or non-Abelian potentials \cite{ZhouNHQC1,ZhouNHQC2,ZhouNHQC3}. Topological
Anderson insulators induced by uncorrelated disorder could also be
explored in similar settings \cite{NHDsod05}.

Overall, we find that both high-frequency and near-resonant driving
fields could be used to trigger, control, and enrich the phases and
transitions in NHQCs, and even create unique Floquet NHQC phases that
are absent in the static limit \cite{ZhouNHFTP09,ZhouNHFTP11}. Non-Hermitian
disordered systems thus provide further playgrounds for the exploration
of new physics that are enabled by Floquet engineering and time-periodic
driving fields.

\section{Conclusion and outlook\label{sec:Sum}}

In this review, we recapitulated some progress we made in the study
of intriguing topological phases in non-Hermitian Floquet systems.
After a general introduction about the backgrounds and motivation,
we first formulated the theoretical frameworks underpinning our study.
These include the Floquet theory applicable in general situations
and some approximation schemes suitable for exploring fast and slowly
driven systems. Efficient means of characterizing the symmetry, topology,
dynamical, and localization nature of non-Hermitian Floquet systems
were then introduced. Equipped with these tools, we discussed prototypical
examples of non-Hermitian Floquet topological phases in insulating,
superconducting, and quasicrystalline systems, where the PT transitions,
topological transitions, bulk-edge/corner correspondences and localization
transitions within these non-Hermitian Floquet phases of matter were systematically
characterized. Therefore, the collection of our works \cite{ZhouNHFTP01,ZhouNHFTP02,ZhouNHFTP03,ZhouNHFTP04,ZhouNHFTP05,ZhouNHFTP06,ZhouNHFTP07,ZhouNHFTP08,ZhouNHFTP09,ZhouNHFTP10,ZhouNHFTP11}
lays a solid foundation for the study of topological phenomena in
non-Hermitian Floquet systems and further uncovers the richness of
topological phases that could emerge due to the interplay between
periodic drivings and non-Hermitian effects.

We note that some other schemes concerning the topological classification
and bulk-edge correspondence of non-Hermitian Floquet matter were
discussed in Refs.~\cite{NHFTP17,NHFTP21,NHFTP23,NHFTP29,NHFTP37}.
The EPs and gapless topological phases in non-Hermitian Floquet systems
were also considered in Refs.~\cite{NHFTP08,NHFTP24,NHFTP27,NHFTP30,NHFTP32,NHFTP36,NHFTP39}.
From the perspective of quantum control, the periodic driving fields
may provide efficient means to stabilize non-Hermitian systems and
control the PT transitions therein \cite{NHFTP03}. Some interesting
aspects regarding the quantum dynamics and anomalous diffusion were
revealed in several chaotic non-Hermitian Floquet systems \cite{NHFTP20,NHFTP26,NHFTP35,NHFTP40}.
Aspects of PT-symmetry in non-Hermitian Floquet phases were also investigated
in Refs.~\cite{NHFTP02,NHFTP09,NHFTP10,NHFTP14,NHFTP25}. On the experimental
side, the quantum walks in cold atoms, photonic setups, and circuit
QED provide useful frameworks to realize and detect non-Hermitian
Floquet topological matter \cite{NHQC04,NHQC05,NHFTP01,NHFTP04,NHFTP05,NHFTP06,NHFTP07,NHFTP15,NHFTP16,NHFTP34}.

Even with all the progress mentioned above, our knowledge about non-Hermitian
Floquet topological matter is still quite limited. This is especially
the case for systems beyond one spatial dimension, with uncorrelated
disorders, and with many-body interactions. Dynamical quantum phase
transitions may exhibit unique critical signatures triggered by the
non-Hermitian Floquet exceptional topology \cite{ZhouNHDQPT1,ZhouNHDQPT2,ZhouFDQPT1,ZhouFDQPT2,ZhouFDQPT3}.
The entanglement and transport properties of non-Hermitian Floquet
systems \cite{ZhouFESEE1} may also deviate significantly from their
Hermitian or static counterparts. The exact connection between the
topological phases of non-Hermitian Floquet Hamiltonians and the Floquet
master equation (or Floquet Liouvillian) of driven open systems is
unclear. More experimental efforts deserve to be made in order to
realize and observe non-Hermitian Floquet phases with large topological
invariants, many topological edge states, and multiple topological
phase transitions. All these facts indicate that the area of non-Hermitian
Floquet matter is still in its infancy, and further substantial developments
are eagerly needed in this area.

\section*{Acknowledgments}
This research was funded by the National Natural Science Foundation of China (Grants Nos. 12275260, 11905211 and 12275155), the Fundamental Research Funds for the Central
Universities (Grant No. 202364008), the Young Talents
Project of Ocean University of China, and the Qilu Young Scholar Program of Shandong University.

\end{document}